\documentclass[]{spie}  

\usepackage{amsmath,amsfonts,amssymb}
\usepackage{graphicx}
\usepackage[colorlinks=true, allcolors=blue]{hyperref}

\title{On-sky Fibre-Target-Alignment of the 4MOST instrument: calibration and performance}

\author[a]{Roland Winkler}
\author[a]{Weijia Sun}
\author[a]{Daniel Sablowski}
\author[a]{Thomas Liebner}
\author[a]{Ole Streicher}
\author[a]{Steffen Frey}
\author[b]{Ingo Stilz}
\author[b]{Alexander Pramskiy}
\author[c]{Scott Smedley}
\author[d]{Thomas Szeifert}
\author[d]{Diogo Rio Fernandes}
\author[d]{Gerard Zins}

\affil[a]{Leibniz-Institut f\"ur Astrophysik Potsdam; An der Sternwarte 16, 14482 Potsdam, Germany}
\affil[b]{Landessternwarte Heidelberg; K\"onigstuhl 12, 69117 Heidelberg, Germany}
\affil[c]{Australian Astronomical Optics, Faculty of Science and Engineering, Macquarie University, NSW 2109, Australia}
\affil[d]{European Southern Observatory, Karl-Schwarzschild-Stra\ss e, 85748 Garching, Germany}

\authorinfo{Further author information: (Send correspondence to Roland Winkler)\\R.Winkler: E-mail: rwinkler@aip.de}

\begin{document} 
\maketitle

\begin{abstract}
The 4-metre Multi-Object Spectroscopic Telescope (4MOST) is a new wide-field, fibre-fed spectroscopic survey facility for the VISTA telescope at ESO’s Paranal Observatory. The instrument enables the simultaneous acquisition of 2436 spectra across a 4.2 deg² field of view, using a tilting spine fibre positioner feeding three dedicated spectrographs.
In this paper, we describe the calibration process, and performance verification of the Fibre-Target-Alignment (FTA) process for 4MOST.\par

We show the complete FTA process, including calibration of the individual hardware- and software components.
Namely the Metrology camera system, the Fibre Positioner AESOP, a spine based Secondary Guiding System, the sky to focal surface projection software, and residual minimization via raster scans.\par

In total, the FTA system required one special tool, a large calibration target for the focal surface, and approximately 1 month of accumulated calibration work on the telescope.
The FTA process reached approx. $24 \mu m$ ($0.4 ''$) RMS distance between fibres and targets on sky about 3 weeks after installation of the final hardware components of 4MOST, which is when 4MOST had its first light event.
By the time of writing this paper, i.e. 6 months later, we reach approx. $16 \mu m$ ($0.27 ''$) RMS.
Currently, we far exceed our requirements in terms of accuracy, and are doing trade-off studies to maximize scientific returns.
\end{abstract}


\newpage
\section{THE METROLOGY SYSTEM IN VISTA/4MOST} \label{sec:metro}

\begin{figure}[h!]
\begin{center}
\includegraphics[width=0.8\textwidth]{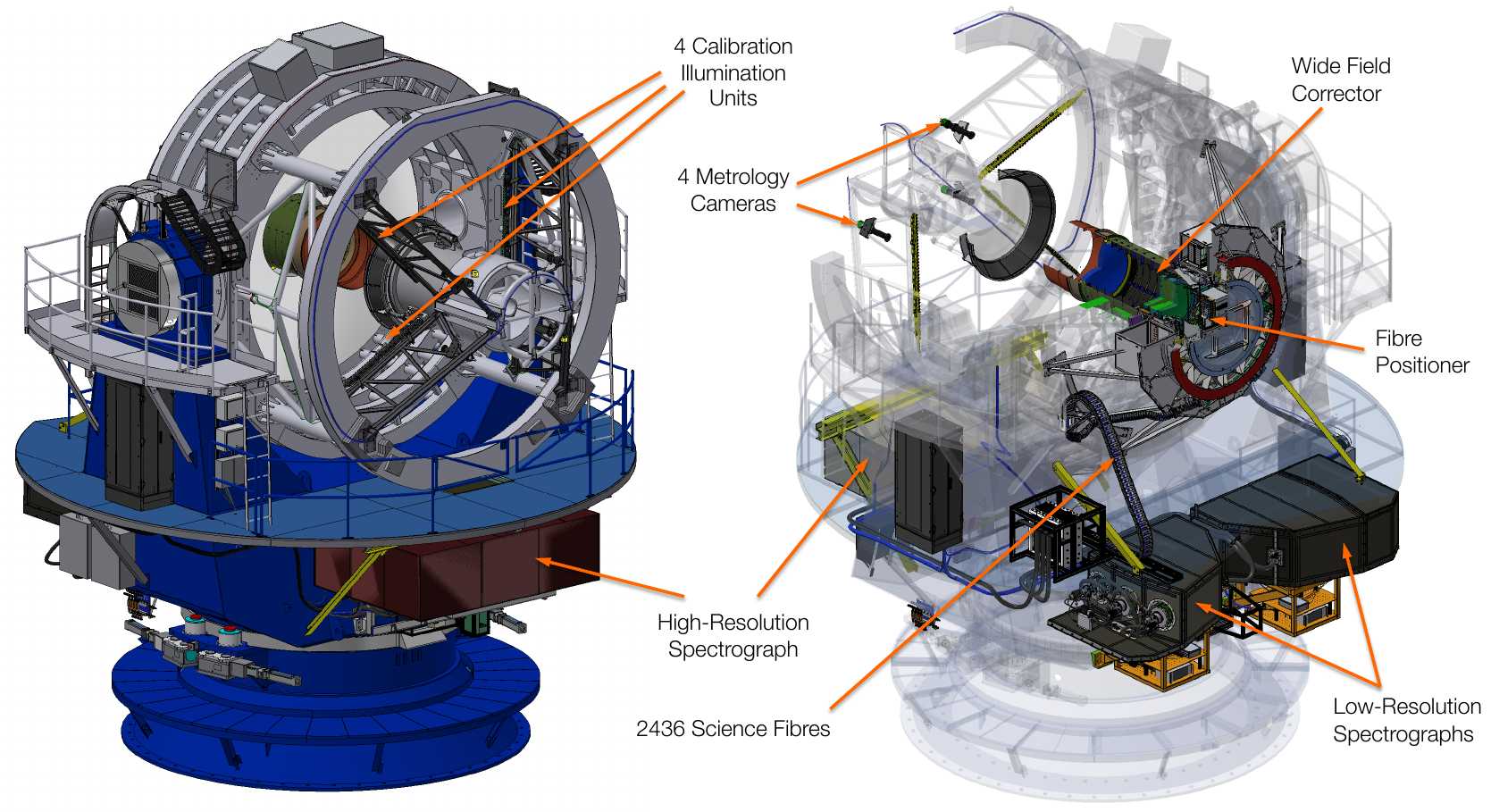}
\end{center}
\caption[4MOST Overview]{
    \label{fig:4mostvista}
    Overview of the 4MOST instrument, integrated in the VISTA telescope. 3 of the 4 Metrology cameras are visible at the Spider.
}
\end{figure}

The 4MOST instrument \citenum{deJong2024} is deeply integrated in the VISTA Telescope, its components are shown in Figure~\ref{fig:4mostvista} with the Telescope being grayed out on the right-hand side.

\subsection{FTA Hardware Components} \label{sec:hardware}

To position the fibres in the 4MOST instrument, the AESOP Fibre Positioner \citenum{brzeski2022aesop} actuates 2436 science fibres and 12 guide fibre bundles of 7 fibres each on 2448 spines.
The AESOP positioner does not contain any encoders to actively control the motion of spine, instead the Metrology System is used to give a closed loop feedback.\par

\begin{figure}[h]
\begin{center}
\includegraphics[width=\textwidth]{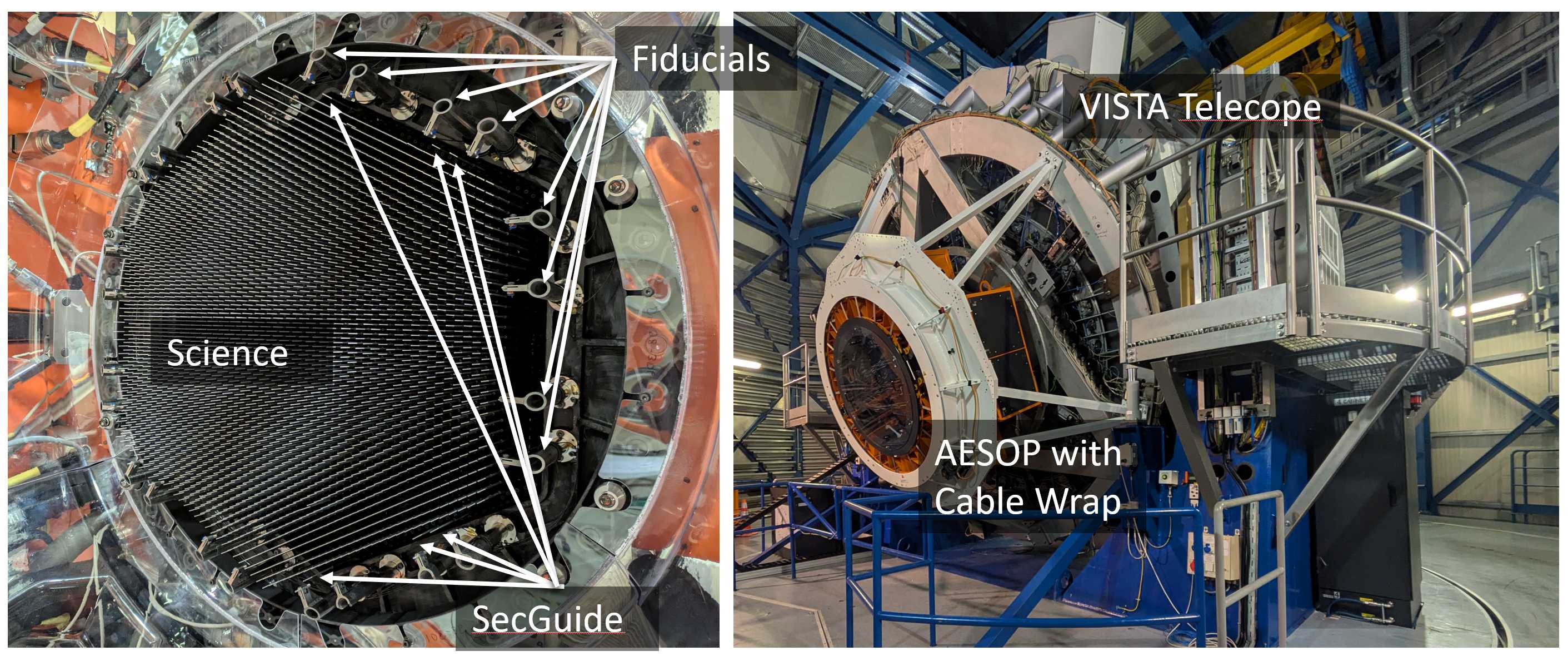}
\end{center}
\caption[AESOP Overview]{
    \label{fig:aesopvista}
    Left: The AESOP fibre positioner. Right: Back of the VISTA telescope with 4MOST.
}
\end{figure}

\begin{figure}[h]
\begin{center}
\includegraphics[height=7cm]{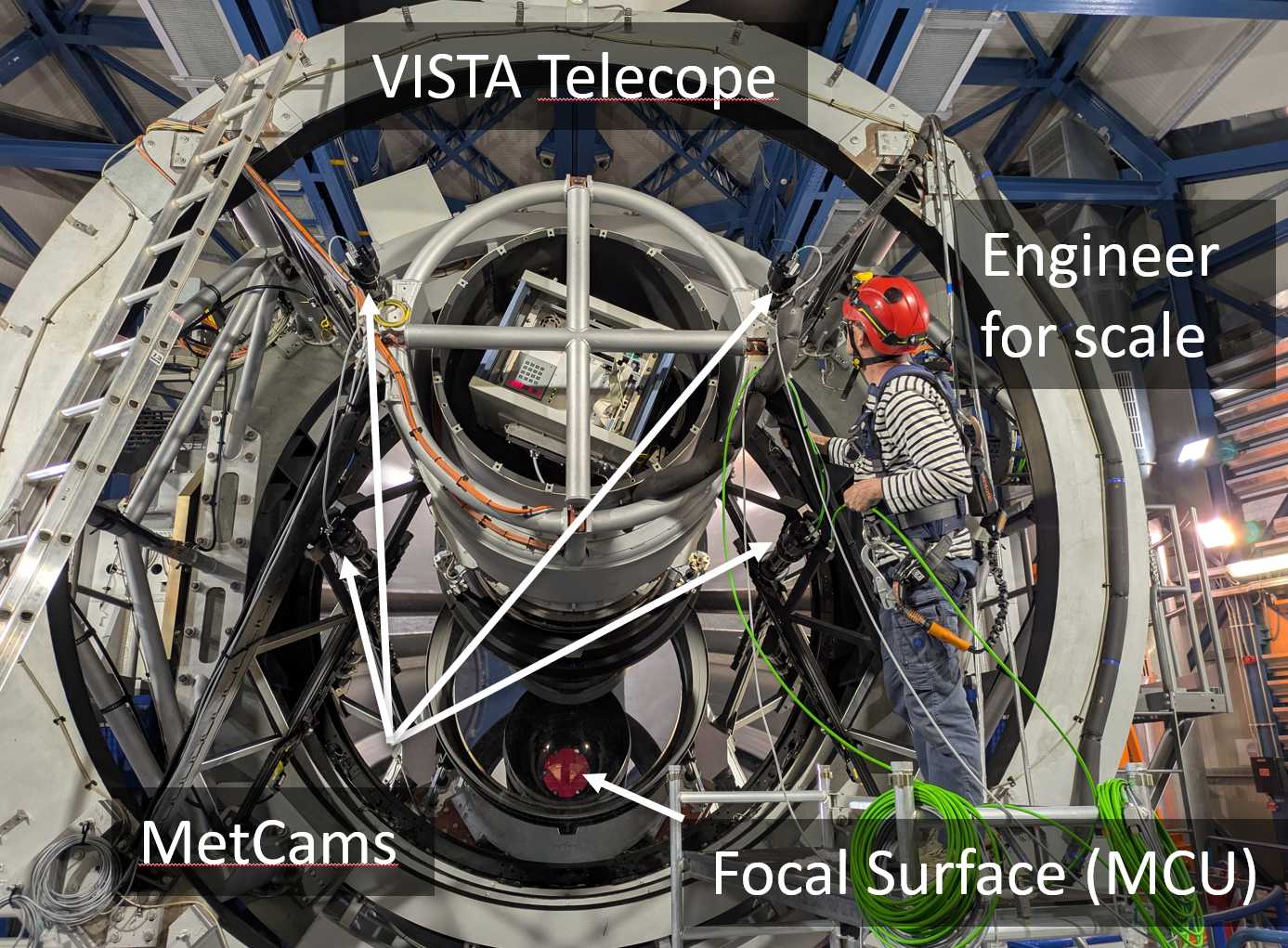}
\end{center}
\caption[MetCams Overview]{
    \label{fig:metcamvista}
    Front view of the VISTA telescope with MetCams during installation.
}
\end{figure}

The Metrology System consists of 4 cameras that are mounted on the M2 spider of the VISTA telescope (see Figure~\ref{fig:metcamvista}), a Back-Illumination unit, and various hardware for calibration.
Each camera observes the entire field of the fibre area, providing 4 independent measurements of all fibres, including 24 additional fiducial fibres which are mounted on fix metal posts that are not movable and provide a local reference frame.
An image of the AESOP positioner is provided on the left-hand side of Figure~\ref{fig:aesopvista}.\par

The fibres have $85 \mu m$ diameter cores which are mounted in spines that have $750 \mu m$ diameter glass ferrule tips.
There is no room for separate identification markers, as for example in the MOONS instrument \citenum{cirasuolo2020crescent}.
Instead, the 4MOST Metrology System relies on measuring the location of the fibre cores directly, which need to be illuminated from the spectrograph side.\par

\begin{figure}[h]
\begin{center}
\includegraphics[height=8cm]{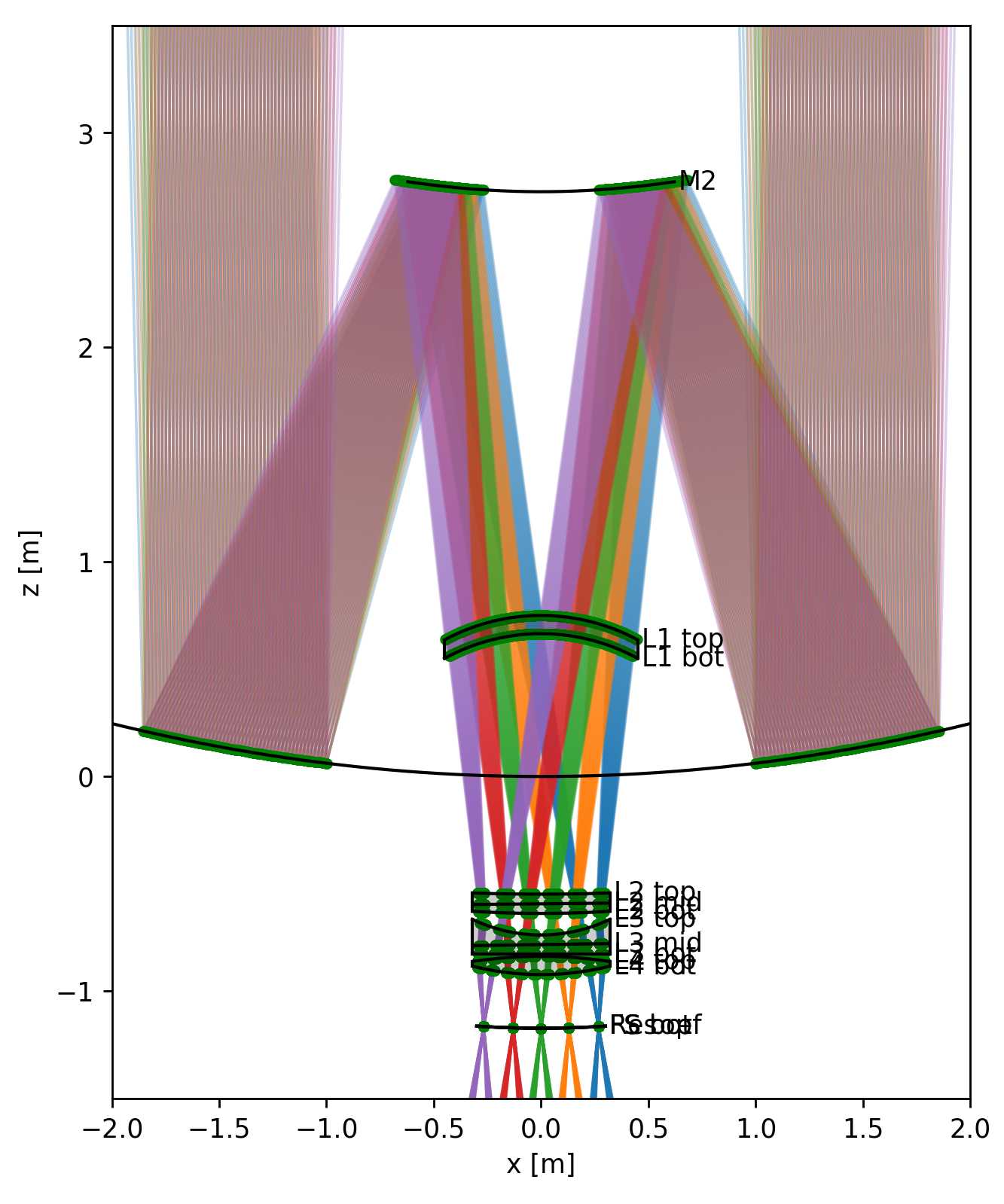}
\includegraphics[height=8cm]{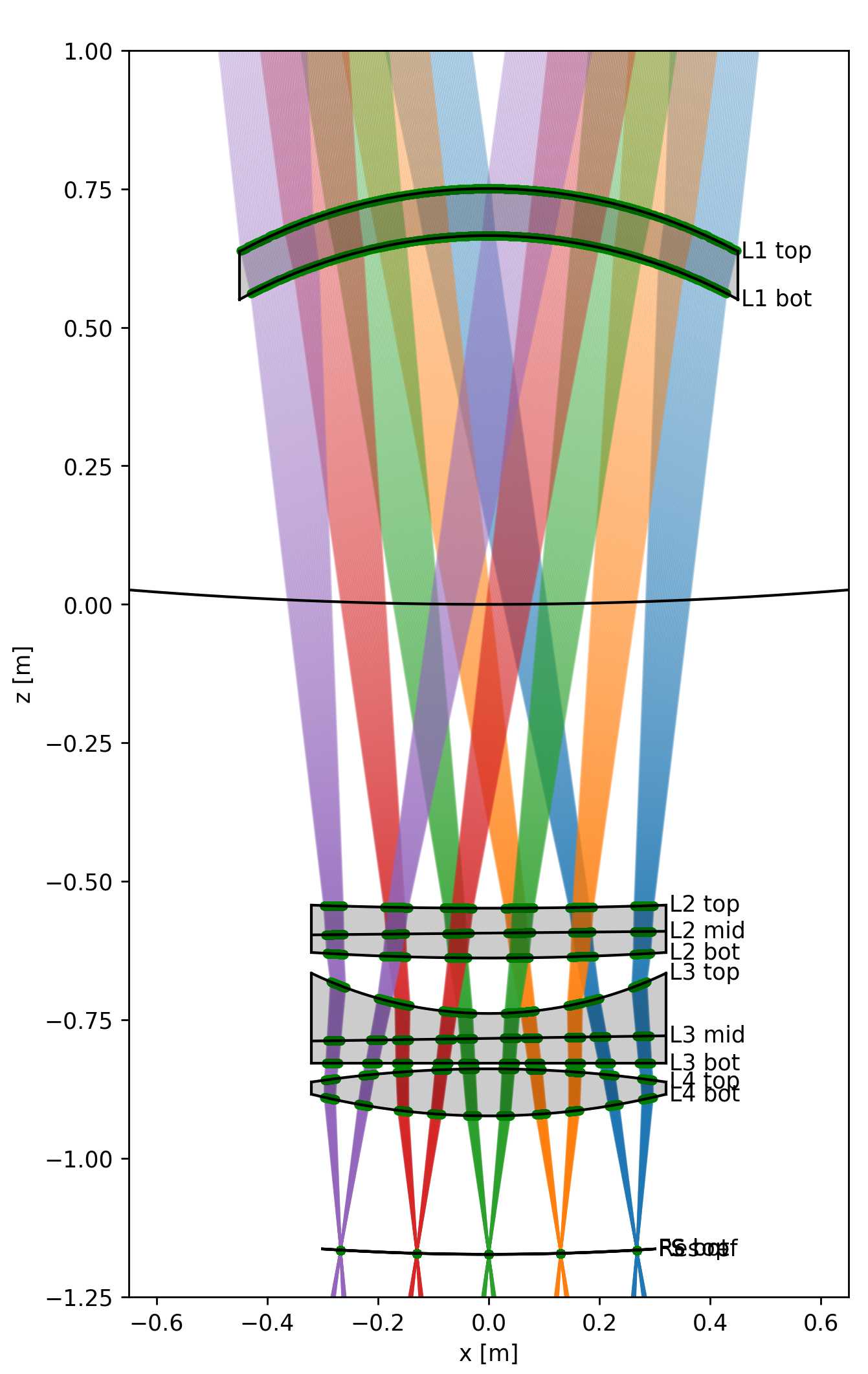}
\includegraphics[height=8cm]{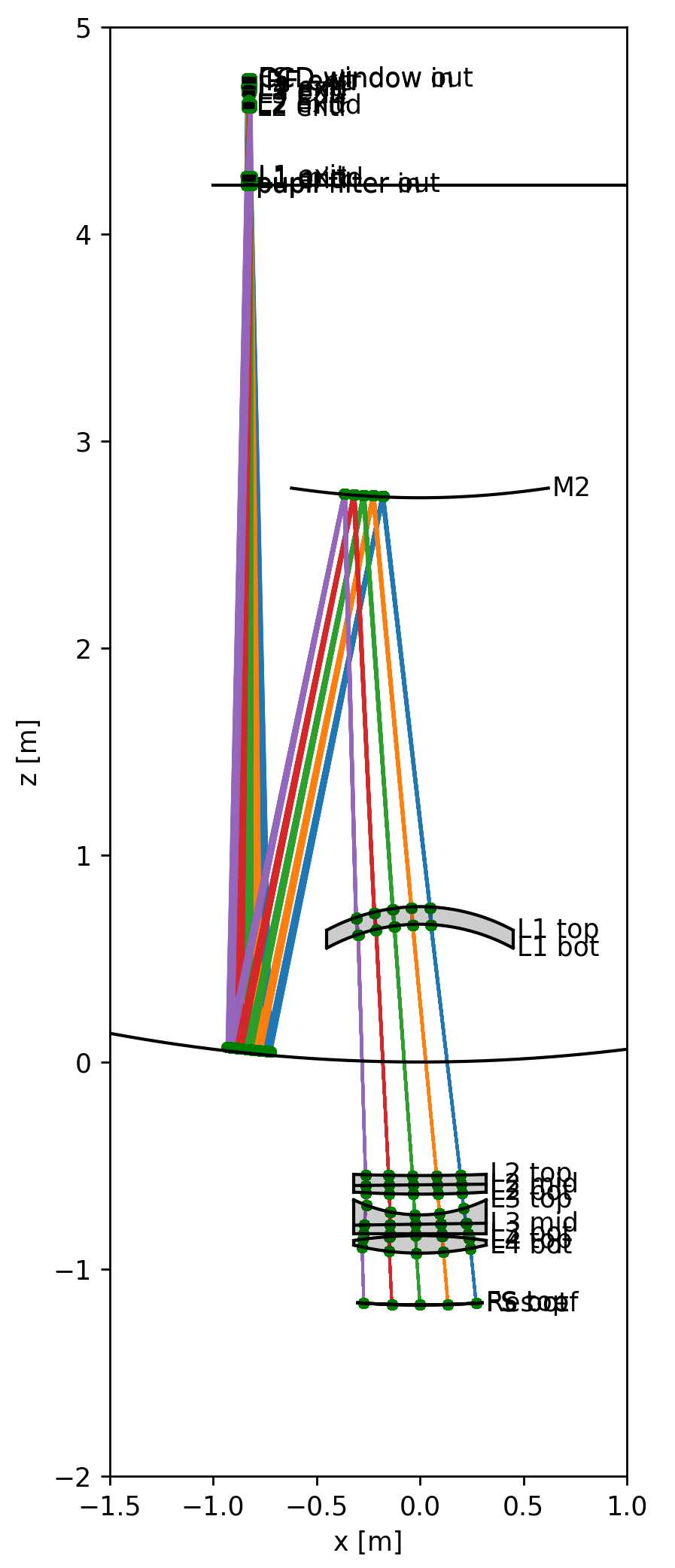}
\end{center}
\caption[Sky and MetCam Raytrace]{
    \label{fig:raytrace}
    Left 2 panels: Ray-trace of parallel light from the sky to the focal surface of the telescope for 5 field positions.
    Right panel: Raytrace of the light from the focal surface to one of the metcams.
}
\end{figure} 

The metcams view the science field through the entire optical train of the telescope, as depicted in Figure~\ref{fig:raytrace}.
This includes the primary and secondary mirrors as well as the WFC/ADC, described in \citenum{cunningham2022wfc}, which has important consequences for calibrating the system discussed later.\par

The opto-mechanical design of the MetCams \citenum{BardenSPIE2016optomechanic} is such, that thermal expansion of various components cancel each other out such that the camera has negligible focus shift due to temperature.
In addition, the cameras are water cooled to prevent any heat dissipation into the air surrounding the cameras, which would disturb science observations by introducing air turbulence.
In addition, the cameras are encased in a metal shroud that prevents wind from shaking the camera body and hides shiny metal parts from the light path.\par

The entrance opening of the MetCam is intentionally undersized, only $25mm$ for the $50mm$ optical components of the camera.
This is an important feature as it sets the diffraction limit of the camera to about $2.3$ pixels FWHM on the $5.5 \mu m$ pixels of the camera detector.\par

\begin{figure}[h]
\begin{center}
\includegraphics[height=4cm]{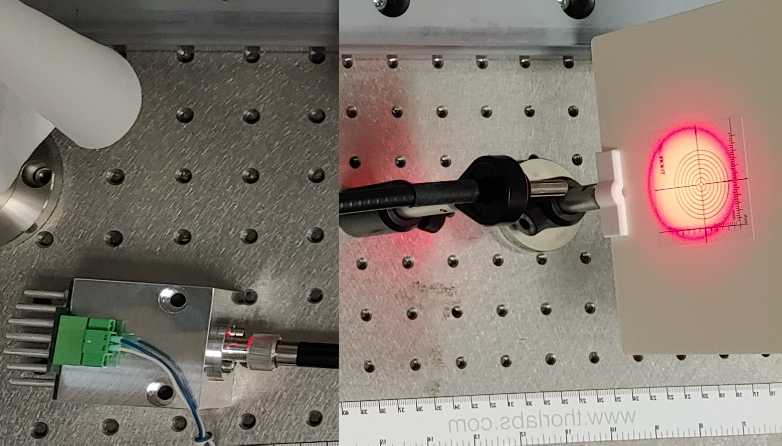}
\includegraphics[height=4cm]{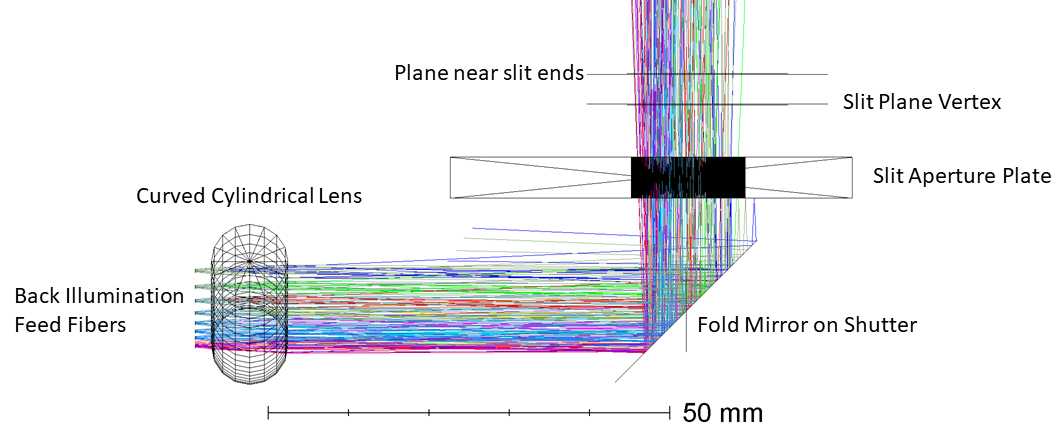}
\end{center}
\caption[Back Illumination Overview]{
    \label{fig:backillumination}
    One of the lamp units is shown on the left-hand side, a schematic overview of the slit unit on the right-hand side.
}
\end{figure} 

There are 5 Back Illumination lamps, one for each of the 3 spectrographs, secondary guiding fibres and fiducial fibres.
The light is channeled by liquid light guides from the lamps to the spectrogrpahs, split into a bundle of 19 optical fibres which are then fed into the science fibres in the spectrograph slits, shown in Figure~\ref{fig:backillumination}.
The pick-up mirror shown in the right hand side of the figure is part of the shutter mechanism and moved out of the way when observing.
The back illumination light is narrow band with a wavelength of $630nm$ to $640nm$, that is small enough to avoid dispersion effects by the telescope ADC but wide enough to prevent any speckle issues with fibres which would occur with monochromatic laser light.\par

All metrology 4 cameras will observe the entire field of fibres \citenum{HaynesSPIE2018fibrefeed} simultaneously, which provides 4 independent measurements for each fibre as opposed to each camera observing only a quarter of the field.
The centroiding performance of 4 full field observations is the same as 4 tiled observations, as the centroiding precision is $2 = \sqrt(4)$ times as good as a single camera, which is equal to the centroiding precision of a single camera observing only a quarter of the field.
The implemented strategy has the advantage that seeing perturbations within the dome have less influence on the result, all cameras see all fiducial fibres, no stitching effects between tiles and the system is fault tolerant as a camera can fail without rendering the system inoperable.\par

\begin{figure}[h]
\begin{center}
\includegraphics[height=4.4cm]{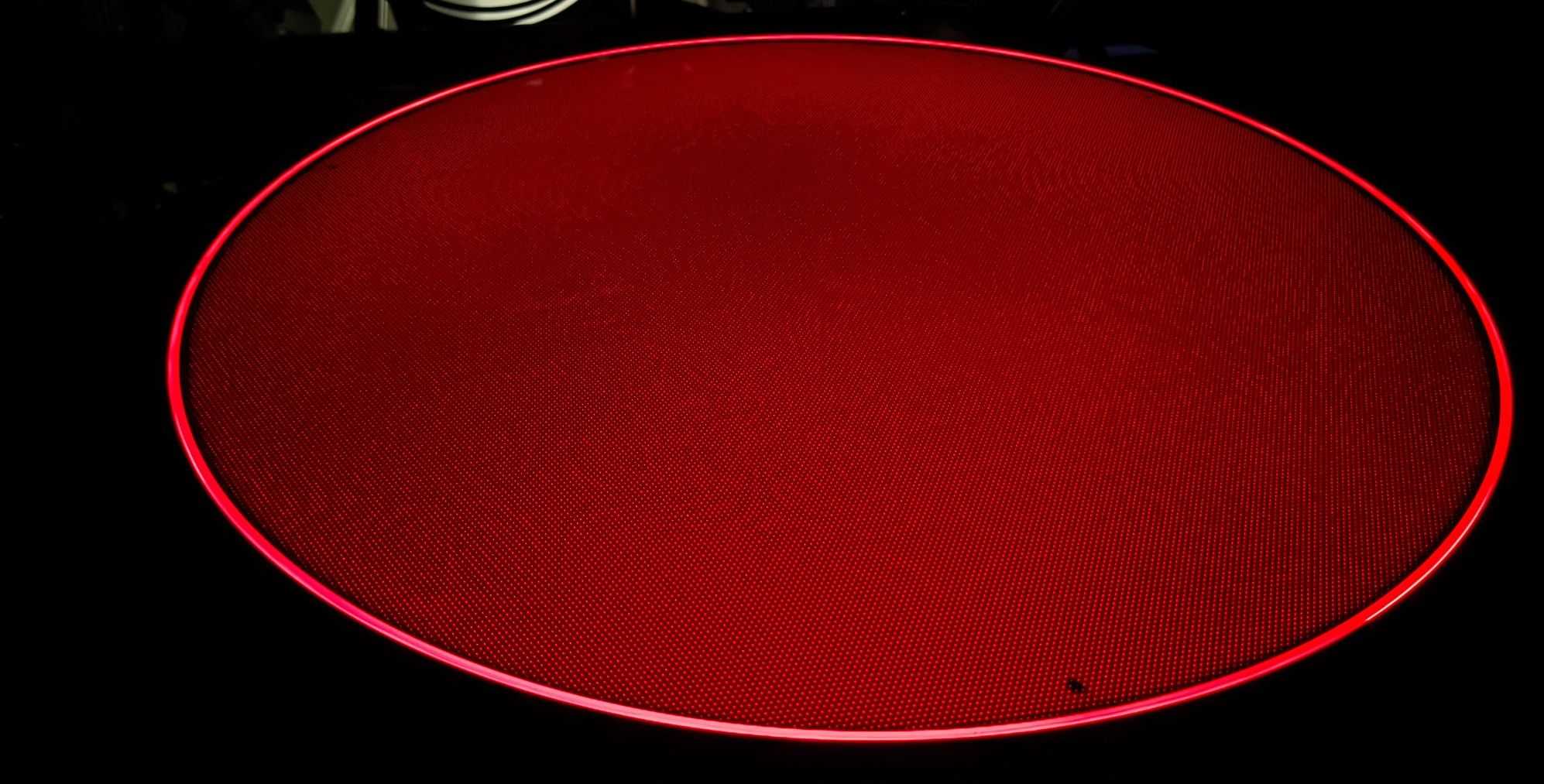}
\includegraphics[height=4.4cm]{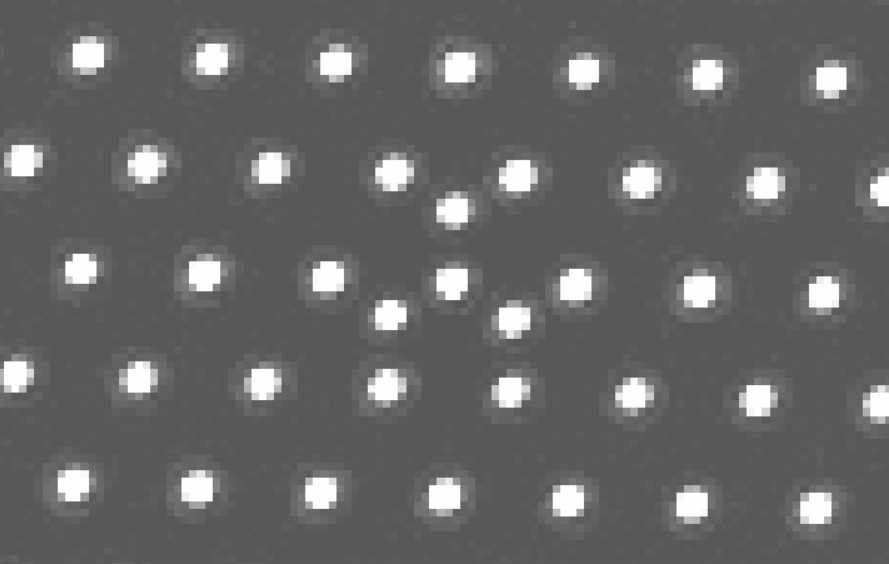}
\end{center}
\caption[MCU]{
    \label{fig:mcu}
    Left: Metrology Calibration Unit
    Right: Center of the MCU as seen from the MetCams.
}
\end{figure}

The footprint of each fibre on the optical component in its path to the MetCams is very small, see the top left of Figure~\ref{fig:raytrace}.
When observing point source targets with a telescope, optical imperfections cause a slight blurring of the image, or widening of the Point Spread Function (PSF).
This is somewhat tolerable for the telescope operation as atmospheric blurring (seeing) is the dominating factor for a wide field instrument like 4MOST.
However, for the metrology system, due to the narrow beam for each fibre, imperfections in the optical path causes a displacement of the fibre position measurement that strongly depends on the position of the fibre in the focal plane.
As a result, the MetCams require a sophisticated calibration procedure, which is discussed in Section~\ref{sec:calibration}.
A calibration plate, called Metrology Calibration Unit (MCU) is required, that is placed in the focal plane instead of the AESOP positioner during instrument commissioning.
The MCU houses about $65000$ small $85 \mu m$ diameter holes in a chrome surface, shown in Figure~\ref{fig:mcu}.
The holes have a pitch of $2mm$, which gives enough sampling for any effects to be measured and modeled.
The glass surface is illuminated by approx. 200 LEDs, the light is uniformly scattered by a diffusion foil at the underside of the glass plate.\par

In the center of the MCU, 4 additional spots are used for alignment when analyzing the image.
An additional hole at about $200mm$ from the center provides rotational alignment.
These alignment spots can be identified in software by analyzing the distances to neighboring holes, the center is shown in the right-hand side of Figure~\ref{fig:mcu} as seen by one of the metrology cameras.\par

In retrospect, without the MCU and SW that we developed in house to measure and calibrate mirror surfaces \citenum{Winkler2024FTA}, 4MOST would not have been operational this quickly (see Section~\ref{sec:calibration}).\par

\begin{figure}[ht]
\begin{center}
\includegraphics[width=0.3\textwidth]{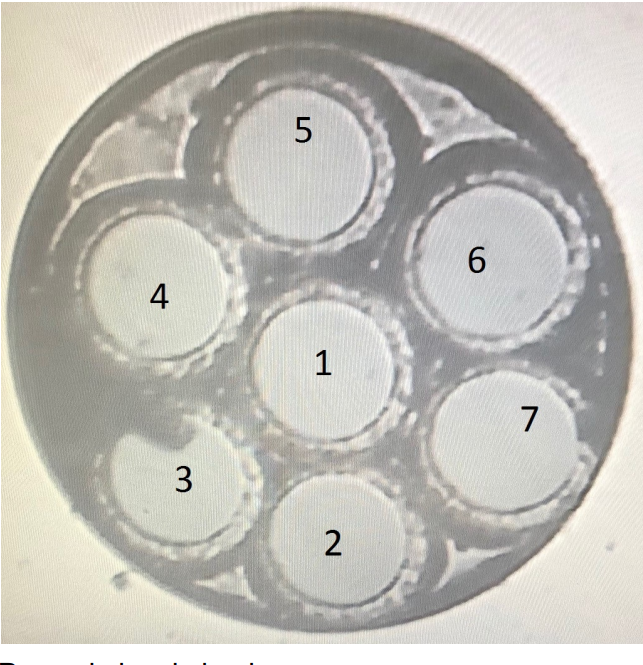}
\end{center}
\caption[sg bundle]{
    \label{fig:sgbundle}
    Microscopic image of one secondary guiding fibre bundle.
}
\end{figure}

In addition to 2436 science fibres, 4MOST also has 12 Secondary Guiding bundles of 7 fibres with $50 \mu m$ cores each.
The fibre bundles are channeled to a dedicated camera, that reads the flux in each fibre.
From the flux values and microscopic measurements of fibre positions within the bundle, a centroid from incoming target light is constructed.
By comparing the centroid with an expected position based on metrology measurements, the SG system is capable of refining telescope pointing in a very limited region.
Each bundle is only about $200 \mu m$ in diameter (edge to edge) and covers a diameter on sky of approx. $3$ arc seconds, giving it a very limited range of refinement.
The SG fibre bundles are mounted in the same spines as the science fibres, subjecting them to the same behavior and influence as science fibres in operation.

\section{METROLOGY: FROM IMAGE TO COORDINATES} \label{sec:data}

The Metrology cameras are designed to image the AESOP positioner and the MCU, which is again designed to mimic the properties of AESOP.
In both cases, the processing steps are largely the same, but the data of the MCU can be used to calibrate the system while images with AESOP cannot.
Here, we will first describe the data processing as if it was fully calibrated to give an understanding of what the calibration needs to achieve.
The calibration bootstrapping it self is discussed in Section~\ref{sec:calibration}.\par

The data processing is performed in 4 steps, image processing, ray-tracing, fiducial correction and data combination.
The first three steps are done individually for each camera image while the last step combines all available results from multiple cameras and/or multiple images per camera.
The software to analyze the images is executed in the 4MOST Instrument Control Software \citenum{MandelSPIE2016control} within the ESO VLT Software framework.

\subsection{Image Processing} \label{sec:image}

The metrology camera is a GT6600 from Allied Vision with a ONSEMI KAI-29050 monochrome detector with 6576 x 4384, 5.5 $\mu m$ pixels.
After taking an image of AESOP with the metrology camera, the fibre images are detected in pixel space.
The optical properties of the camera are chosen such that a fibre image has a size of about 2.3 to 2.5 pixels FWHM.\par

\begin{figure}[ht]
\begin{center}
\includegraphics[width=\textwidth]{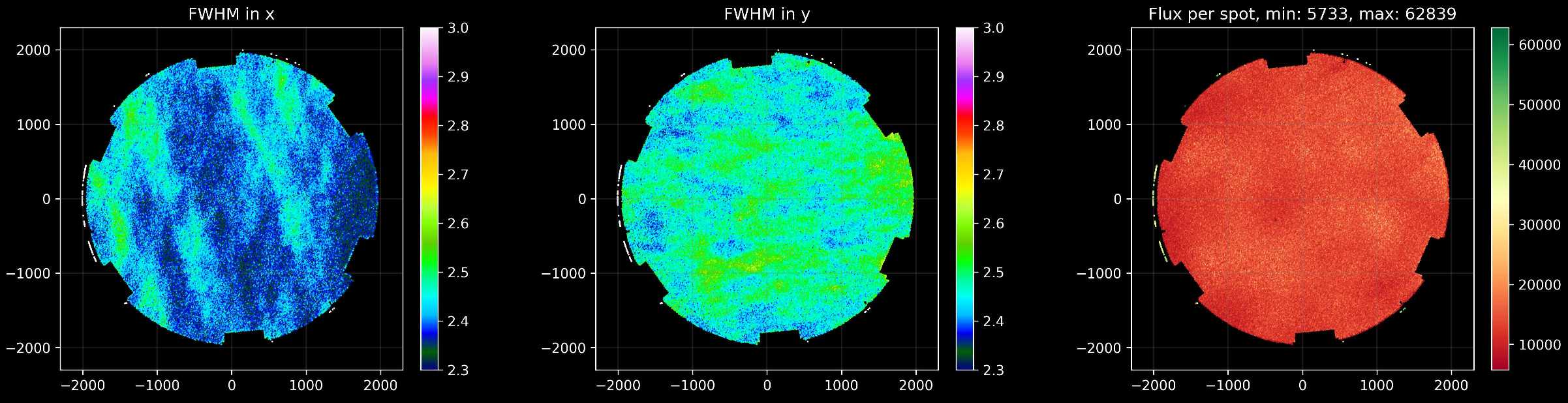}
\end{center}
\caption[MCU Spot Properties]{
    \label{fig:spotprop}
    A visualization of the horizontal, and vertical spot FWHM as well as spot brightness for an image of the MCU.
}
\end{figure}

The dominating profile of the fibre image is an airy disk, which is close enough to a 2D Gaussian profile, which is why we fit this instead.
We use a 2D Gaussian profile fit as opposed to two 1D profile fits of the marginal distributions because some pixels might hold invalid data and need to be masked out, which can be done in 2D, but not 1D fits.\par

\begin{figure}[ht]
\begin{center}
\includegraphics[width=\textwidth]{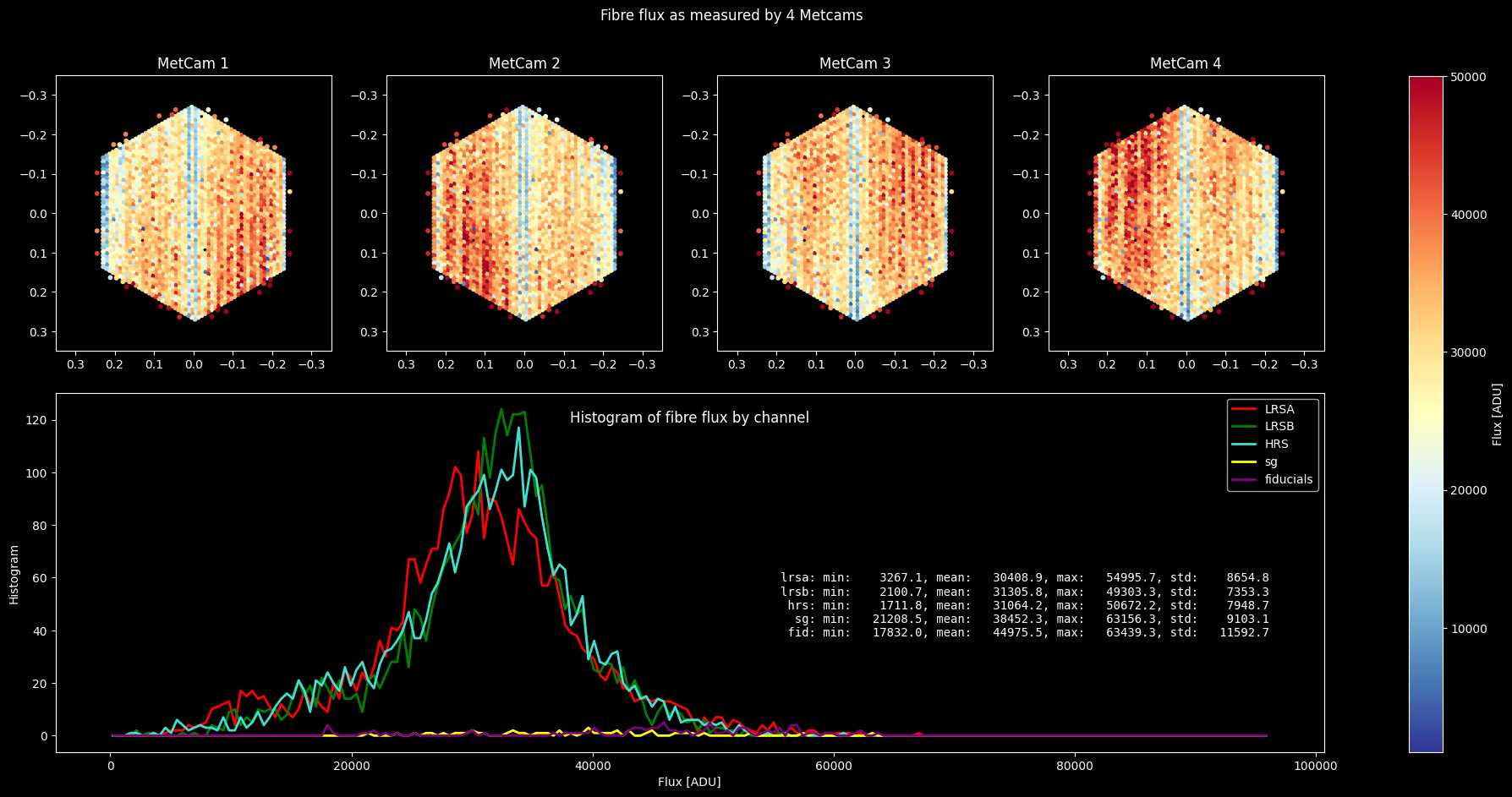}
\end{center}
\caption[AESOP Back Illumination flux]{
    \label{fig:biflux}
    Flux of the individual channels in the back illumination system as seen by the 4 metrology cameras.
}
\end{figure}

Fibres can come as close as 1 mm, or 7.5 pixels center-to-center.
Centroiding is done in a window of 7 x 7 pixels, to avoid the influence of close by neighbors while keeping the fitting window as large as possible.
For comparison, the central spots images in Figure~\ref{fig:mcu} are 1.4 mm / 10 pixels apart.
The image of a fibre can be blurred by motion of the Cassegrain rotator in the telescope, vibration of the camera or optical perturbations in the camera but still fits safely within the centroiding window.
A visualization of the horizontal, and vertical spot FWHM as well as spot brightness for an image of the MCU is shown in Figure~\ref{fig:spotprop}.\par

The properties of the AESOP fibres will be very similar, as the illuminated holes in the MCU are the same size as the fibre cores.
Though in case of AESOP, the fibres are illuminated by 5 different channels in different optical arrangements, making it necessary to carefully adjust the brightness of each spot in each channel.
A dynamic range of a factor of 10 made it challenging but possible to find and centroid all fibres, see Figure~\ref{fig:biflux}. \par

Due to defects in the image, unexpected reflections of lab illumination or cosmic rays, it is possible that the centroiding algorithm detects more spots than expected.
At the same time, it is possible that part of the spines get obstructed from the guiding and wave front sensing cameras, or that fibres do not transmit light, are otherwise dark for some reason and invisible to the MetCams.
Missing fibres as well as spurious detections are problems that are handled during spine identification, see Section~\ref{sec:aesopspotidentification}.
When AESOP is observed, and camera images are processed immediately, the SW is implemented in C and part of the CLIP library by ESO.

\subsection{Ray-tracing} \label{sec:raytracing}

We developed the ray-tracer in house, which takes fibre image centroids provided by the CLIP library (or offline image processing) as input and provides focal surface coordinates for the optical fibres as output.
A visualization of its rays is shown in Figure~\ref{fig:raytrace} for parallel rays from the sky and for rays from the focal surface to the metrology cameras.
The ray-tracing step is composed of several sub-steps to align the optical model and identify spots.
When rays are traced from the camera detector to the telescope focal surface or vice versa, the rays are guided through the center of the MetCam entrance window.
Only one ray per fibre is required for the coordinate transformation because the footprint of any fibre image on the Telescope optics is very small.\par

Any systematic error that might occur from that strategy is calibrated out by the calibration process, discussed in Section~\ref{sec:calibration}.
The guiding of rays is done with a least squares approach where rays are initialized with a rough solution and refined in a few iteration steps.
In the ray-guiding step, the rays end at a virtual plane surface, parallel to the MetCam pupil entrance.
The distance of the rays to the center is minimized, where the x and y component of each ray are handled independently.
This is not necessarily true, but the dependency is so small, that the process works and is much faster compared to dependent x and y components.
This process always terminates with a good solution as there are no local minima and the initial guess is can be estimated quite well.\par

\subsection{Data and Model Preparation} \label{sec:datapreparation}

To prepare the data for the following processes, the spot measurements are stored in a kd-tree, which is created in $O(n \cdot log(n))$ and can query k-nearest neighbors in $O(k \cdot log(n))$.
This data structure makes it possible to execute most of the below described algorithms efficiently.\par

The optical model that is used for the ray-tracing is an idealized view on reality with some non-ideal aspects that are fix over time and some dynamic aspects.
The static aspects are calibrated using the MCU, which is discussed in Section~\ref{sec:calibration}, and loaded together with the optical model.\par

Some dynamic aspects can be added to the model beforehand, like the rotation status of the Cassegrain rotator, the MCU rotation (which can move separately in the telescope and lab) and the ADC.
Other dynamic aspects like the shape of the M1 mirror, the M2 mirror orientation or the exact pointing of the MetCam (which can flex in its mount) need to be estimated using the data in each image.

\subsection{MCU Alignment Spot Identification} \label{sec:alignmentidentification}

The first step in the MCU processing is to identify the alignment spots of the MCU in the image.
The process is different for the MCU vs. AESOP since both have completely different spot structures, so we discuss the two processes separately.\par

In the MCU, we have the alignment spots, that can be identified by their relative location to neighboring spots and their general location in the image.
The center group of spots can be identified because there are 4 spots which each has 3 neighbors closer than 12 pixels and is located roughly in the center.
The angle alignment spot can be identified because it is in between 2 regular spots, so it has two neighbors that are closer than 10 pixels and at a distance roughly 1500 pixels distance from the center.\par

\subsection{MCU Spot Identification} \label{sec:mcuspotidentification}

The identified of MCU spots must happen on detector level because there are no dedicated fiducial spots and the software outlined below would not work without knowing fiducials, the alignment spots alone are not sufficient.
Instead, some of the normal spots on the MCU are declared to become fiducial spots, but to use them, the entire pattern has to be identified first.\par

The alignment spots, discussed above are used to identify the MCU spot pattern.
However, they are not numerous enough and not distributed enough to use for the fiducial correction algorithm or to use for camera alignment, see \ref{sec:metcammodelalignment}.
Any number and distribution of spots can be declared fiducial, but to be consistent with AESOP, 24 spots near the edge of the MCU plate are declared fiducials for performance measurement.
For the MCU, after declaration of fiducial spots, steps discussed in Section~\ref{sec:metcammodelalignment} are executed with declared fiducials instead of alignment spots.\par

The MCU has a very dense set of holes, with a pitch of only 2 mm, or 16 MetCam pixels.
Before optics characterization, difference between physical reality and optical model is strong enough, that the spots may appear shifted by more than 1 mm.
Locally, the spots pattern is of course obvious, but over longer distances (approx. 1/5th of the MCU diameter), the straight forwards approach leads to systematic mis-identification.\par

\begin{figure}[h]
\begin{center}
\includegraphics[height=4cm]{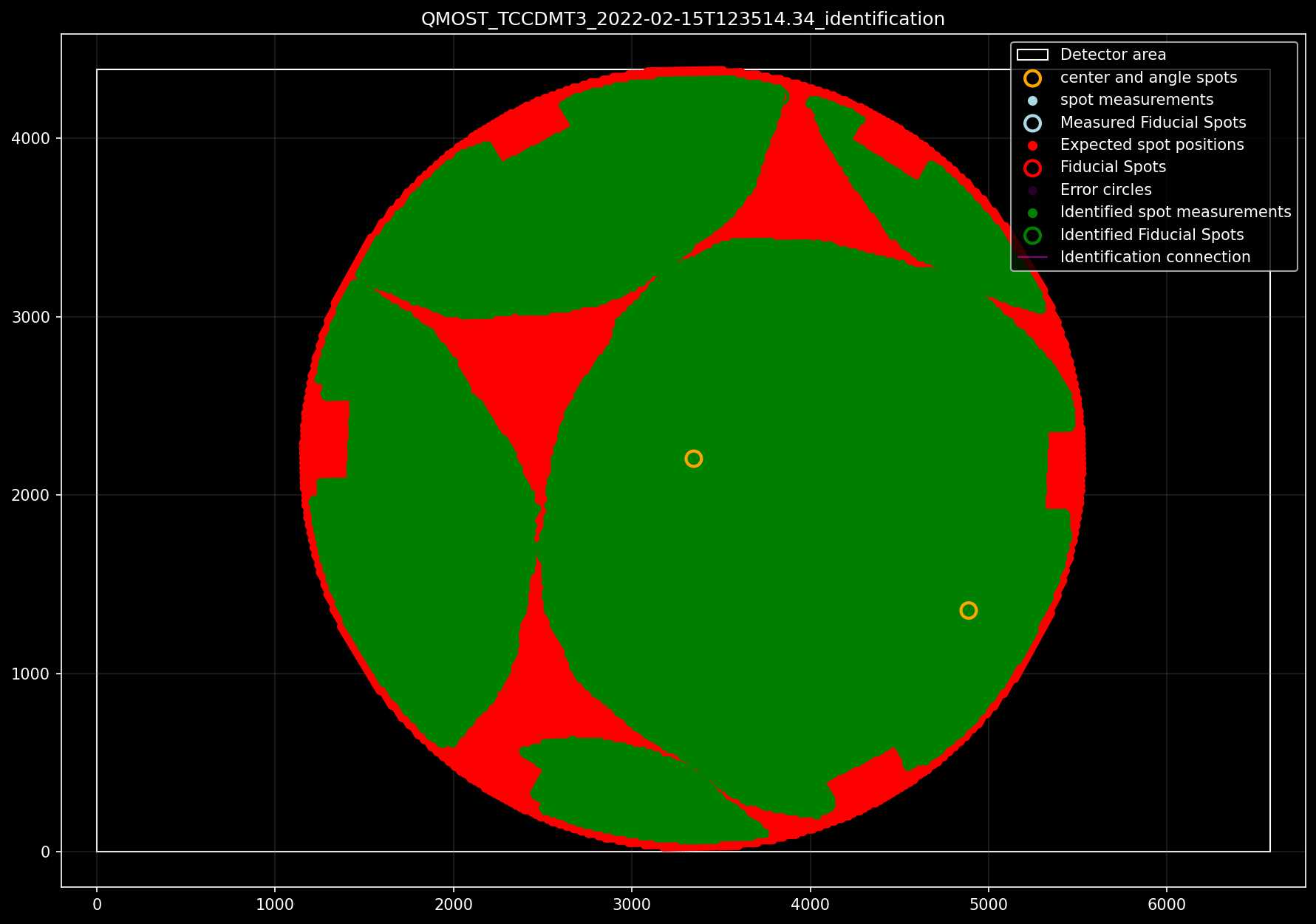}
\includegraphics[height=4cm]{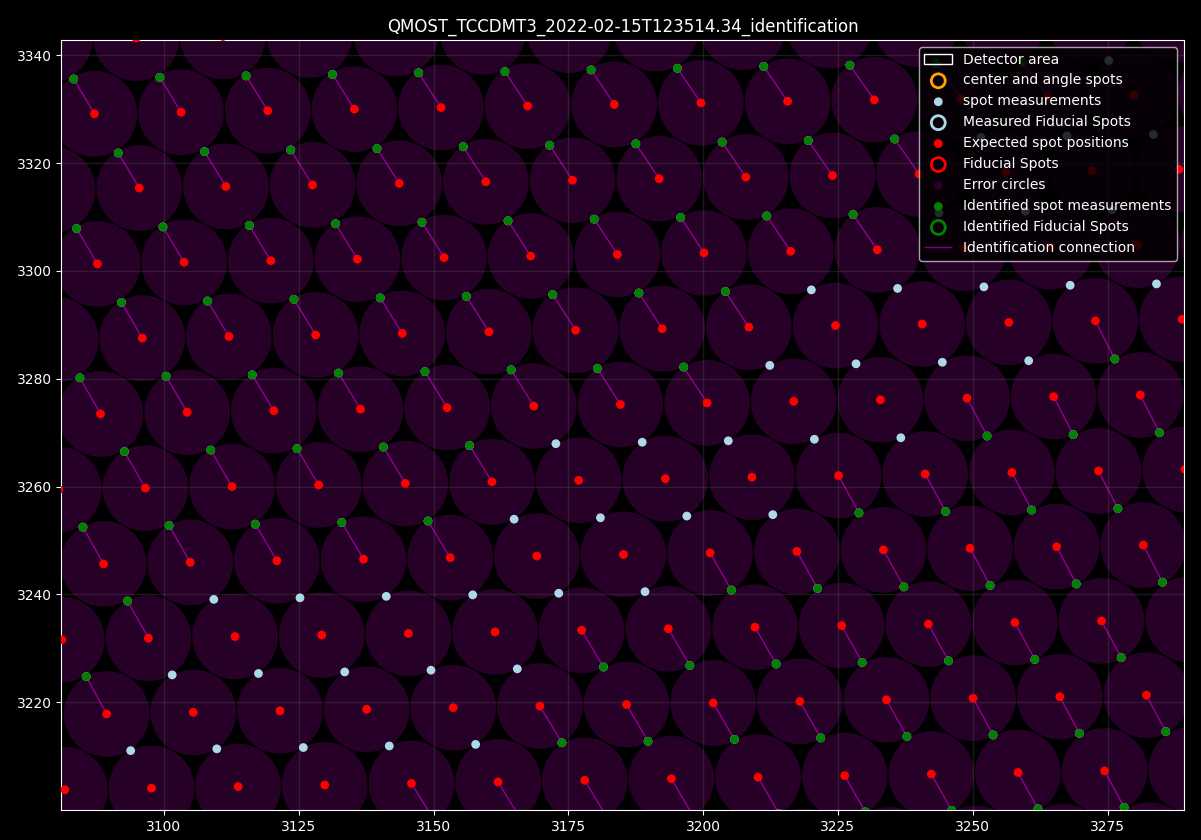}
\includegraphics[height=4cm]{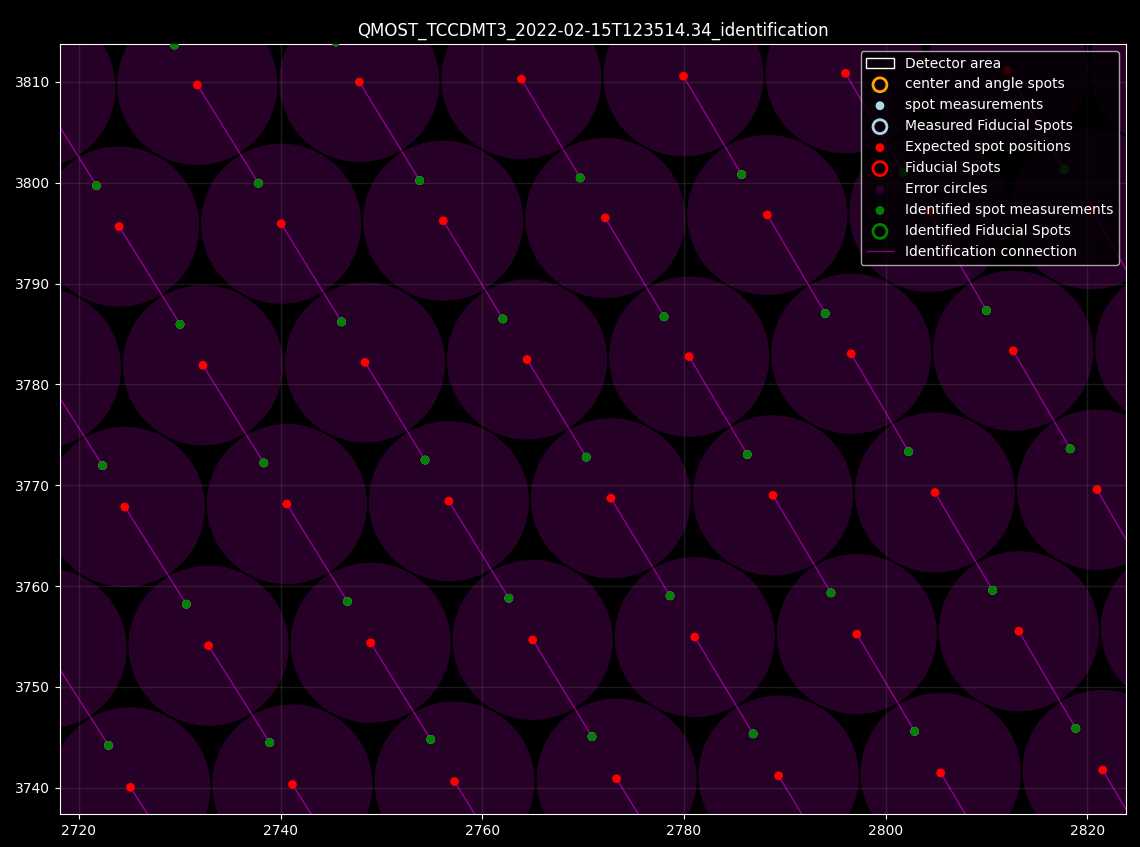}
\end{center}
\caption[Distortion Propagation Algorithm]{
    \label{fig:misidentification}
    Left: Overview of the distorted alignment with green patches of mis-identification
    Center: Zoomed-in transition section.
    Right: Zoomed-in section that was previously false identified.
}
\end{figure}

When spots are misidentified, any optical characterization must fail, see Figure~\ref{fig:misidentification} for an example.
The left hand side shows an overview of the entire MCU while the middle shows a zoomed in section with a transition between correctly associated spots, unidentified spots and false associated spots.\par

The center of the MCU can easily be identified and we can orient the expected spot pattern using the angle alignment marker.
Locally, surrounding the center marker, a few rings of spots can also be identified.
From these few center spots, a piecewise linear spline interpolation is constructed that corrects the spot positions near the center and is extrapolated outwards.
This step grows the region of spots that can be identified with certainty.
The region of this spline interpolation is grown iteratively as more and more of the surface can be correctly associated until all spots can be successfully identified.\par

We call this algorithm the distortion propagation algorithm.
It is only used for spot identification, described in Section~\ref{sec:mcuspotidentification}.
The effect of that algorithm can be seen in the third panel of Figure~\ref{fig:misidentification} where the spot is correctly identified even though it is clearly closer to another expected spot location.
Once the optical model is closer to reality, this problem goes away and the distortion propagation algorithm is not required any longer.
It is however a very useful tool for the initial situation.\par

\subsection{AESOP Fiducial Spot Identification} \label{sec:fiducialification}

In contrast to the MCU, AESOP does have dedicated fiducial spots, which are identified on CCD level by their location and distance to other spots.
This is not trivial because it is not possible to predict their location with very high accuracy and the image rotates and shifts relative to the metrology cameras.\par

We can estimate where the fiducial spots should be by ray-tracing the predicted fiducial position to the MetCam detector surface.
Then, we use a least squares algorithm to match the theoretical fiducial positions with the observed spot cloud.
This process usually terminates perfectly, but it has the potential to land in a local minima, where some normal fibres are misidentified as fiducials.
This can be detected because a match should give an average error of less than 10 pixels while a failure has much larger errors.\par

If we detect that the fitting has failed, we repeat the step with slightly different initial conditions at the expense of some computation time, i.e. rotate the expected pattern by $3^\circ$ or shift the pattern by a few pixels.
This approach was always successful as both effects are probably due to Cassegrain rotation and the flex of the MetCam in their mounts.
The identification algorithm only adjusts shift and rotation for the purpose of fibre identification on pixel level, it does not change the optical model.

\subsection{Dynamic Optical Model Alignment} \label{sec:metcammodelalignment}

Once the fiducial fibres of AESOP or fiducial spots of the MCU are matched, we apply a model optimization step for MetCam orientation, i.e. rotation of the MetCam optical model around all 3 axis.
The center of rotation is the MetCam entrance window vertex, which is the interface point for the rays between the internal MetCam model and the external telescope or lab setup model.
This is done again by ray-tracing the expected alignment spot locations to the focal surface of the camera and minimizing the difference between observed spots and modeled spots.
This process always terminates as there is a unique optimal solution and no local minima within the possible parameter ranges.

\subsection{Fibre Projection} \label{sec:projection}

After applying all dynamic effects to the optical model, the detected spots are projected towards the instrument in 3D space using the raytracer.
While this section is only a short statement, it contains a lot of processing in the background.
The raytracing compensates for surface irregularities (described in Section~\ref{sec:calibration:normal}) and also does not suffer from usual transformation artifacts that would require a radially polynomial correction.\par

In case of the MCU, the rays are terminated at the modeled MCU surface, which is a flat surface.
For AESOP, this is a bit different.
Spines move laterally by tilting around a rotating center at the base of the spines.
This induces a focus shift of the spine as well, which can be up to 280 $\mu m$ between highest and lowest point.
Since the MetCams observe AESOP at an angle, this can cause up to 15 $\mu m$ systematic error due to the parallax.
For the moment, the errors due to parallax are ignored and solved after fibre identification.

\subsection{Expected Spot Positions} \label{sec:expectedspots}

The fibre image spots have no intrinsic property that one could use to decide which spot in the image belongs to which fibre/hole in the observed object, they are identified by their location.
This is a circular argument of sorts, as the Metrology System is used to measure the fibre locations in the first place.
However, it is enough to know the expected spot location on the detector, with an error small enough to uniquely identify each spot.
Depending on the proximity of spots, this error can be quite large.\par

In case of the MCU, the model of the MCU dictates where to expect the spots and they are on a fix relative position.
For AESOP, this is more dynamic as the spines can move.
Provided AESOP knows where the spines are before they move, and then it moves the spines, it can predict with some certainty where the spines will end up.
The distance of the move is roughly linear with the size of the movement error.
AESOP will move spines in such a way, that after each move, error circles do not overlap such that a 1:1 assignment of spines and spots should be possible.\par

\begin{figure}[h]
\begin{center}
\includegraphics[width=\textwidth]{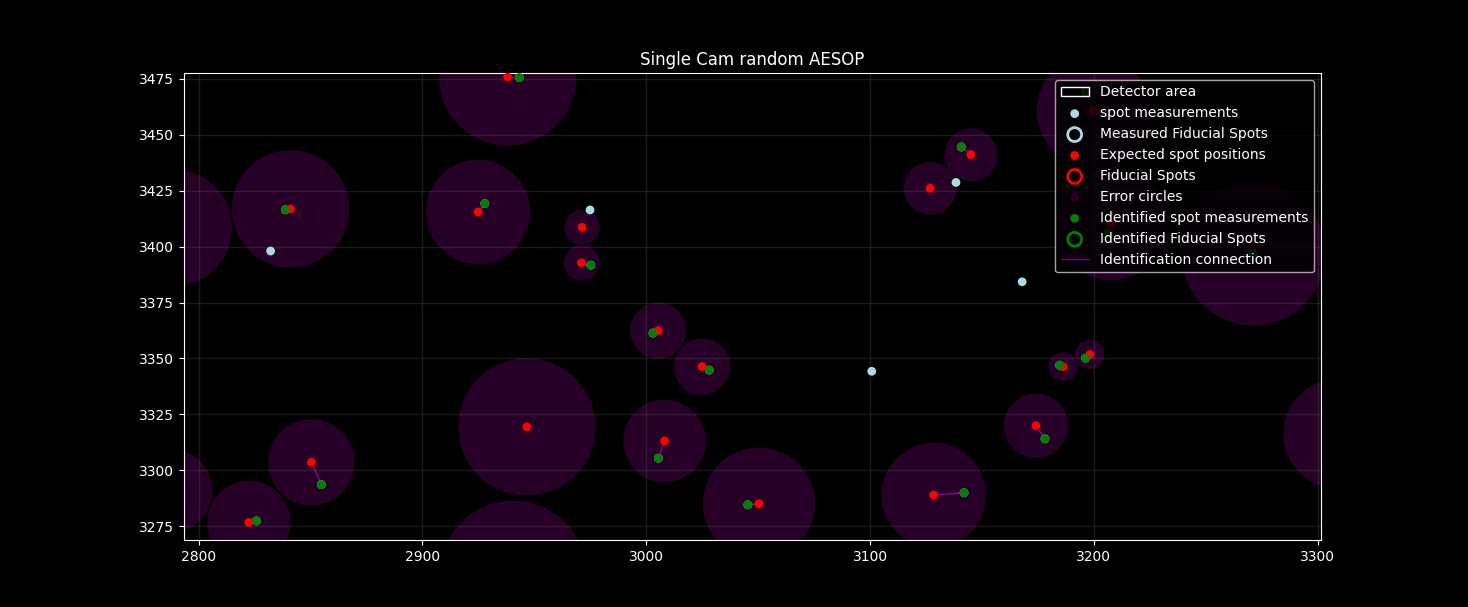}
\end{center}
\caption[AESOP spot identification]{
    \label{fig:spotident}
    Visualization of an image section of AESOP of spot identification. Blue: unidentified spots, red: expected spot location, green: identified spot associated with a red dot. Purple halo: error circle centered on expected spot locations.
}
\end{figure}

Either from the MCU or AESOP, the location of observable holes/fibres is an input value to the Metrology system in 3D space on the telescope focal surface.
The spots that are detected with the metrology cameras are projected to the focal surface using the raytracer, where they are identified.

\subsection{AESOP Spot Identification} \label{sec:aesopspotidentification}

To identify the spots on the focal surface, another kd-tree of projected fibre candidates is constructed, which is querried with all expected spot locations.
Each expected spot position is associated with a list of spot candidates.
Each candidate gets a probability assigned based on proximity and size of the error circle provided by AESOP or MCU.
Since the areas of error circles around expected spot positions are mutually exclusive, ideally one and only one spot is present within one error circle.\par

However, this is not always the case as false observations are possible as well as spots can be missing and spines might move unexpectedly.
All edge cases are handled seamlessly throughout the entire process.
A visualization of the spot identification is shown in Figure~\ref{fig:spotident}.\par

The identification happens by constructing a N x M matrix where N is the number of expected spots, (i.e. 2472 in case of AESOP) and M the number of detected spots.
The matrix contains 0 for all cases where a spine cannot physically reach the location of a detected spot.
The Jonker-Volgenant algorithm implementation of scipy \citenum{CrouseDavid2016} (i.e. the scipy implementation of it) is used to find the best global solution of spine associations with detected spots.\par

This approach is not entirely deterministic, but given the uncertainties in the system between measurement errors, air turbulence and movement uncertainties, it provides the best compromise.
Initially we tried a deterministic system where only fibres within their error circles are identified, with a second pass of fibre association outside error circles to the closest candidate.
This works in most cases, but there are always corner cases where the spine association fails spectacularly.
These corner cases are very rare using the Jonker-Volgenant algorithm instead.\par

\begin{figure}[h!]
\begin{center}
\includegraphics[width=\textwidth]{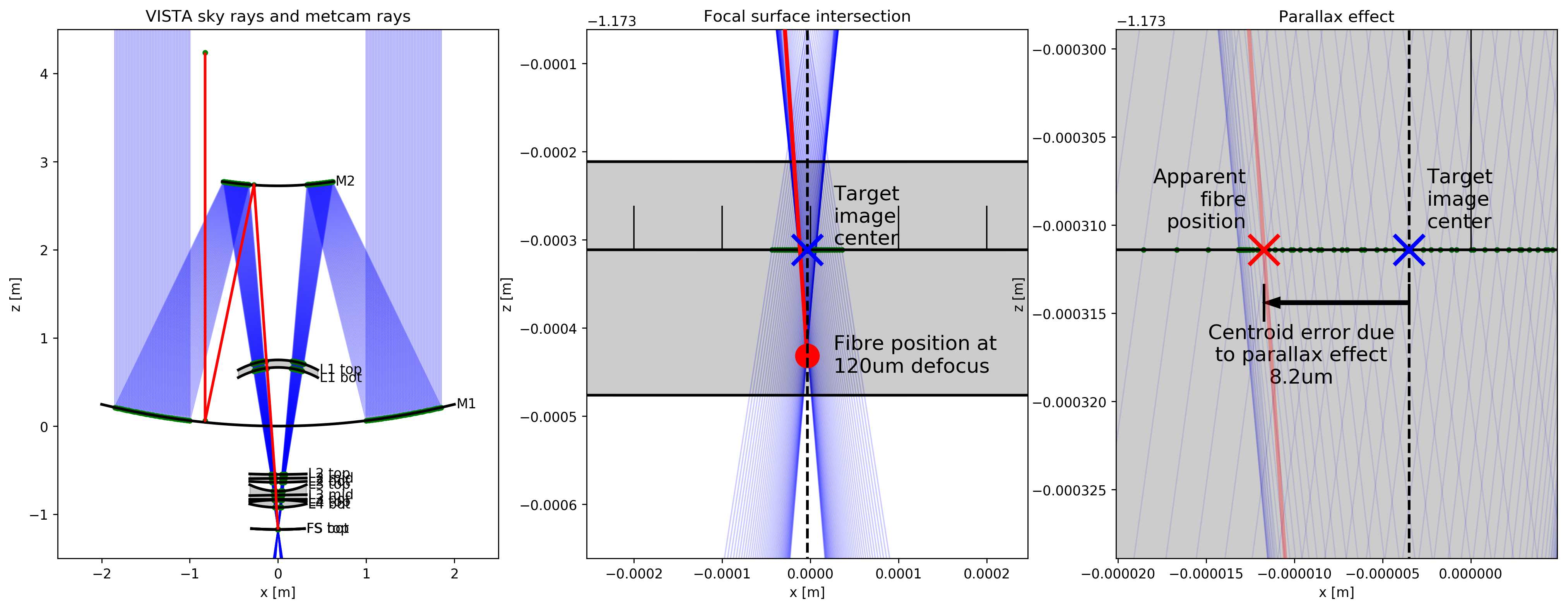}
\end{center}
\caption[Fibre focus induced parallax]{
    \label{fig:parallax}
    Parallax induced error for a fibre out of focus. Blue rays represent light from a target on sky, red rays represent the measurement by a metrology camera. The horizontal offset in focal surface shows the measurement error as a consequence of the fibre focus error.
}
\end{figure}

\subsection{Parallax correction} \label{sec:parallax}

Due to the nature of the instrument, when a spine of AESOP moves a fibre, it rotates around a fix point, 250mm below the focal surface.
This rotation around the fixed point causes a focus shift $\pm 135\mu m$.
In addition, a fibre might have a non-zero offset from its ideal focus position due to its physical mounting.
However, the metrology cameras can only see the flat projection of fibres onto the focal surface.
Once a fibre is identified, its position position in 3d space can be calculated and the true 2D projection onto the focal plane can be calculated using geometry.
A visualization of this effect is shown in Figure~\ref{fig:parallax}.\par

\subsection{Fiducial Correction} \label{sec:fiducial}

\begin{figure}[h]
\begin{center}
\includegraphics[width=\textwidth]{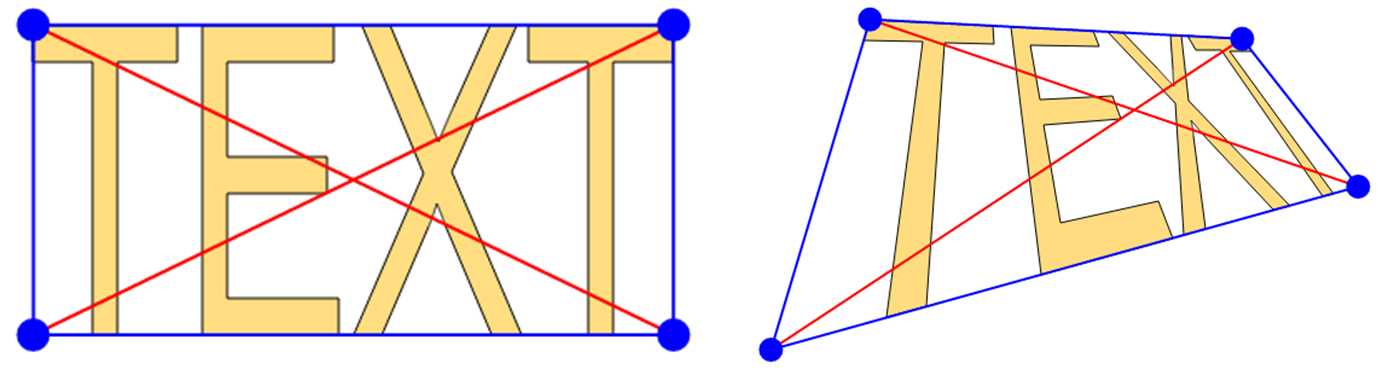}
\end{center}
\caption[Fiducial transformation]{
    \label{fig:trapezoid}
    Principal trapezoid transformation, used for fiducial correction. Images source: \url{https://stackoverflow.com/questions/12919398/perspective-transform-of-svg-paths-four-corner-distort}.
}
\end{figure}

The now projected location of the spot position in 3D space is still subject to inaccuracies of the opto-mechanical model, atmospheric turbulence and other systematic errors.
The 3D position of all 24 fiducial fibres are known from lab measurements, while their metrology measured location is subject to the same errors as the science fibres.
These errors can be minimized by fitting a trapezoid transformation function, which has 8 parameters, described by the corners of a rectangle, see Figure~\ref{fig:trapezoid}.\par

Mathematically, only 4 fiducial points are necessary to find a unique solution for the transformation.
Instead, we use a least squares approach to find the best parameter set to match the 24 fiducial fibres, which improves the solution by reducing the statistical error by a factor of $\sqrt(6)$.
The effect of the fiducial correction can be visualized by comparing the centroiding performance in Figures \ref{fig:metcam_normalmaped} and \ref{fig:metcam_fiducialed}.
Depending on the severity of the residual distortion, the fiducial correction can correct large errors, but is limited due to its linear nature.
Non-linear effects would not be possible to correct with only 24 fiducials at the edge of the observed area, a distribution of many more fiducials throughout the field would be required.

\subsection{Dynamic Fiducial Model} \label{sec:dynfiducial}

Though we had to do some pointing with metrology cameras to compensate for flexure in the spider vanes, we assume that the telescope can be represented using is a rigid body model.
That is not truly the case in reality as the telescope pointing, tilting and field rotation introduce flexure that leads to deformation of the fiducial reference frame.
It is nearly impossible to model this based on the 3D model of the telescope, so we did not attempt it.\par

During commissioning and operation of AESOP, it became clear that a flexure model is required as the fibre coordinates relative to the sky are distorted depending on the telescope altitude angle (i.e. how high the telescope is pointed), and Cassegrain rotation angle (i.e. rotation of AESOP around the telescope pointing axis).
To solve this issue without a complete FEM, a data driven dynamic model for the location of fiducial positions is used.
The characterization of the dynamic fiducial model is done using raster scans, described in Section~\ref{sec:rasterscans}.
Because the fiducials correction (see Section~\ref{sec:fiducial}) can only correct affine transformations, the dynamic fiducial model can also only compensate errors that can be represented by an affine deformation of the fiducial positions.
Despite these limitation, the residual is small enough to fit in the accuracy budget, see the FTA performance measured by raster scans in Section~\ref{sec:rasterscans}.\par

The dynamic fiducial model is implemented by learning a 2D spline interpolation for each of the 8 trapezoid function parameters, each as a function of altitude and cass angle.
The parameters are queried for the true telescope state during execution, and applied as a shift to the lab-measured fiducial positions prior to executing the correction algorithm described in the last subsection.
So the system automatically changes the reference frame for metrology system measurements based on the data we created during raster scans, described in Section~\ref{sec:rasterscans}.

\subsection{Data Combination and Centroiding Performance} \label{sec:combine}

\begin{figure}[h]
\begin{center}
\includegraphics[height=7.cm]{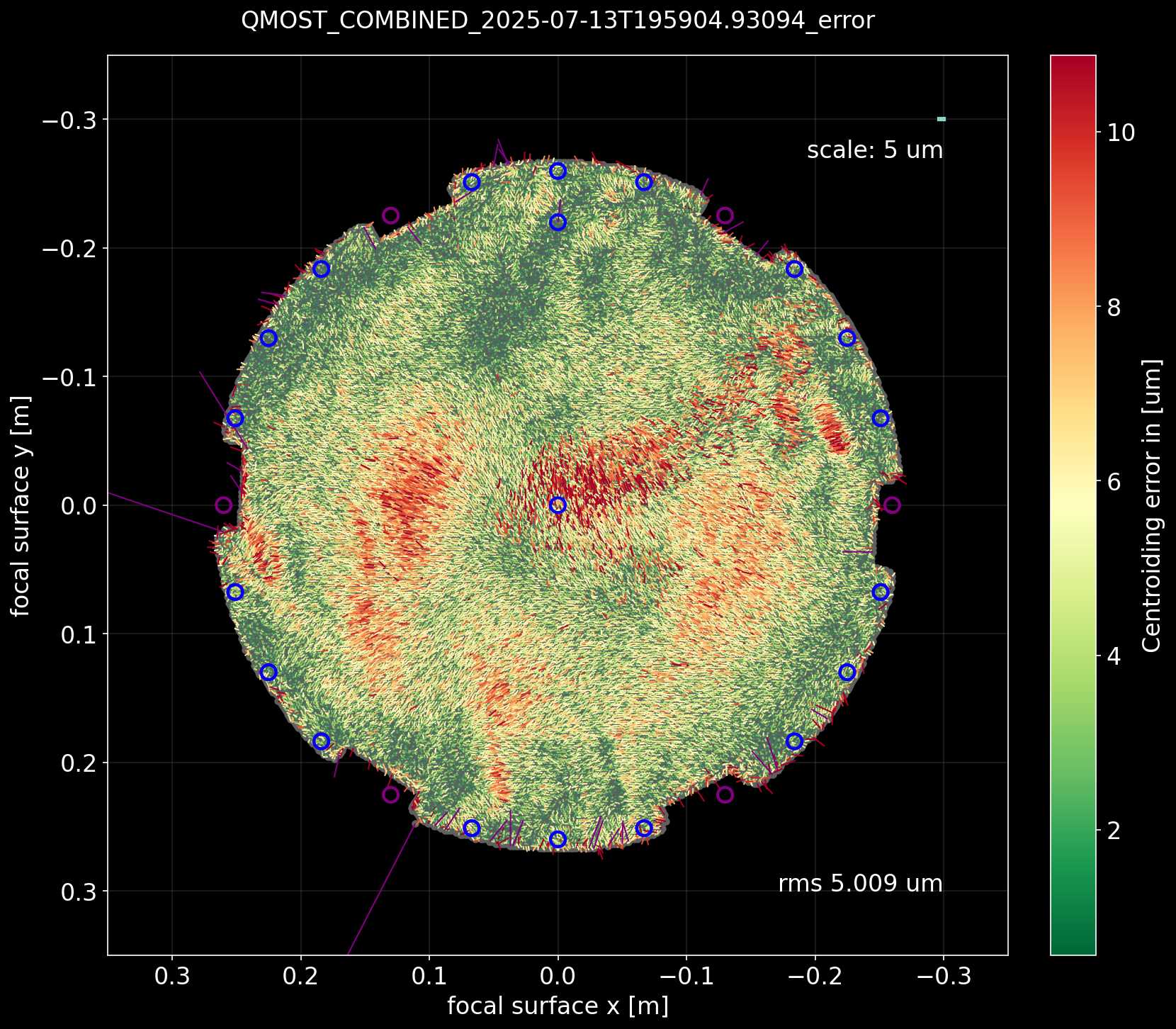}
\includegraphics[height=7.cm]{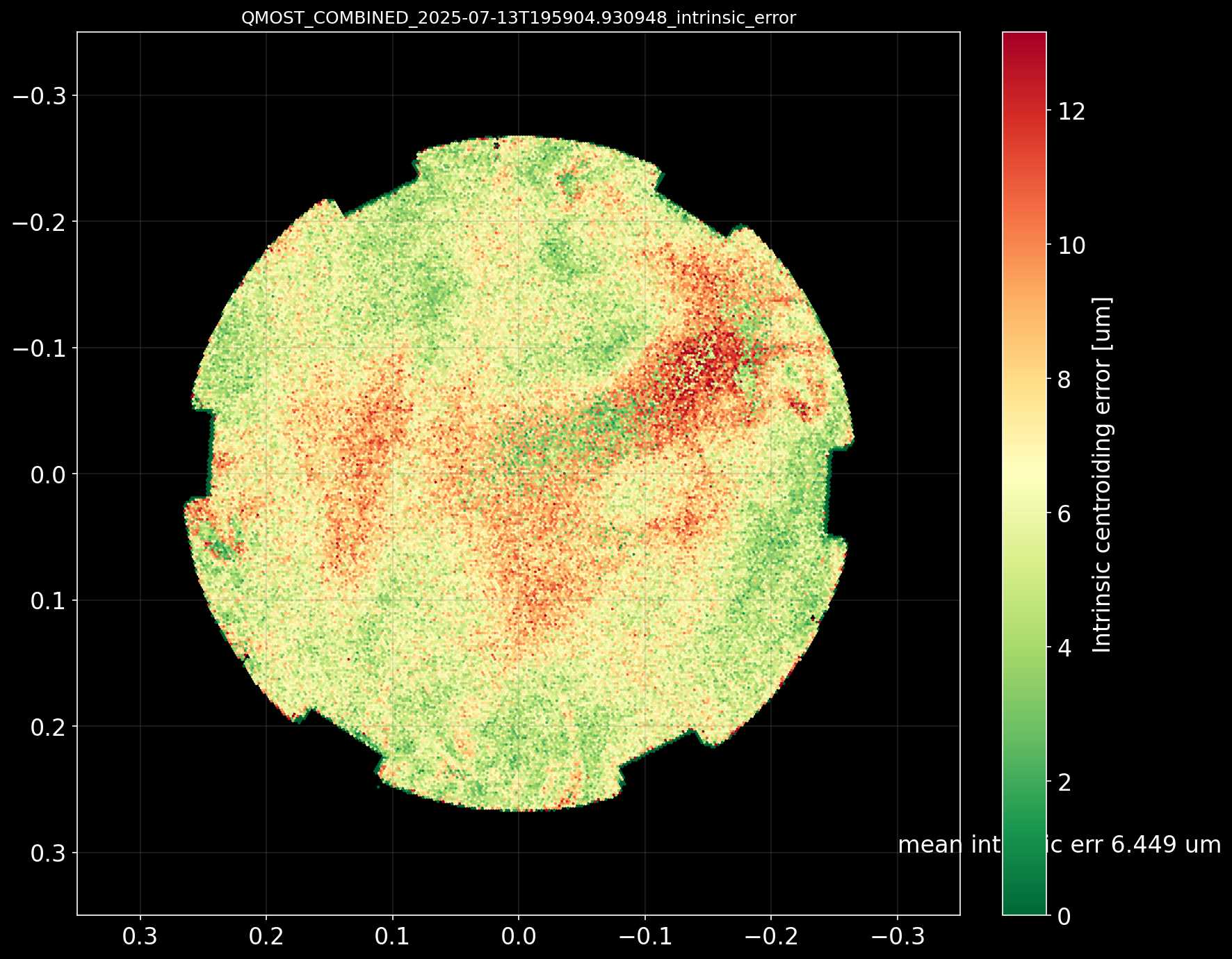}
\end{center}
\caption[Metrology Performance]{
    \label{fig:combined}
    Left: Performance of the metrology system after combining 4 MetCams using the MCU as reference.
    Right: Estimated variance for each fibre location measurement, without using the known position of the MCU.
}
\end{figure}

So far, all steps were performed on an individual frame of one MetCam.
Since there are 4 MetCams, we will receive 4 measurement sets, additionally, it is possible to take multiple frames with the same camera.
These measurements are combined by using a weighted sum of each available measurement set.
The weights are determined by the average centroiding error in the fiducials, see the left panel of Figure~\ref{fig:combined}.\par

Since this approach of averaging is agnostic to the number of measurements available, it provides fault tolerance as the metrology system would function (with degraded performance) with only 1 camera, but can also work with any number of frames taken in a series.
During commissioning, a trade-off was be made between number of frames per camera (which cost time) and accuracy of the result.
In its final positioning iteration, the location error of fibres is dominated by turbulence of air inside the dome.
However the time it takes to increase the number of metrology camera frames was determined to be not worth it.
This might change at a later point.\par

As a byproduct of combining 4 or more measurement results for each fibre, it is possible to compute an intrinsic measurement error for each spot.
This is simply the weighted mean distance of each spot from its combined position, i.e. a level of confidence, see the right panel in Figure~\ref{fig:combined}.\par

This level of confidence is likely a pessimistic approximation of the error since it is very likely that some errors of opposite cameras cancel out due to measurement from a different perspective.
For example, a focus error of a spine is somewhat compensated by the fact that opposite MetCams will create a systematic error in opposite direction.
Also, this uncertainty estimation does not improve with more metrology frames, which is intentional as we cannot exclude systematic errors, for instance from uncharacterized surface deformations.\par

It occasionally happens that one metrology camera is effected by air turbulence more than others.
In these cases, the centroiding performance is improved by adding another step to the combining algorithm.
For each spot, a second combining pass is done where spots that are further than 1-sigma away from the original centroid are excluded.
This strategy guarantees that at least 3 of 4 results are taken into account for each fibre in the default case of a combination with 4 cameras and one frame each.
When the MetCams are used with more than one frame each, always strictly more than half the measurements are used in the centroiding.

\section{CALIBRATION} \label{sec:calibration}

For the results above, we used a fully calibrated system.
In this section, we discuss how the calibration has been performed on the telescope.
For executing the calibration steps, data is taken with the MetCams while the instrument executes a sequence of movements with the MCU being installed.
While image processing of AESOP images is done on the fly, image processing of the MCU is done after data is taken and stored.
This section describes the process.\par

\subsection{Image Calibration} \label{sec:calibration:image}

The offline image pixel calibration (MCU) is different from the online image calibration (AESOP).
AESOP processing is very time critical and image calibration steps are kept to be minimal, only the detector bias is removed in the raw data.
For the MCU images, centroiding performance is more critical, and both bias and background are removed.
In both cases, the Bias is measured using 0s exposure time images and removed before centroiding.\par

In case of the MCU, the background is estimated by using a median filter 11x11 kernel, which guarantees that the median value is indeed a background pixel as the number of pixels associated with fibre images in any 11x11 window are always less than the number of background pixels.
After that filter, the background is smoothed with a Gaussian filter with sigma 7 pixels.
This creates a smooth background, which leaves very little artifacts when removed from the image.\par

For centroiding, again a 2D Gaussian function is used which ignores hot pixels.

\subsection{Uncalibrated Centroiding Performance} \label{sec:nocal}

\begin{figure}[h]
\begin{center}
\includegraphics[height=7cm]{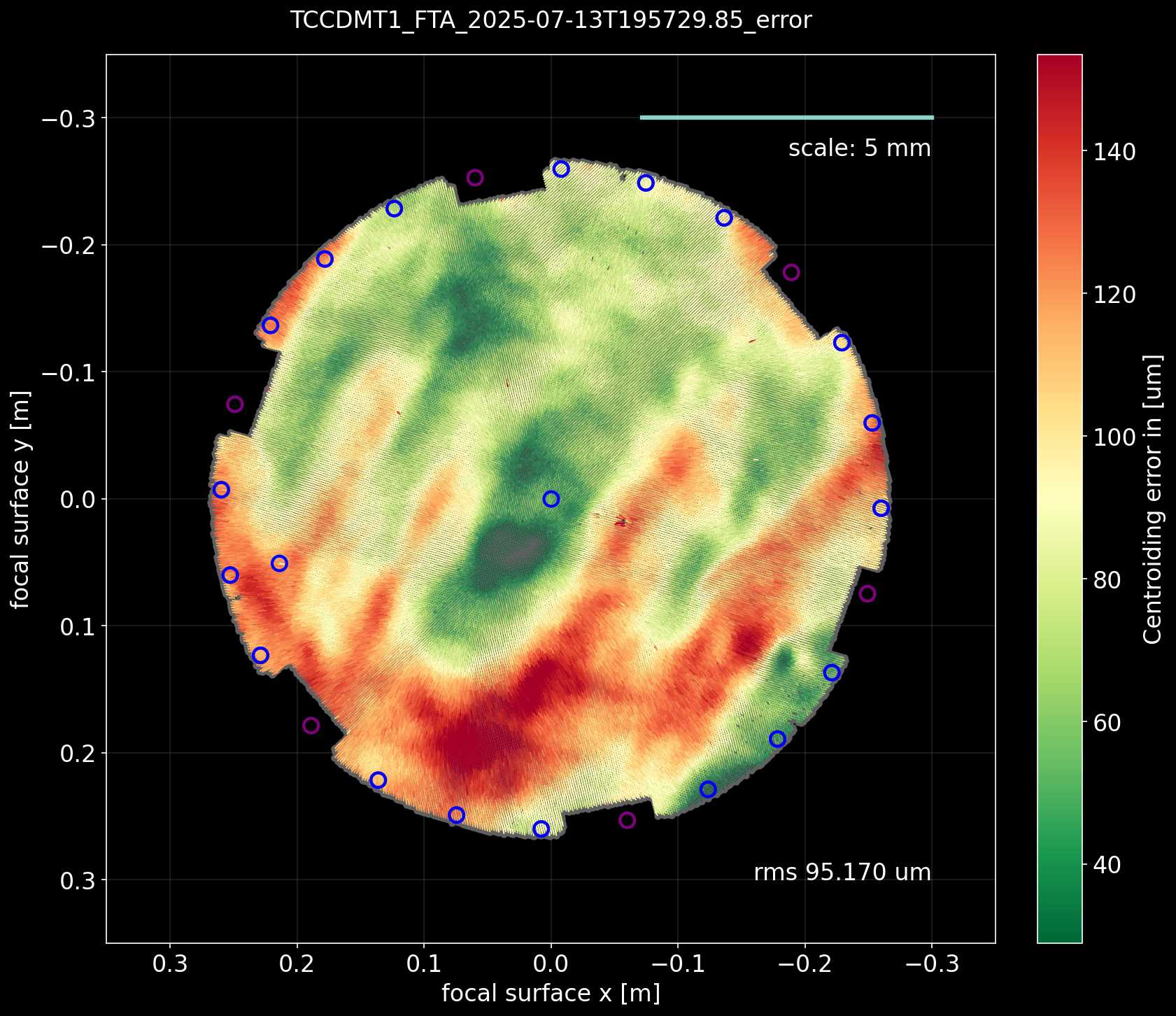}
\includegraphics[height=7cm]{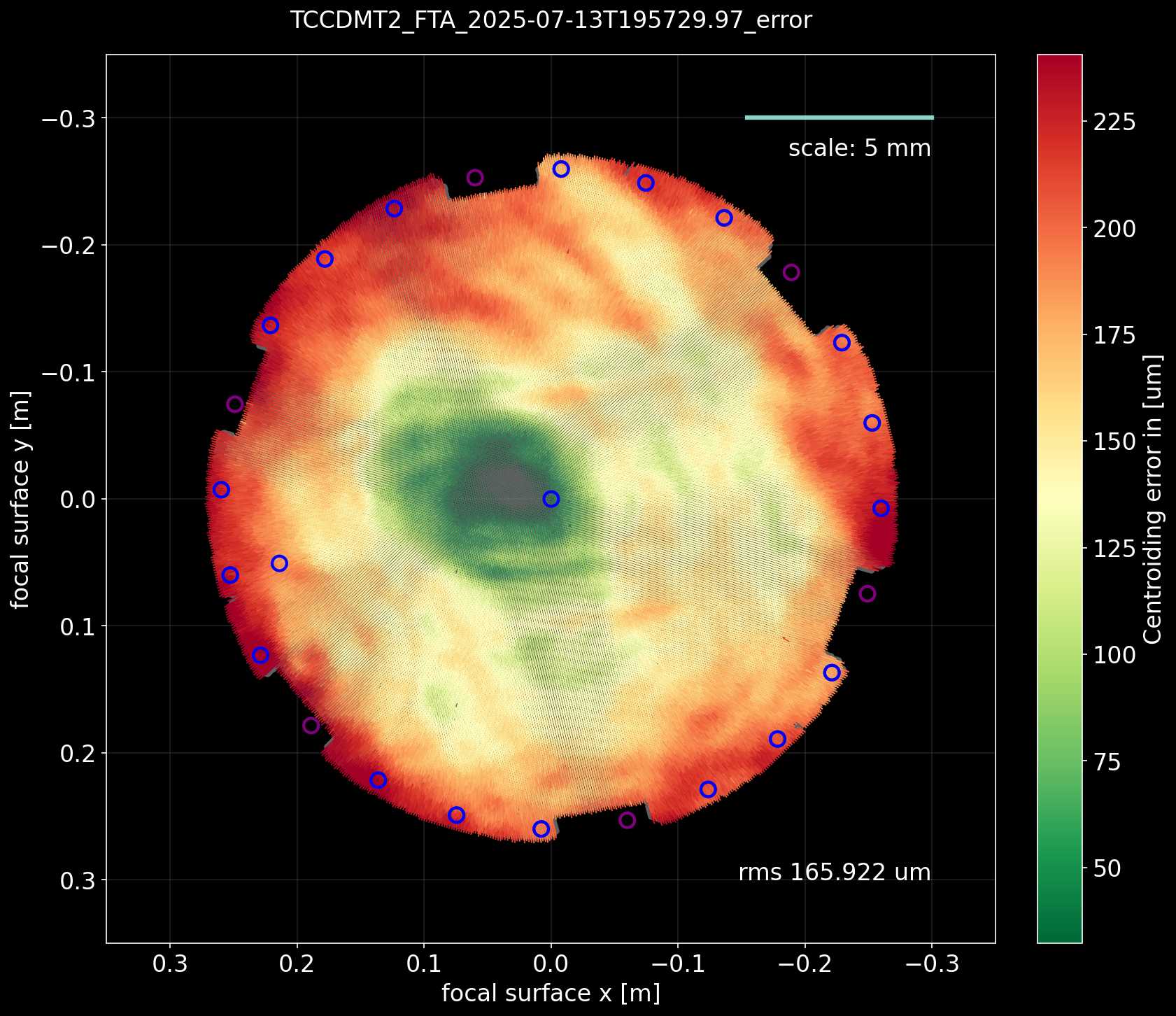} \\
\includegraphics[height=7cm]{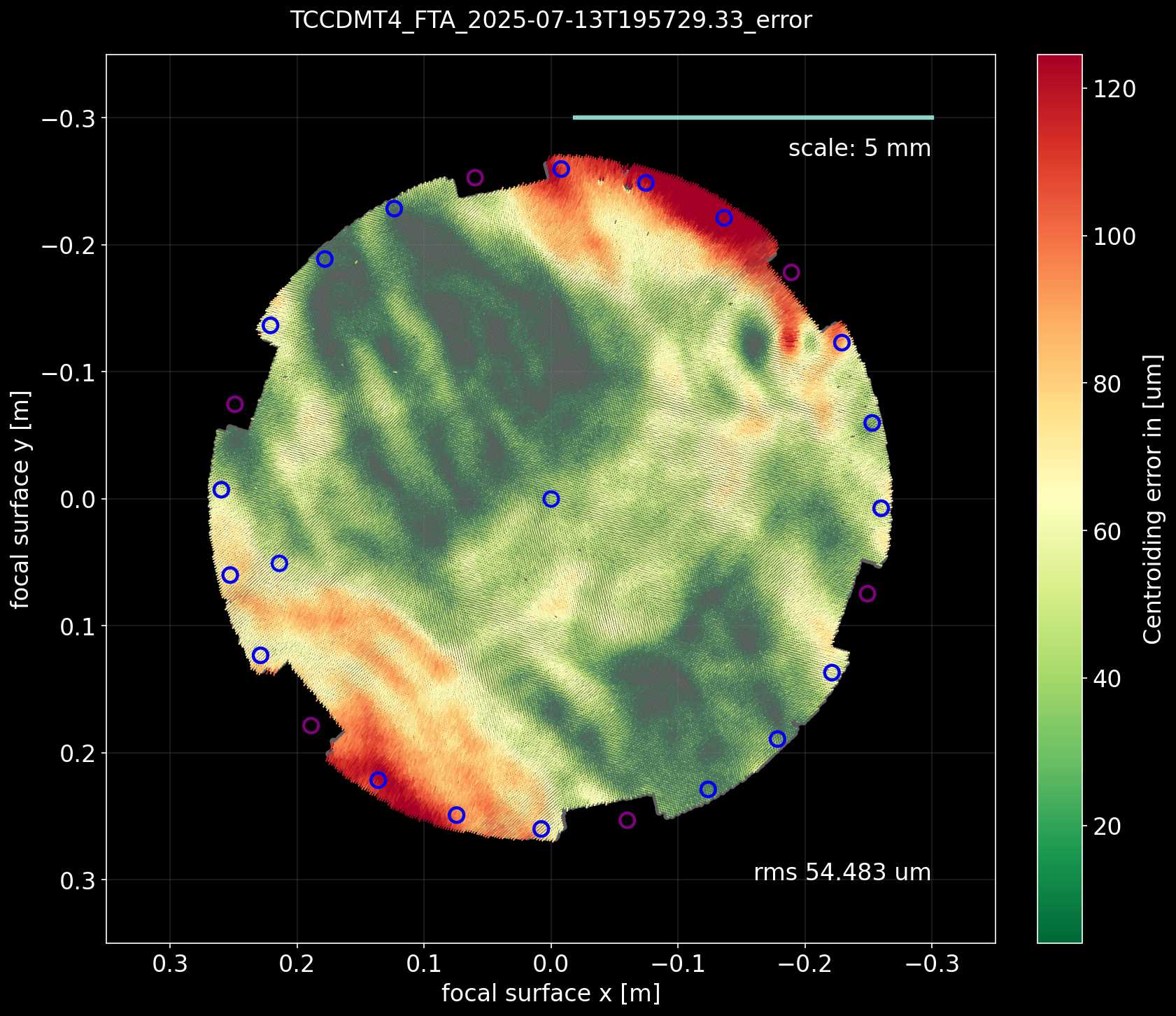}
\includegraphics[height=7cm]{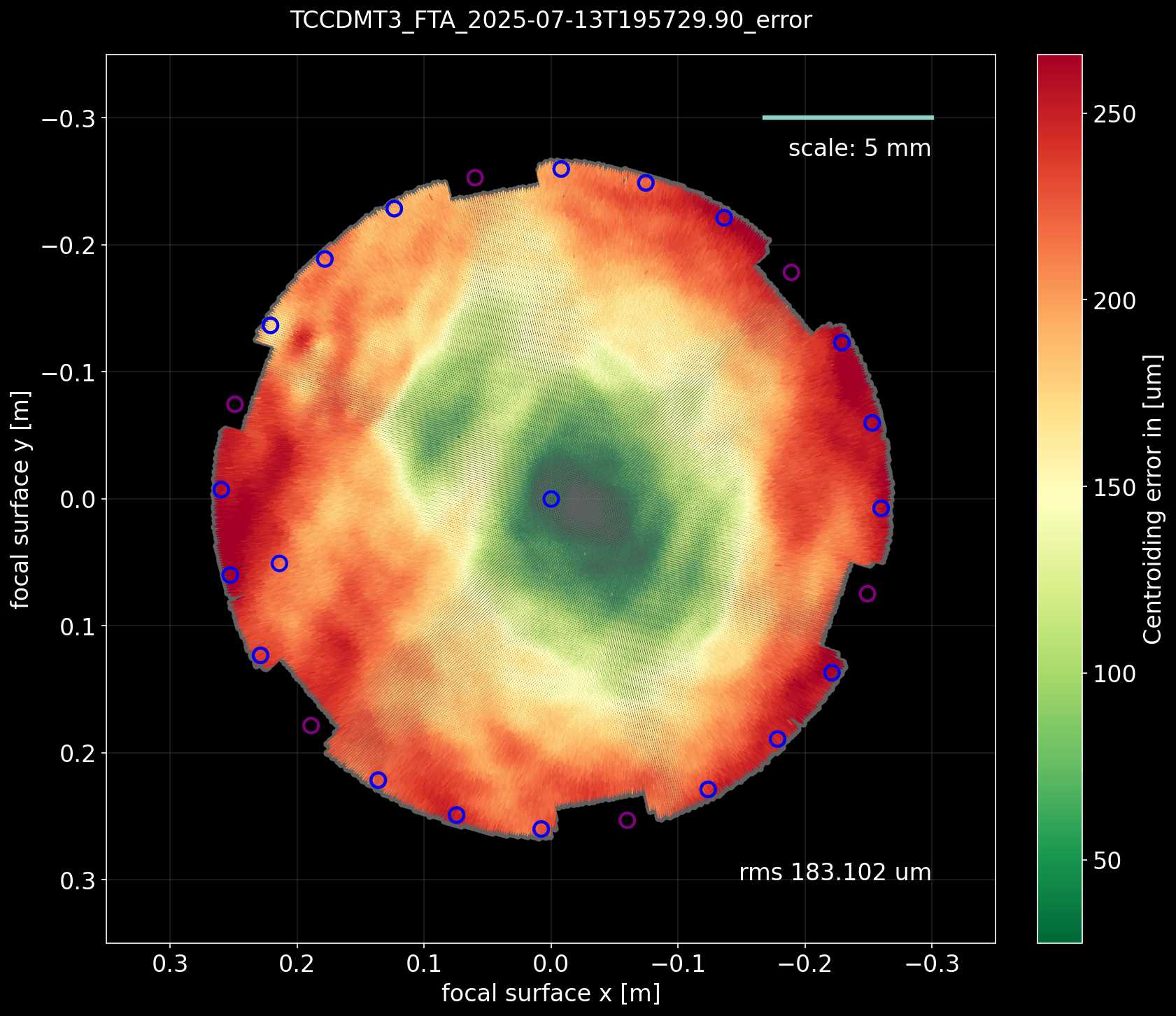}
\end{center}
\caption[No Calibration Centroiding Performance]{
    Centroiding error pattern without any optical calibration for all 4 cameras. The color scale is different for each image.
    \label{fig:metcam_nocal}
}
\end{figure}

Figure~\ref{fig:metcam_nocal} shows the centroiding performance of all 4 cameras without any calibration.
To see the error pattern without distortion, these error patterns are derived without using the fiducial correction algorithm discussed in Section~\ref{sec:fiducial}.
Please note that the color scale is different for each image.
The RMS error in these examples ranges from 50 to 200 $\mu m$, about 40 times larger than we can tolerate.

\subsection{Optical Model Parameters} \label{sec:calibration:optics}

With a correct spot association, the next step is to refine the parameters of the optical model.
This includes lens and surface positions as well as tilts in 3D space, radius of curvature, conic constant and other a-spherical parameters.
It does not make sense to optimize all available parameters though.
For example a lateral shift of a spherical lens is optically identical to a tilt of the surface.
All surfaces in the MetCams are either flat or spherical.\par

\begin{figure}[h]
\begin{center}
\includegraphics[height=7cm]{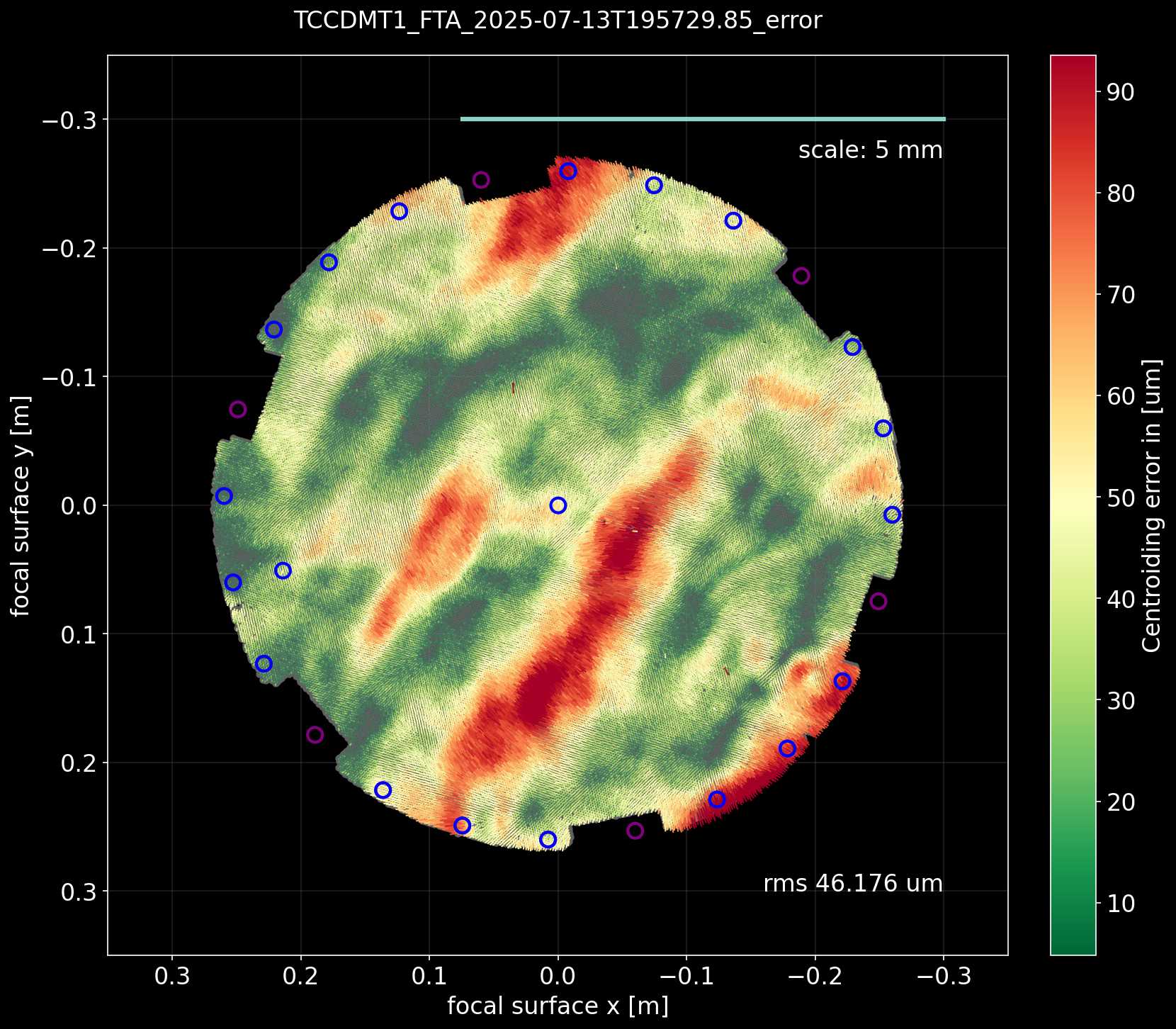}
\includegraphics[height=7cm]{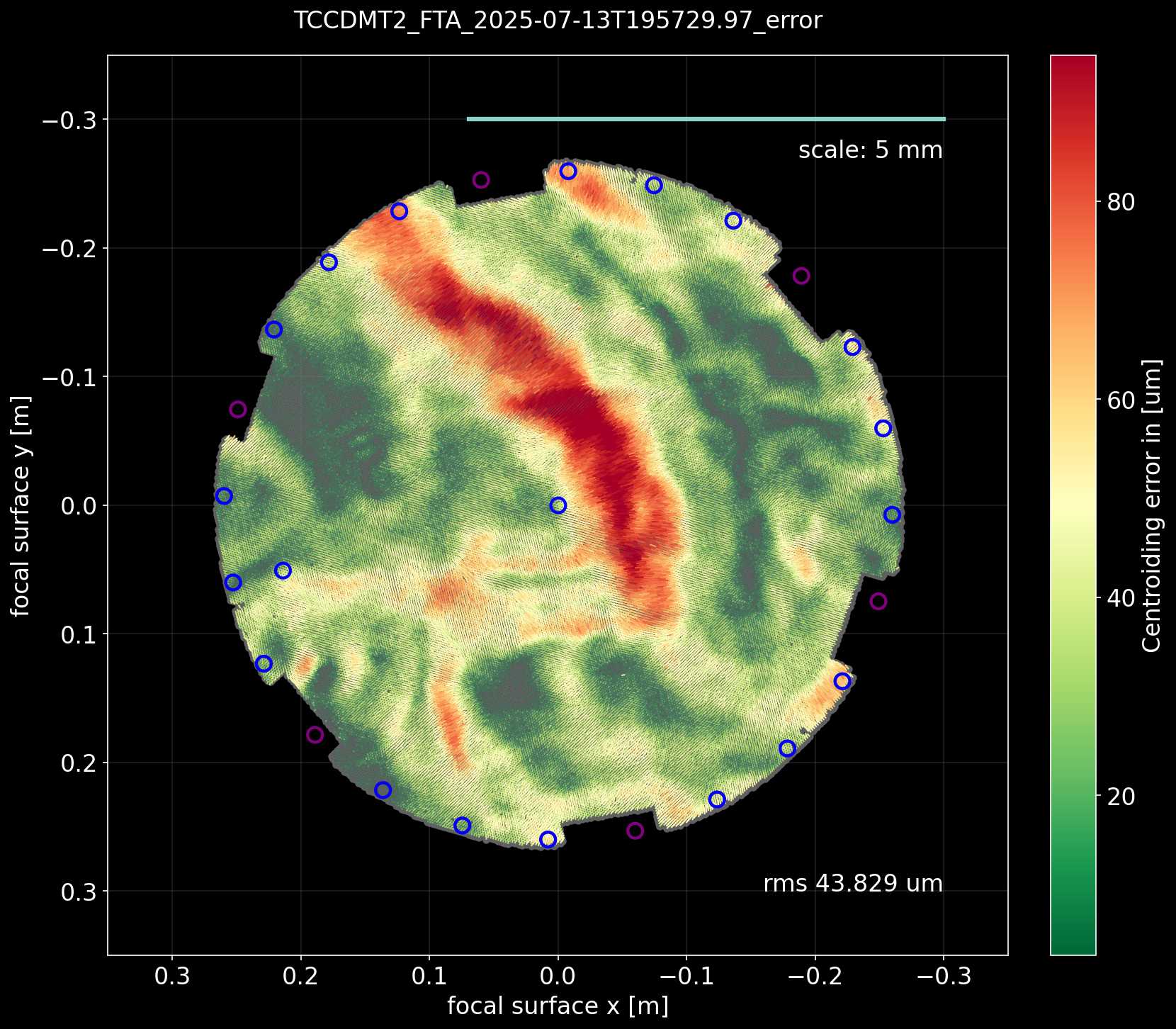} \\
\includegraphics[height=7cm]{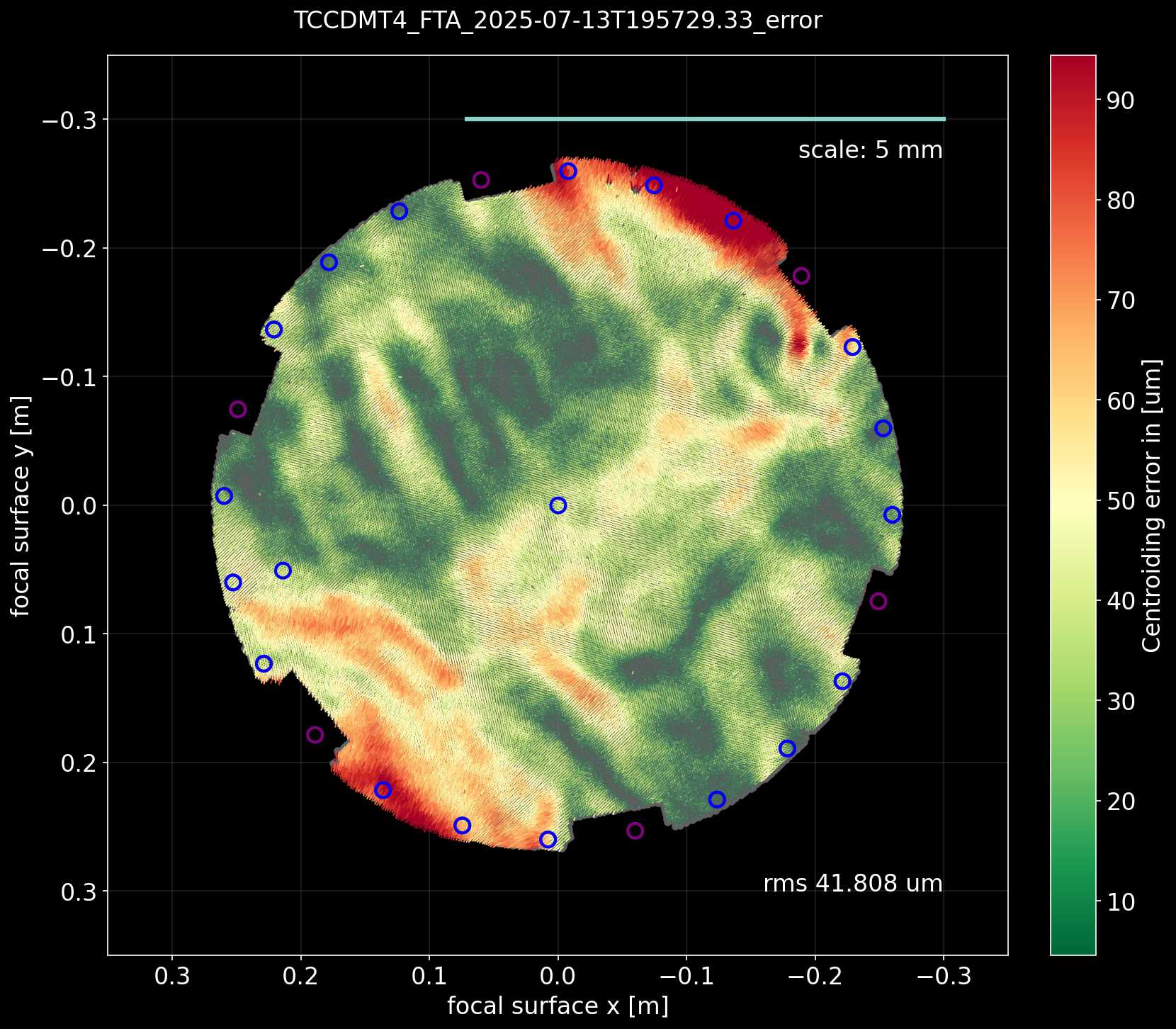}
\includegraphics[height=7cm]{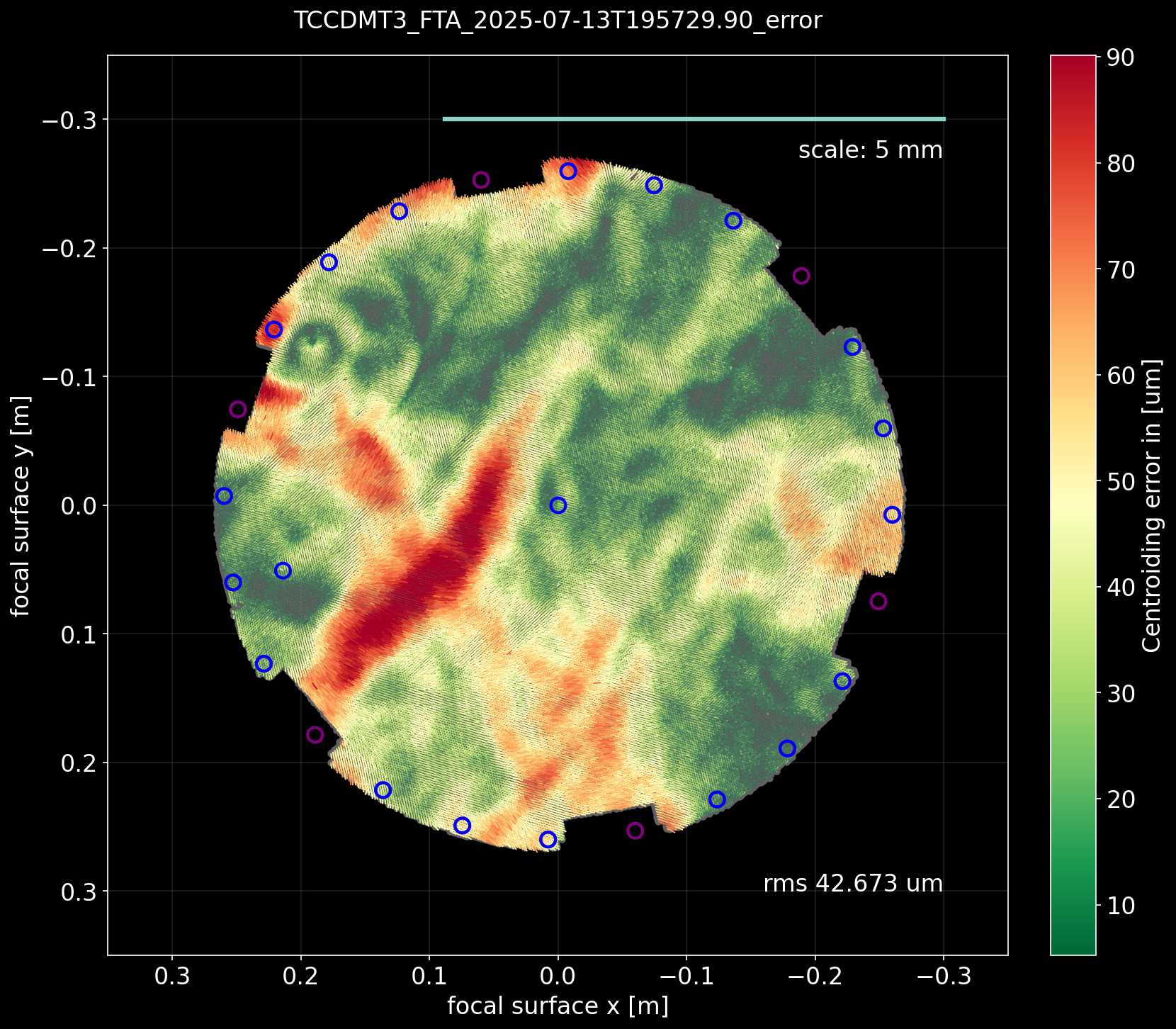}
\end{center}
\caption[Aligned Optical Model Centroiding Performance]{
    Centroiding error pattern after optical alignment for all 4 cameras. The color scale is different for each image.
    \label{fig:metcam_aligned}
}
\end{figure}

In our lab tests, we optimize for many parameters and there was a strong concern how to calibrate MetCams when they need to be replaced.
However, on the telescope, the alignment of lenses was less impactful and it was possible to restrict calibration one parameter per camera.
That parameter controls the scale of the image, see Figure~\ref{fig:metcam_aligned} for a visualization of the same data as in Figure~\ref{fig:metcam_nocal}, but including the optical alignment.
This parameter can also be aligned using the fiducials of AESOP, meaning the Metrology system can be re-calibrated without re-installing the MCU in case of a MetCam failure.\par

After optical alignment, the RMS centroiding error improved by to approx. 40 $\mu m$ to 50 $\mu m$, which is still at least a factor of 10 too large.

\subsection{Surface Normal Map} \label{sec:calibration:normal}

The distortortion depicted in Figure~\ref{fig:metcam_aligned} is stable when rotating the MCU, or the entire WFS including the lens group L1 to L4 (see Figure~\ref{fig:raytrace}).
We know that the distortions cannot come from the MetCams because when replacing the cameras at their mounting location, the distortions remain approximately stable.
It must follow therefore that either M1 or M2, or a combination of both, are causing the observed effect.
We have seen this issue before in the lab (but with different structure), as presented in \citenum{Winkler2024FTA} and software to solve these issues is available.\par

\begin{figure}[h]
\begin{center}
\includegraphics[width=0.49\textwidth]{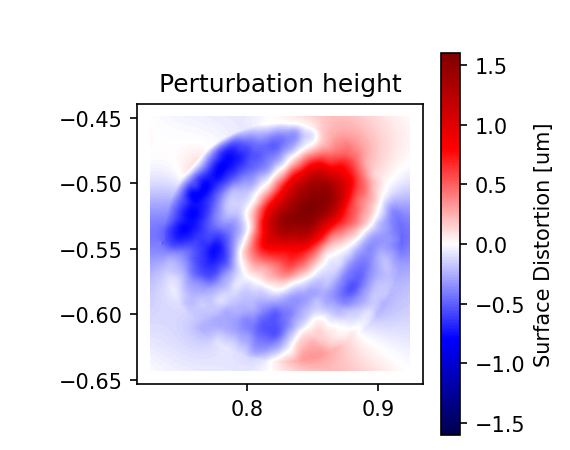} 
\includegraphics[width=0.49\textwidth]{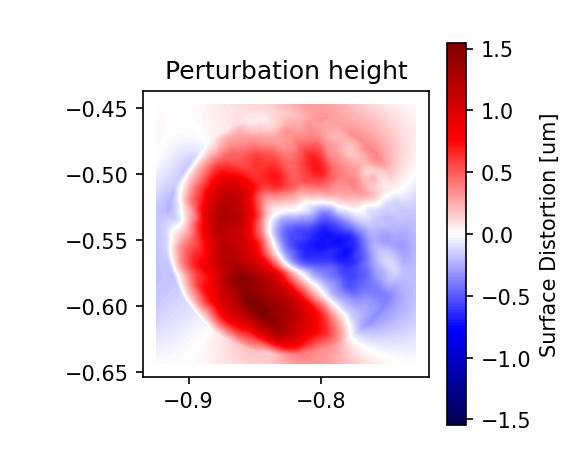} \\
\includegraphics[width=0.49\textwidth]{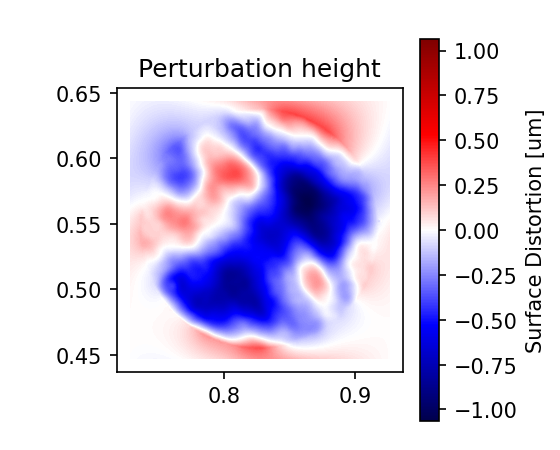} 
\includegraphics[width=0.49\textwidth]{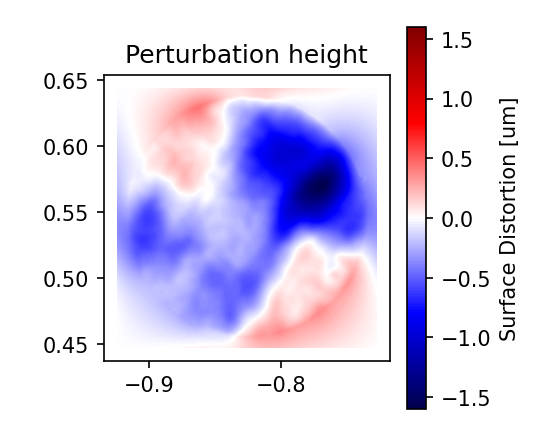}
\end{center}
\caption[Mirror Surface Profiles]{
    Height maps and surface normals of all 4 sections of the primary mirror.
    \label{fig:heightmap}
}
\end{figure}

We chose to attribute the distortion to the surface of M1, implying that the polishing of the mirror (at least at the Section observed by MetCams) is suboptimal, though as said above, at least part of it could originate from M2.
Using the MCU, it is possible to estimate the surface profile of M1 for the sections that are visible by the metrology cameras, see the right panel in Figure~\ref{fig:raytrace} that indicates the coverage of a MetCam on M1 in the lower left region of the image.\par

Our solution is to represent the surface shape of M1 in our optical model that would cause exactly the error pattern we observe.
By using the ray-tracer, we can compute rays that go from the MetCam detector to M1 and know where they hit the mirror surface.
At the same time, we also know where the rays should end up on the MCU, giving us an observed reflection on the mirror that is different from the ideal surface.
Since a reflection on the mirror surface is entirely described by the surface normal vector of the mirror surface, we can construct a normal map that defines the slope of the mirror surface for each ray.\par

\begin{figure}[h]
\begin{center}
\includegraphics[height=7cm]{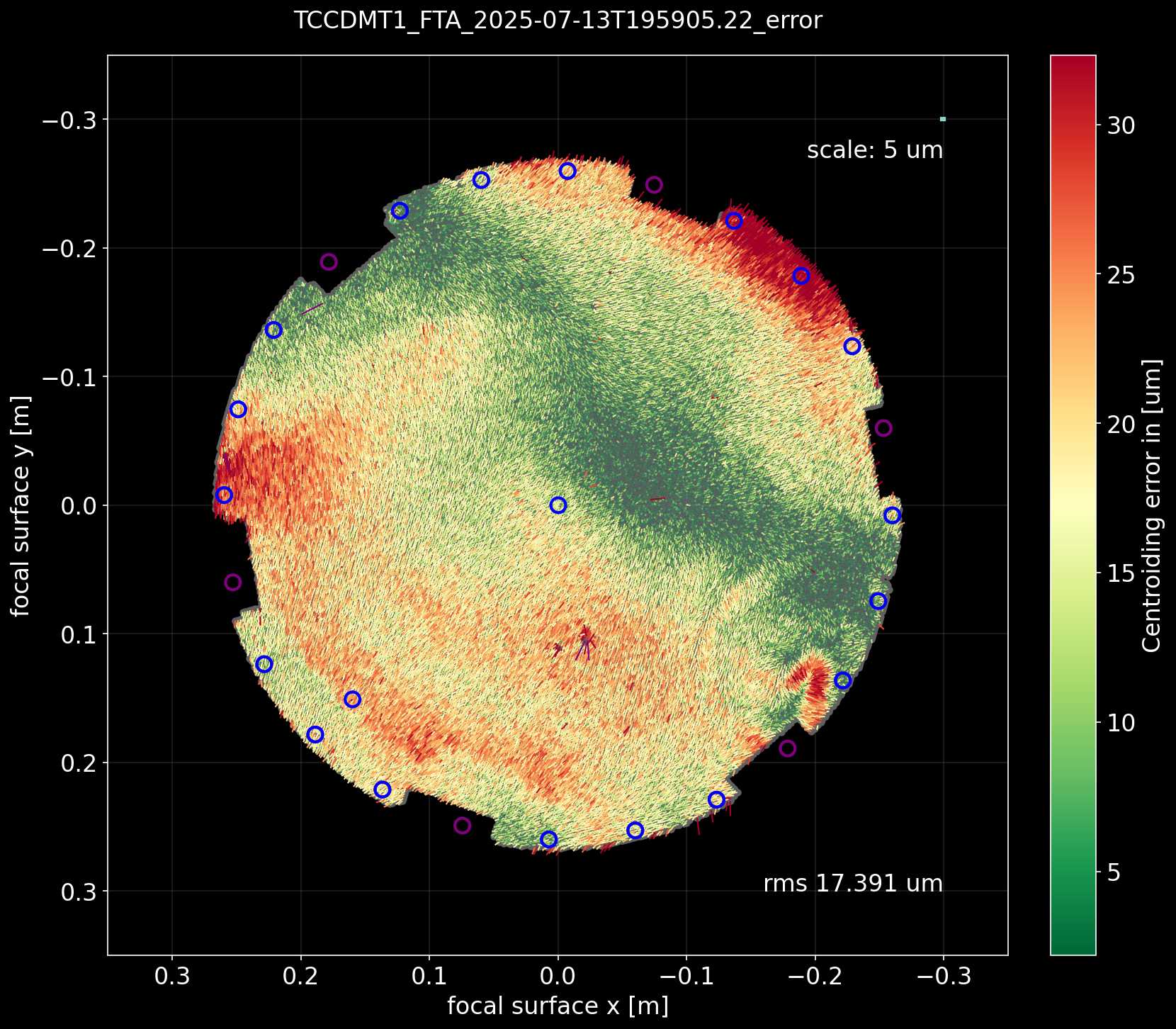}
\includegraphics[height=7cm]{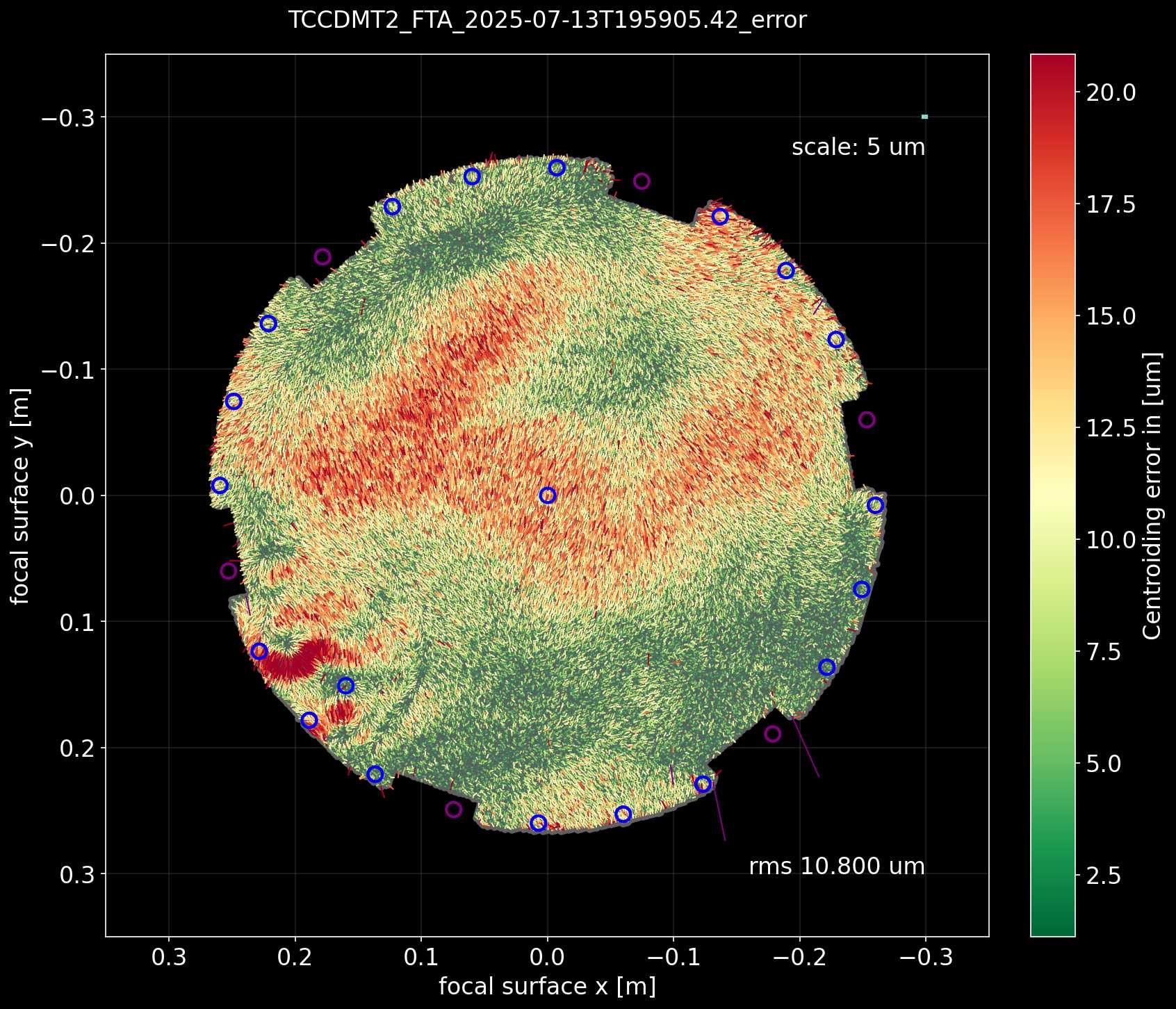} \\
\includegraphics[height=7cm]{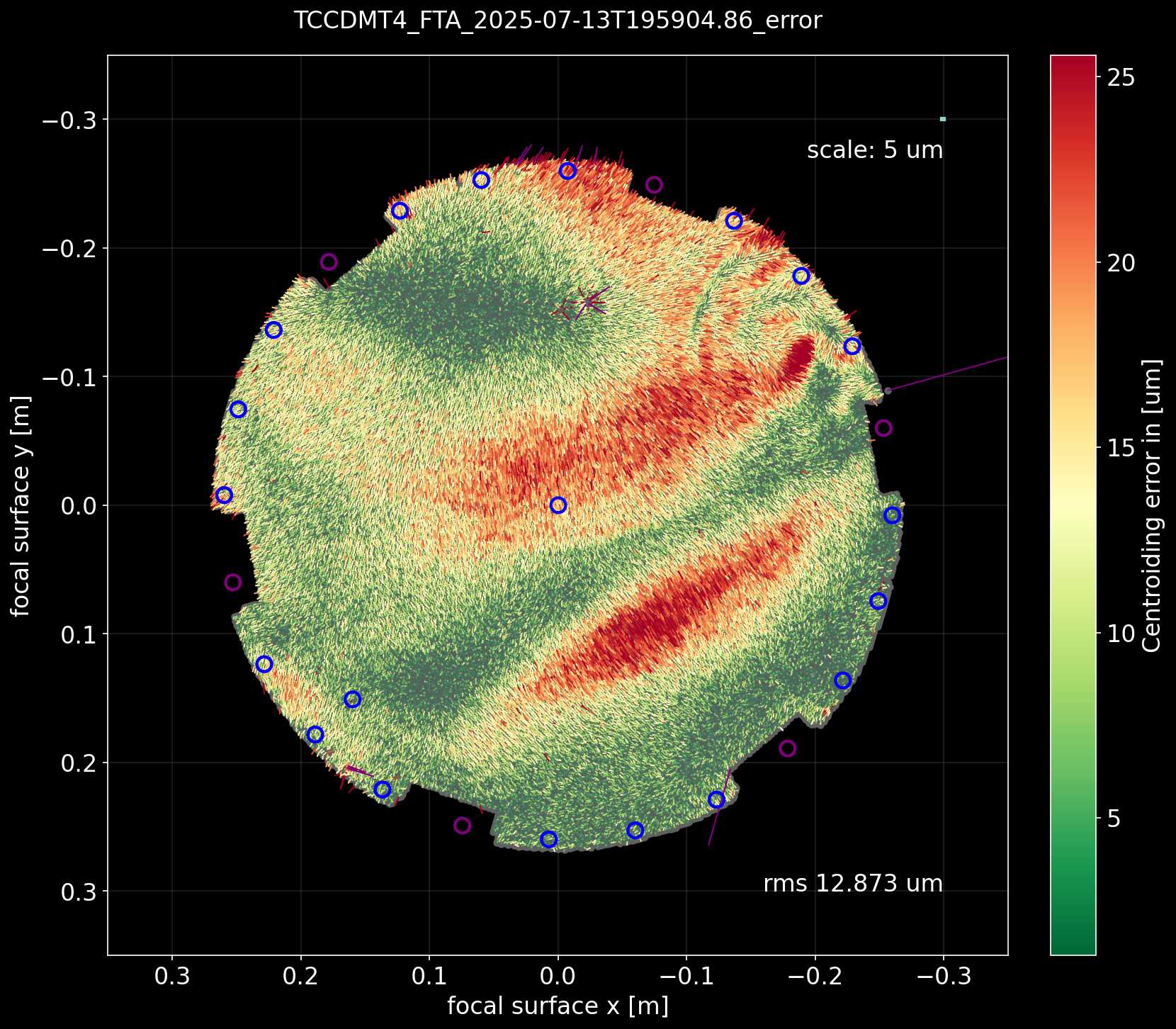}
\includegraphics[height=7cm]{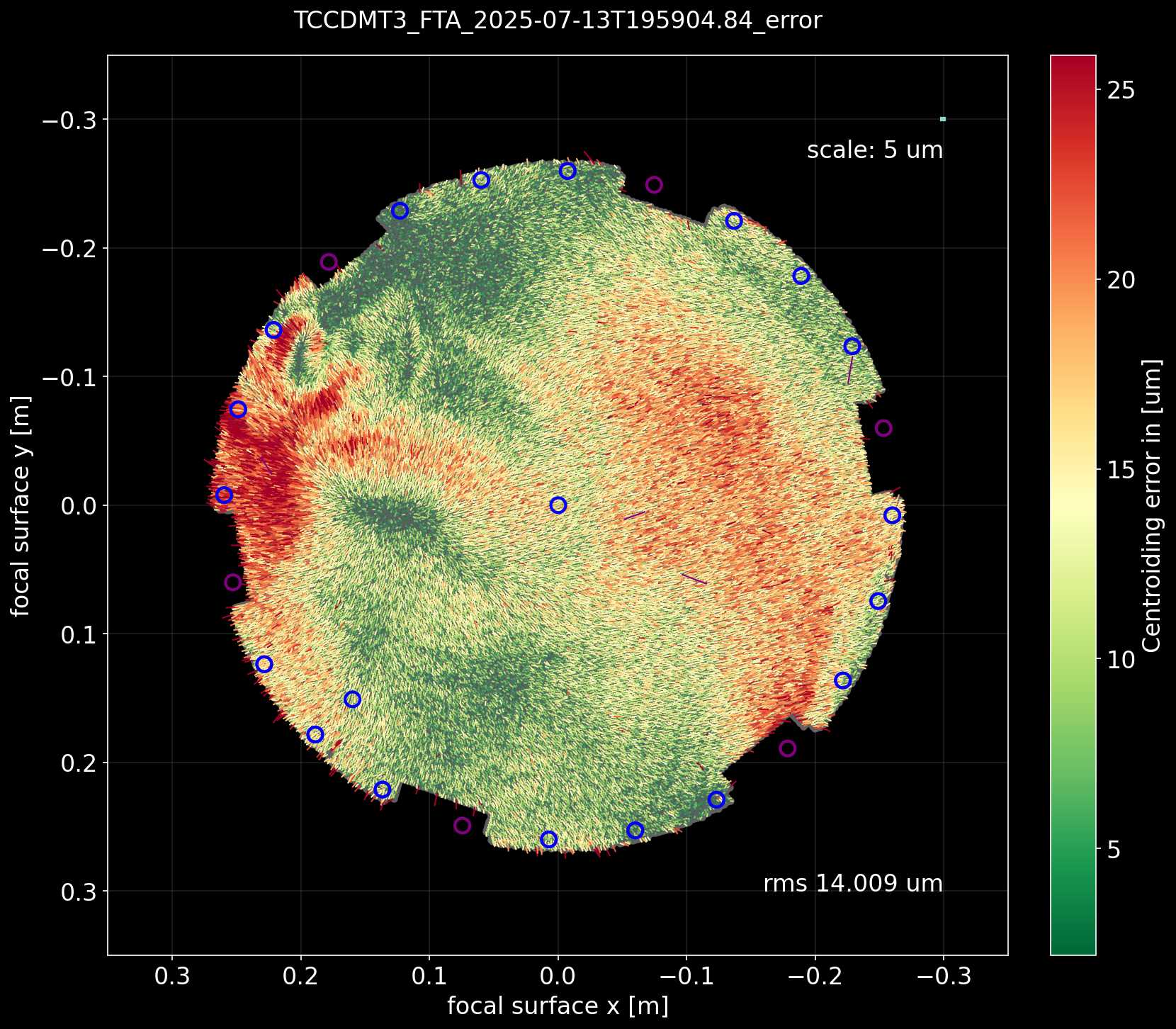}
\end{center}
\caption[Normal Map Calibrated Centroiding Performance]{
    Centroiding error pattern after optical alignment and correction by the normal map mirror surface. The color scale is different for each image.
    \label{fig:metcam_normalmaped}
}
\end{figure}

The normal map is modeled by a 2-dimensional cubic spline interpolation, which is equivalent to the first derivative in 2 dimensions.
In the ray-tracer, we use that normal map directly.
However, for visualization purposes, we derive a height map by de-convolution of the normal map on the mirror surface, which is shown in Figure~\ref{fig:heightmap}.\par

\begin{figure}[h]
\begin{center}
\includegraphics[height=7cm]{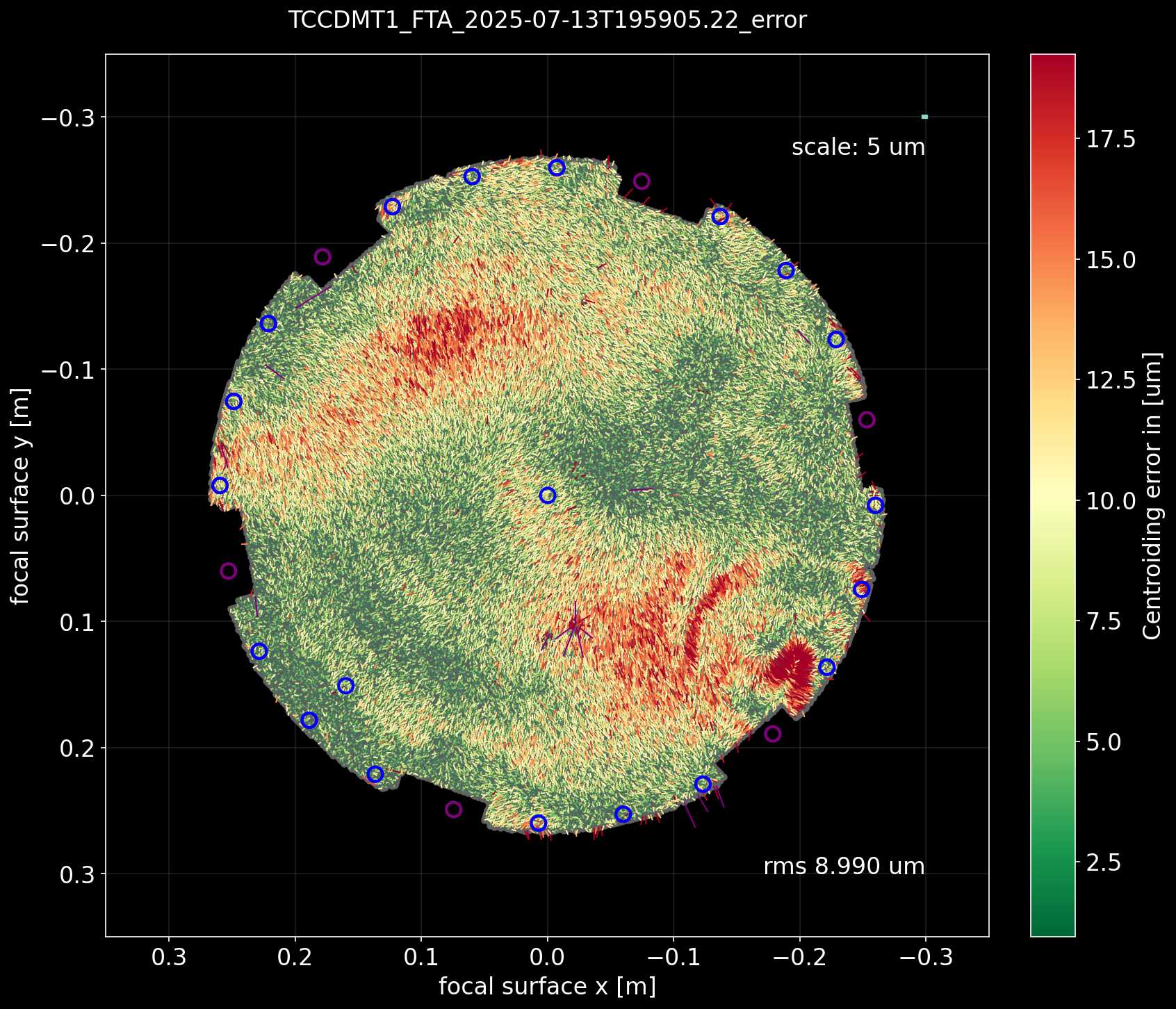}
\includegraphics[height=7cm]{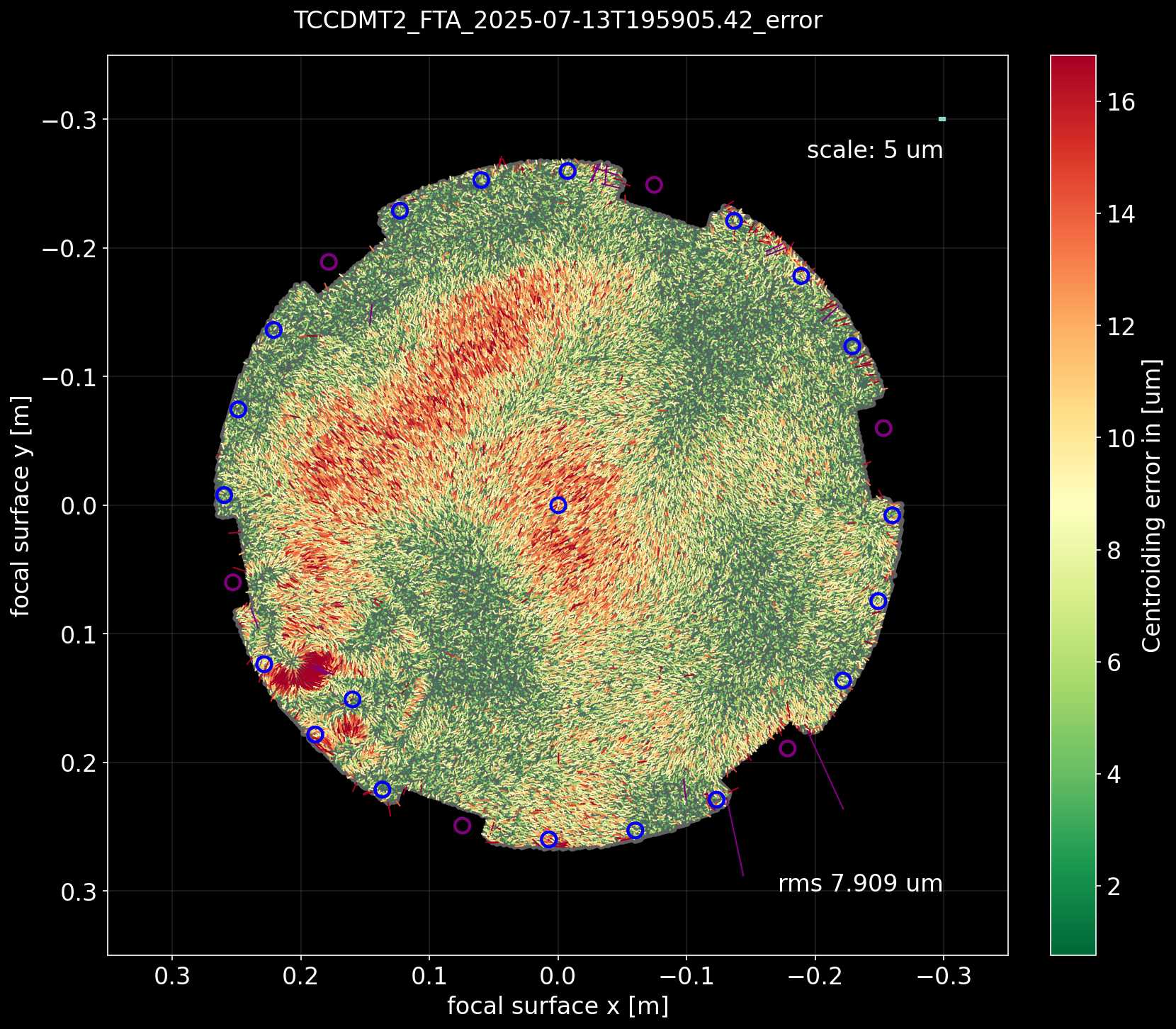} \\
\includegraphics[height=7cm]{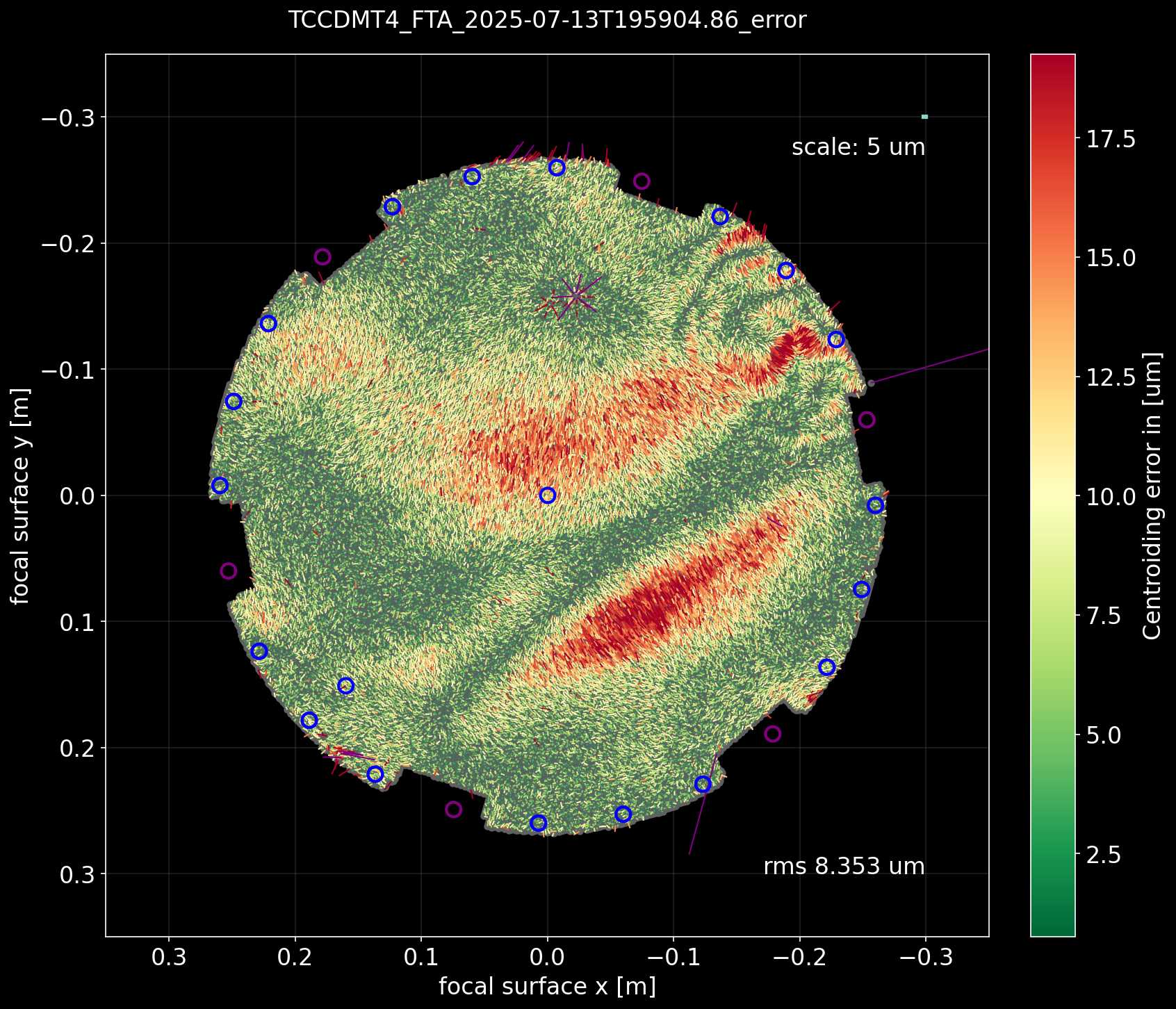}
\includegraphics[height=7cm]{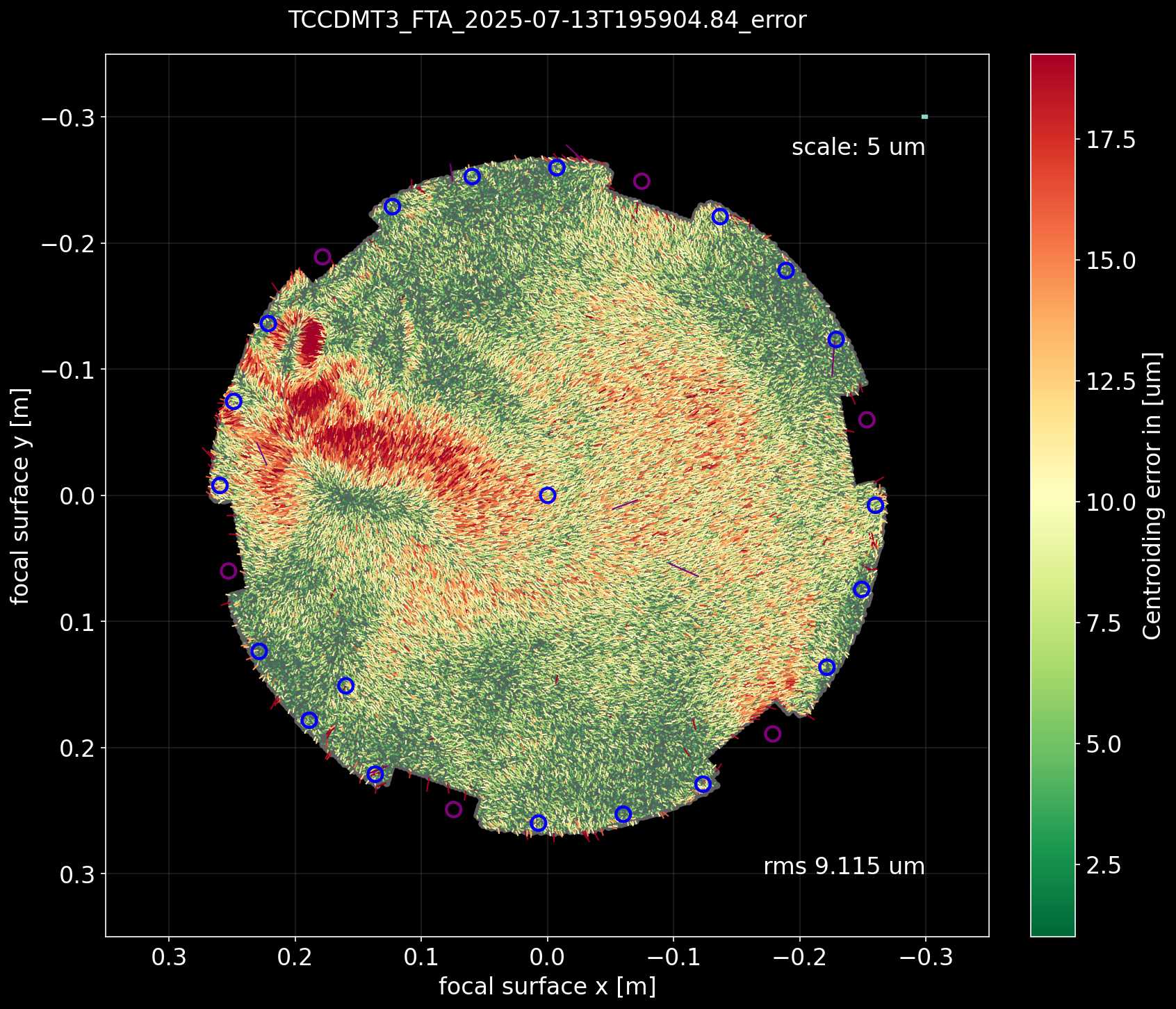}
\end{center}
\caption[Residual Calibration Performance]{
    Centroiding error pattern after optical alignment, correction by the normal map mirror surface and fiducial correction. The color scale is different for each image.
    \label{fig:metcam_fiducialed}
}
\end{figure}

It is also possible to use the same method to measure surface irregularities on lenses, and we did this for the back side of L1 (the top most lens).
The images in Figure~\ref{fig:metcam_normalmaped} show a residual swirl in the corner of each metcam image, that is not explained by the mirror surface.
The feature is in a different corner in each MetCam measurement because each camera sees the center of L1 from a different perspective and the rays pass through the same section in a different angle.
The feature is most likely originates from polishing errors on the backside of L1, which has a non-spherical shape that is prone to this type of manufacturing defacts.
We did improve this location but were not able to eliminate it entirely. \par

In the raytracer, we are using the normal map directly (without the conversion to a height profile) because it is less restricted and provides better performance in correcting the centroiding errors.
The residual error after modeling the mirror surface with the normal map but without applying the fiducial correction is shown in Figure~\ref{fig:metcam_normalmaped}.
The magnitude of all RMS values is now considerably reduced to below 20 $\mu m$ and the optical calibration is approaching its limit.\par

It is hard to state conclusively how accurate this method actually is in measuring the surface profile of the mirror.
Judging from the residual error pattern however, the relative surface height might be within 5 to 10 nm, though it is possible that the uncertainty is larger as we cannot exclude that part of the error originates from M2.
A calibration plate like the MCU could be a valuable tool in optics characterization in other applications as well.

\subsection{Residual Errors} \label{sec:calibration:residual}

After the above calibration, air turbulence is the primary impact on measurement location.
While we cannot remove this completely, we can mitigate its impact.\par

After applying the fiducial correction method, using the declared fiducials on the MCU, indicated by small blue circles, the residual error pattern is shown in Figure~\ref{fig:metcam_fiducialed}.
The remaining errors drop to between 8 and 10 $\mu m$ in the examples shown here, but in general the performance is better at around 5 to 6 um.
We chose to show less accurate images here as they provide better material for explaining the concepts.
When combining these results, a residual centroiding error between 4 and 5 $\mu m$ is achieved, see the left panel of Figure~\ref{fig:combined}.
This falls short of the intended (and proven in the lab) performance of 3 $\mu m$ RMS, though the performance is good enough to have minimal impact on the operation of the instrument. \par

\begin{figure}[h]
\begin{center}
\includegraphics[height=3.5cm]{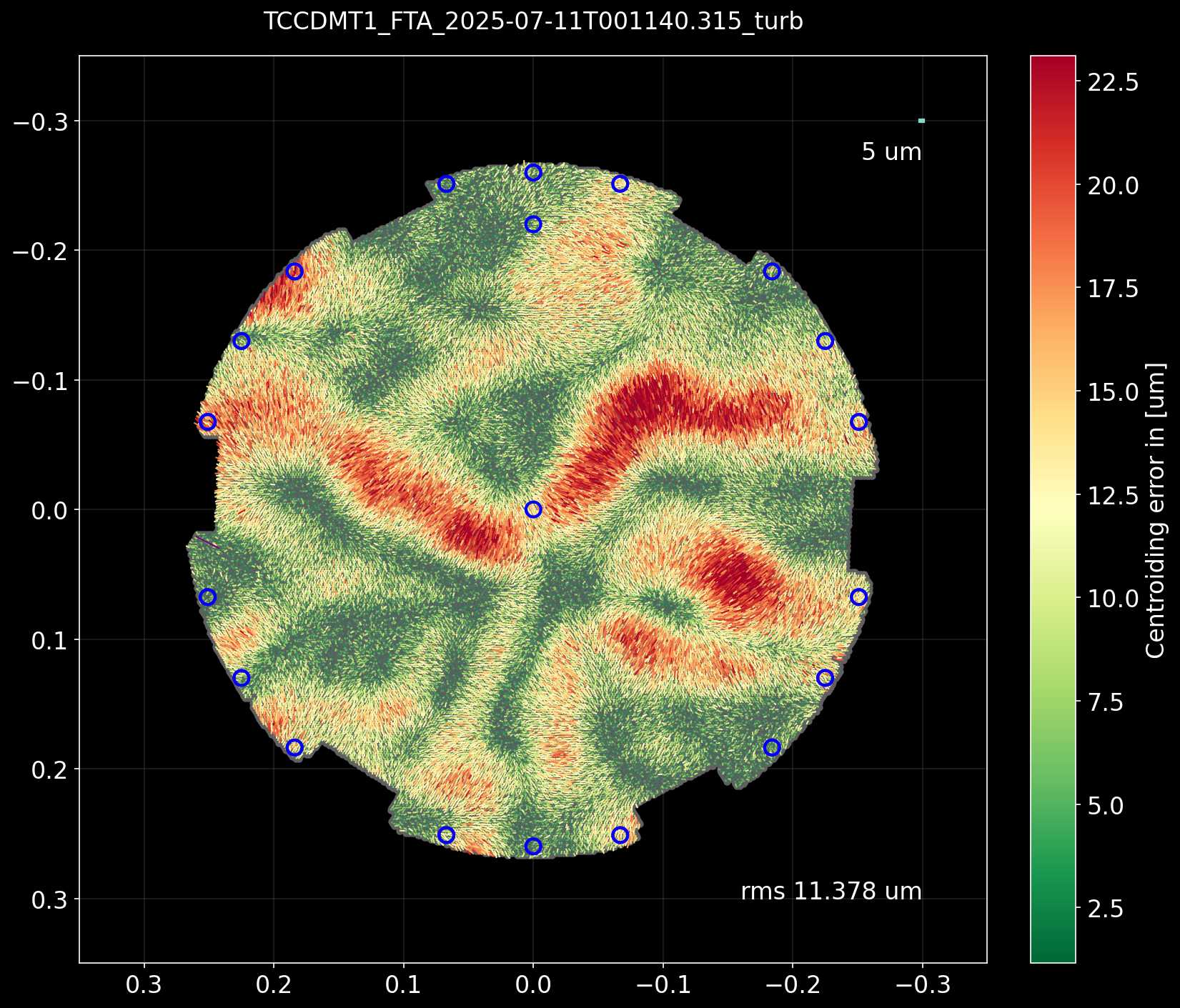}
\includegraphics[height=3.5cm]{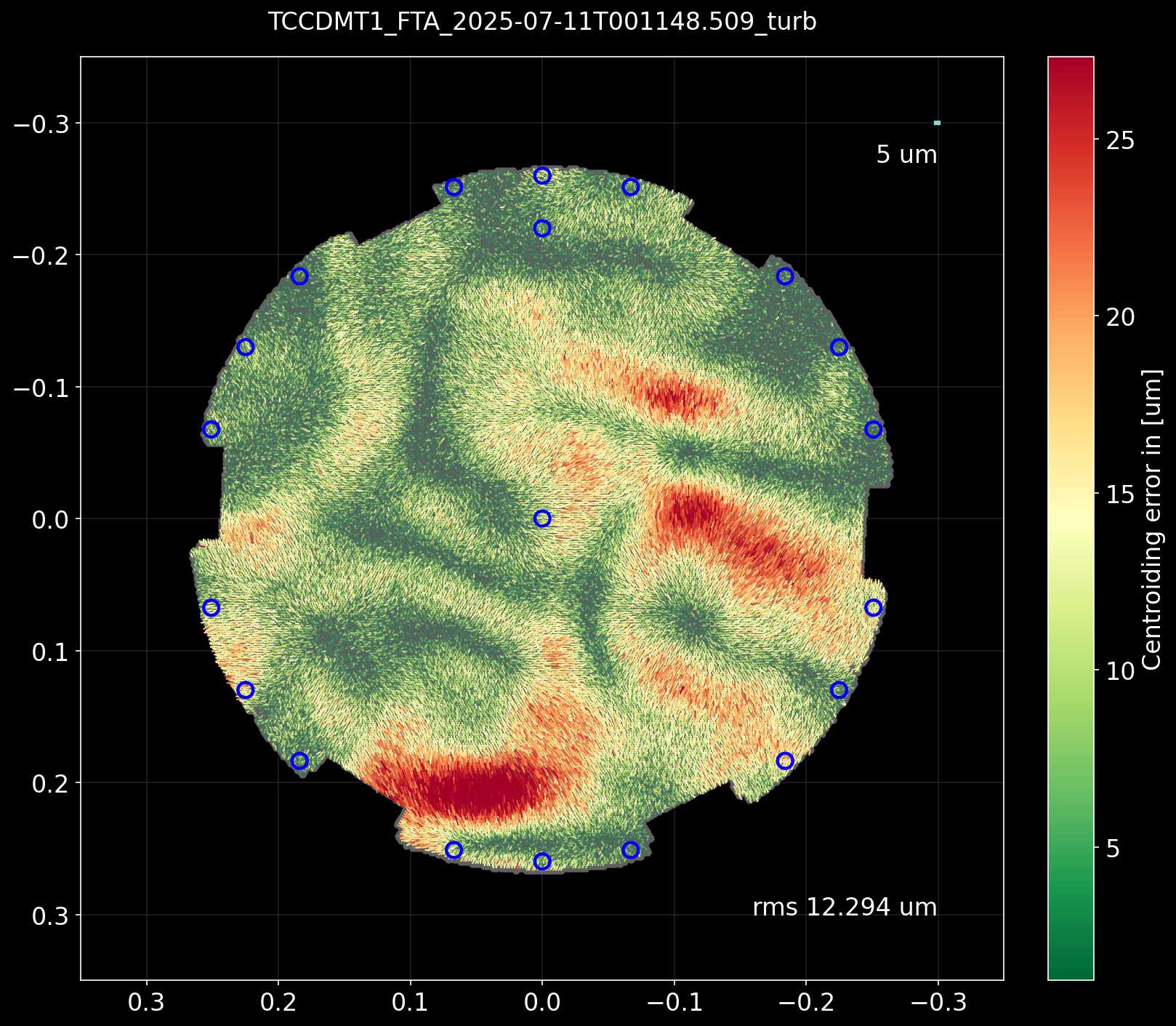}
\includegraphics[height=3.5cm]{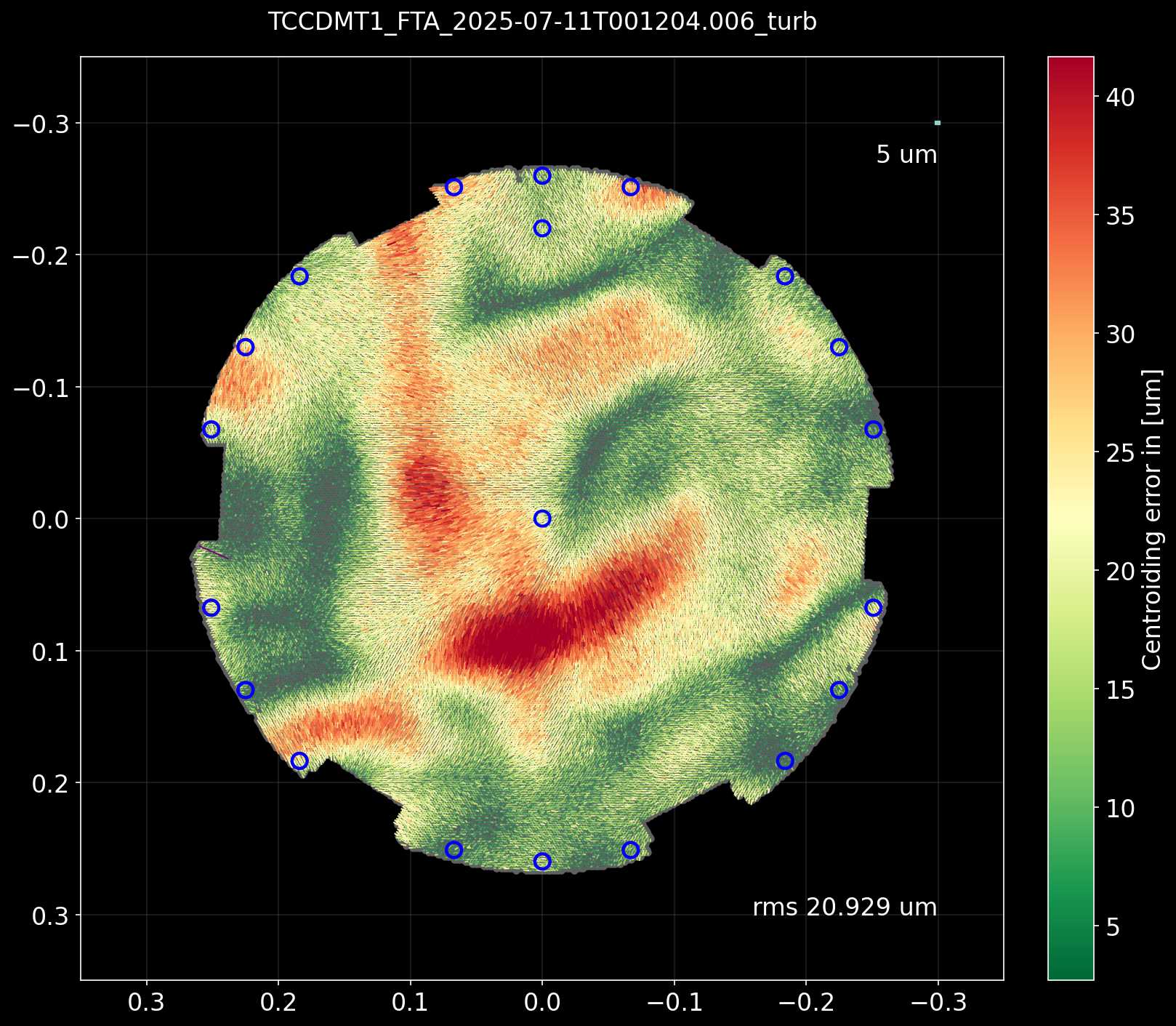}
\includegraphics[height=3.5cm]{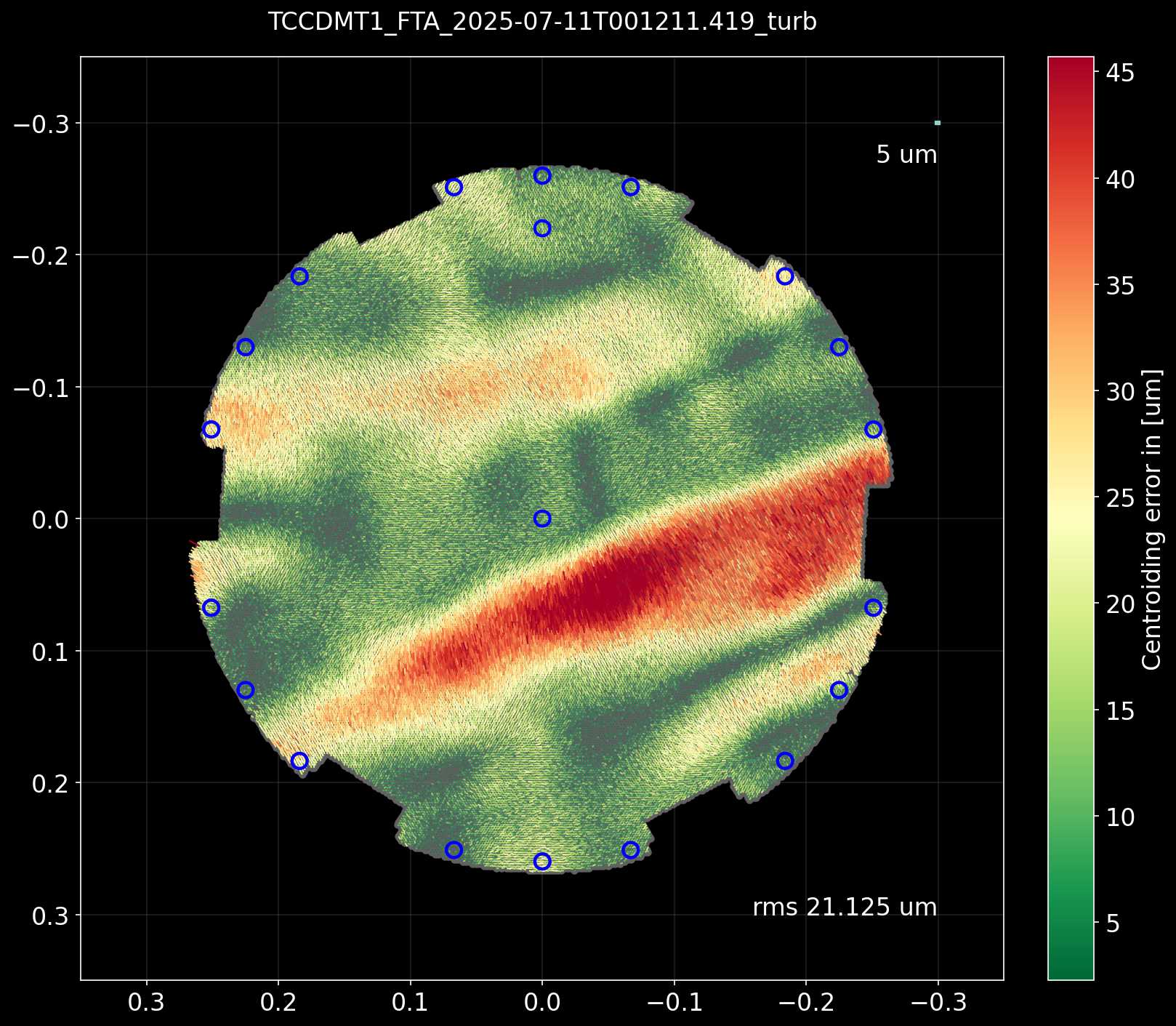}\\
\includegraphics[height=3.5cm]{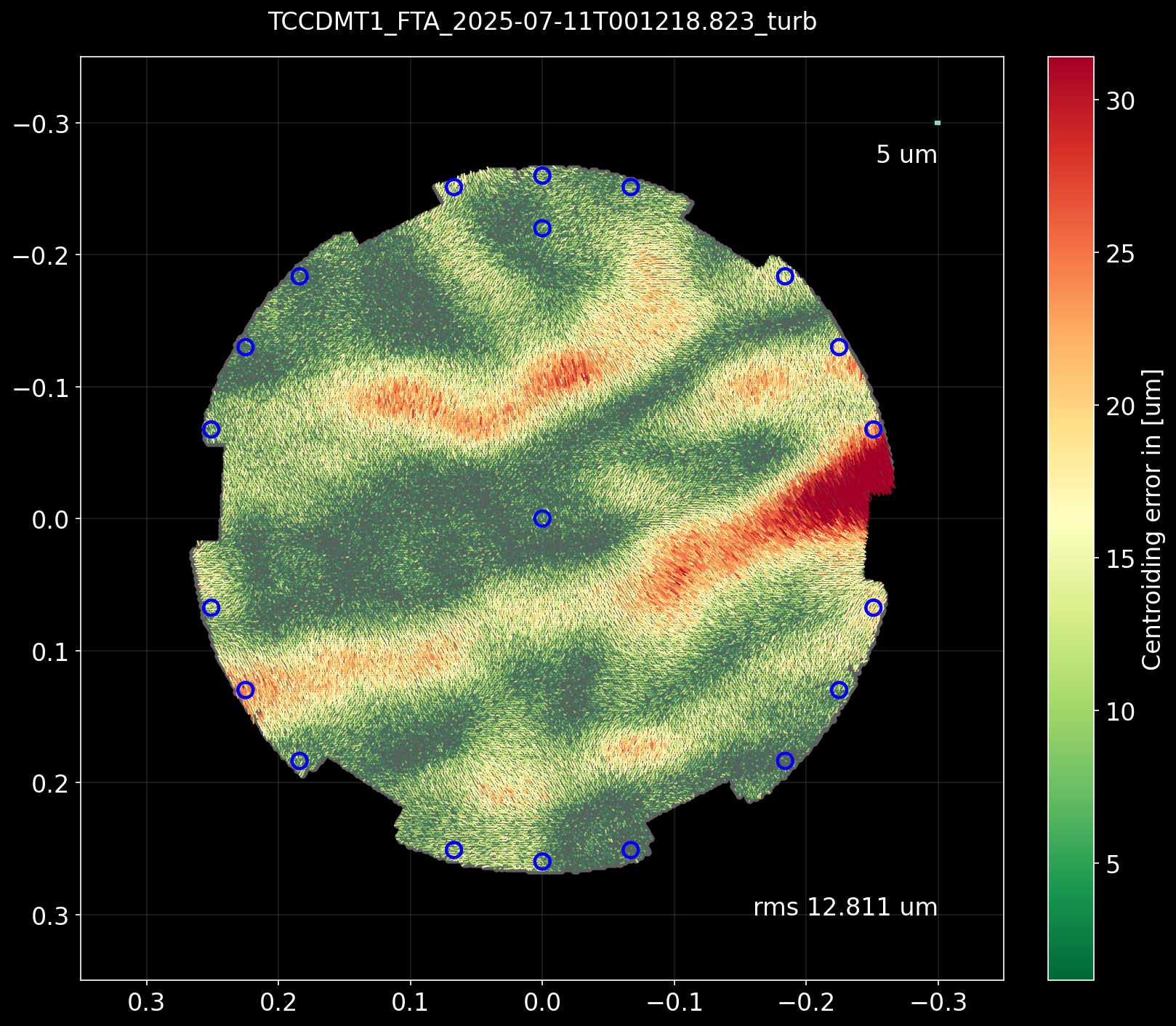}
\includegraphics[height=3.5cm]{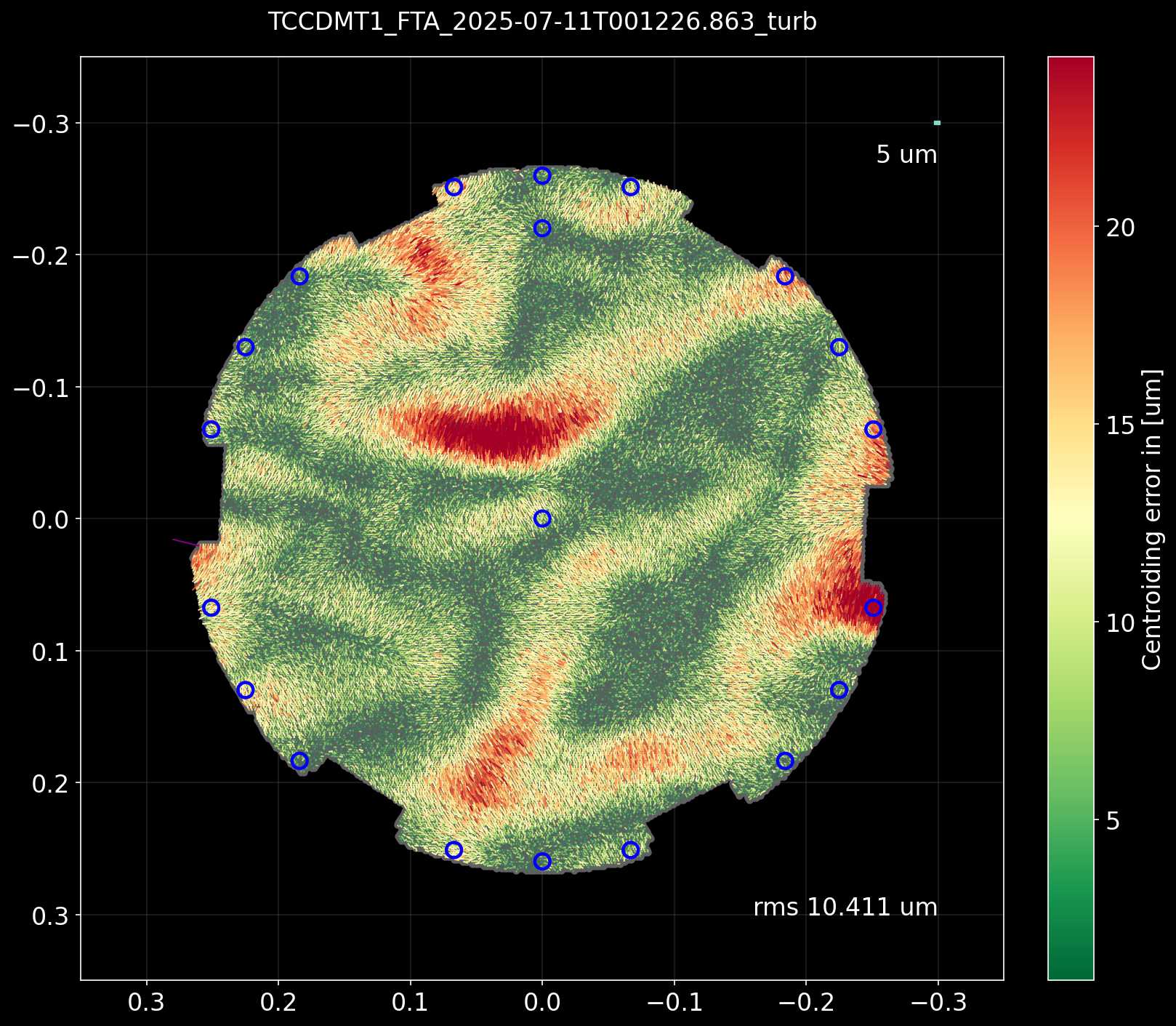}
\includegraphics[height=3.5cm]{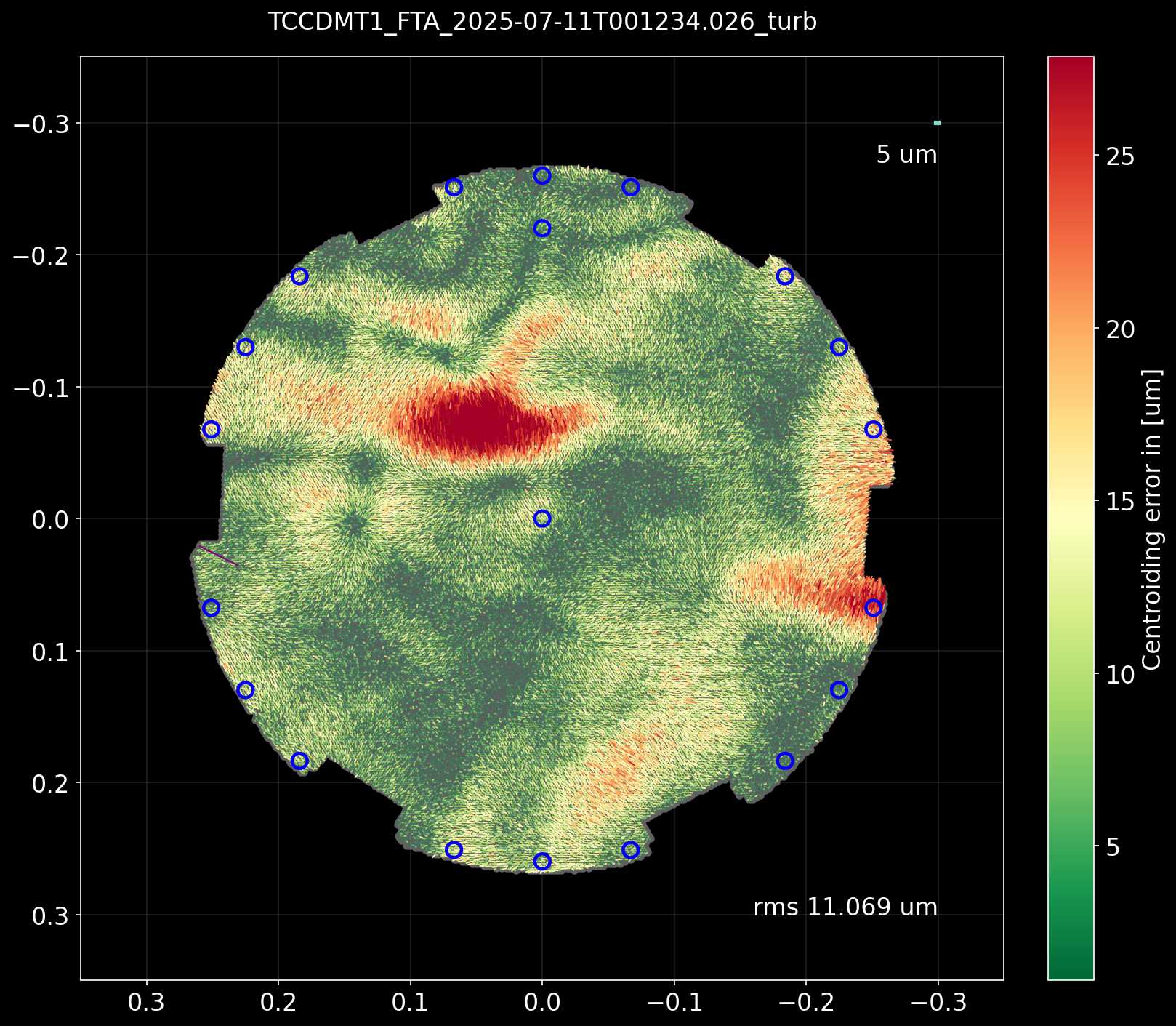}
\includegraphics[height=3.5cm]{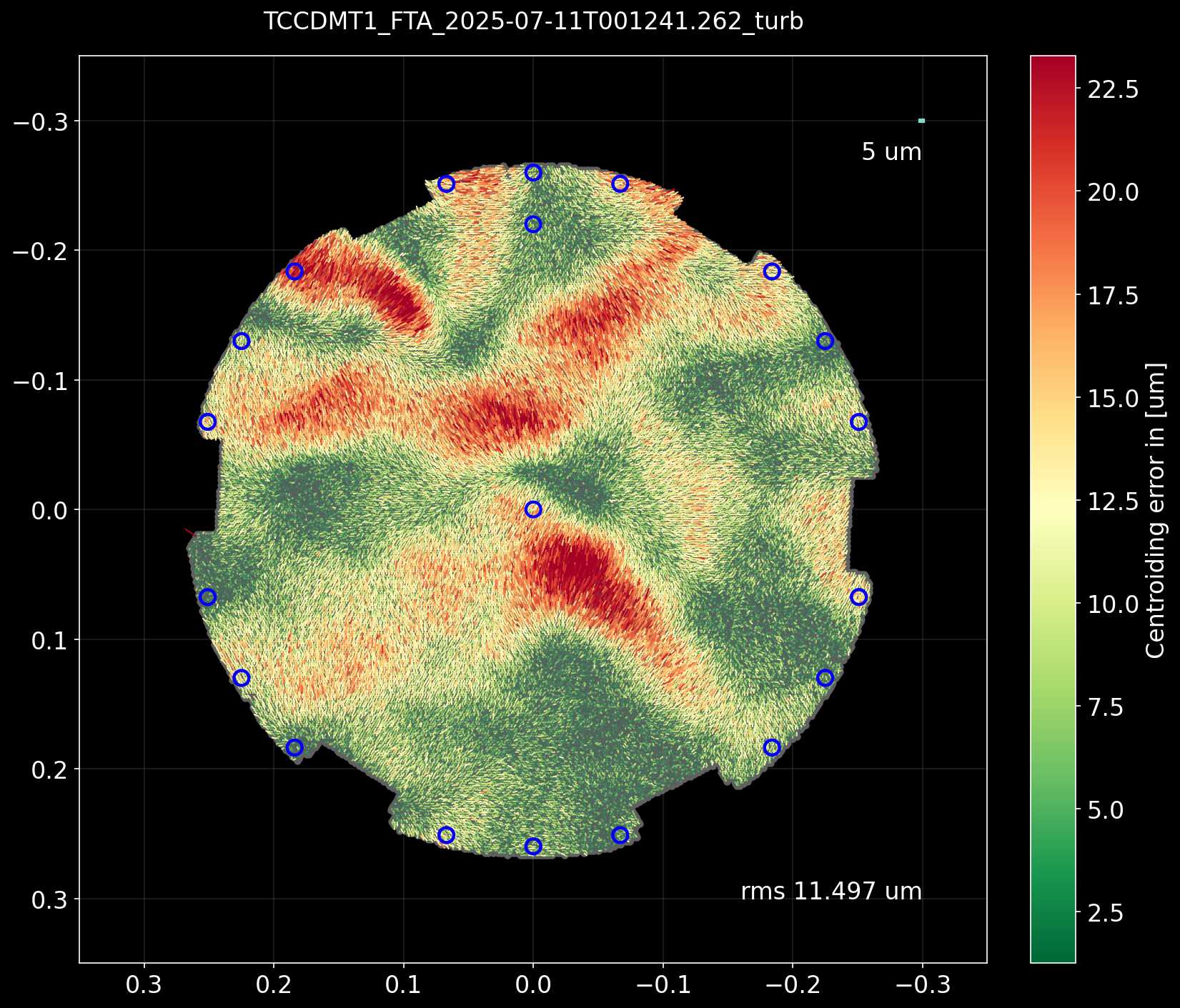}
\end{center}
\caption[Centroid differences in a sequence of images]{
    Centroiding error pattern that emerges from visualizing the differences between images of the same camera, taken every 8 seconds.
    \label{fig:metcam_sequence}
}
\end{figure}

When looking at a sequence of images, and visualizing the difference of spot measurements in between images, the impact of the seeing pattern emerges.
The fiducial correction system does a good job removing the linear component of the observed error pattern, but the non-linear component remains.
These image sequences in Figure~\ref{fig:metcam_sequence} show that it is very advantageous to have multiple cameras observing the same fibres as the errors inside the optical path are different to each camera.\par

\begin{figure}[h]
\begin{center}
\includegraphics[height=4.4cm]{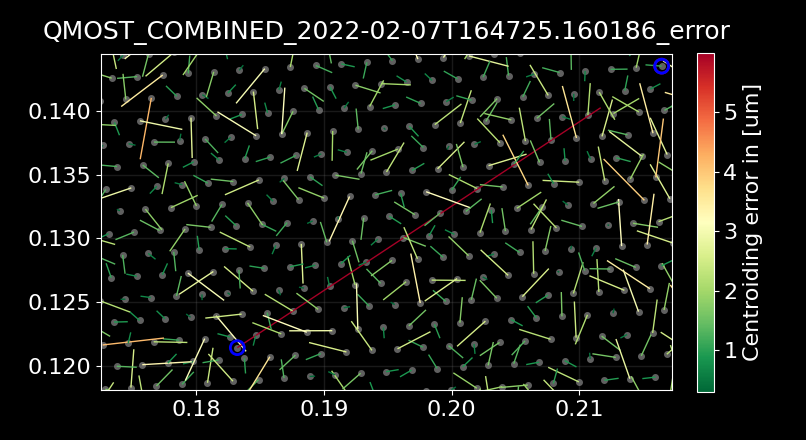}
\includegraphics[height=4.4cm]{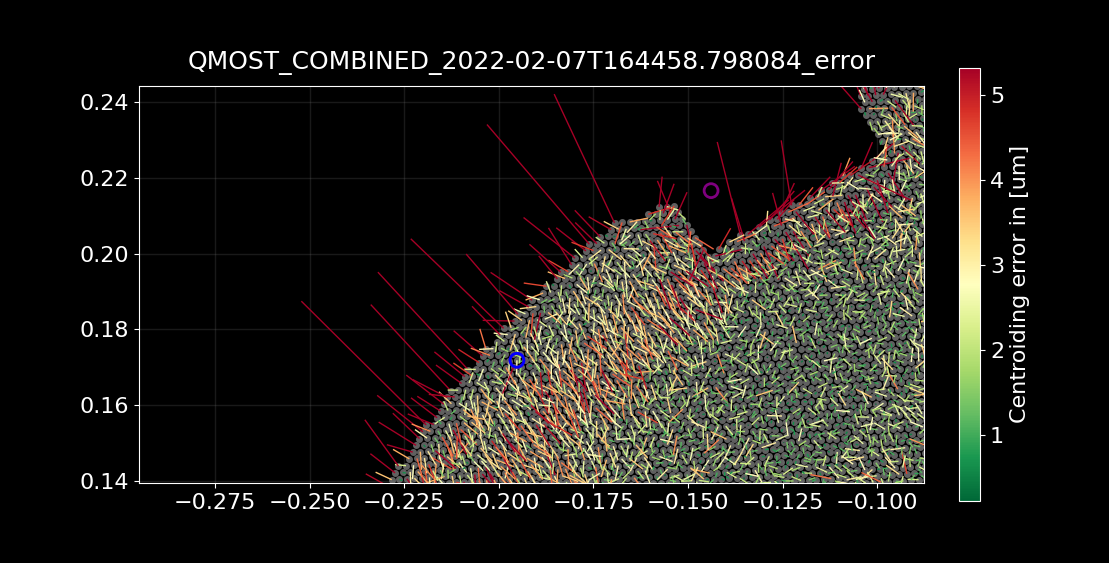}
\end{center}
\caption[Angle Spot Error]{
    Left: Systematic position error of the angle alignment spot. Its error is far outside the color scale, see orange spot errors which are about 4 $\mu m$.
    Right: Error at the edge of the MCU, caused by the limits of spline interpolation of surface normals at the edge of the MCU covered area.
    \label{fig:anglespoterror}
}
\end{figure}

So far, we have not addressed any potential manufacturing defects in terms of spot location on the MCU.
In our analysis, both on the telescope and in the lab, we found no evidence that any spot (other than alignment spots) was systematically misplaced.
We show the error son the angle alignment spot in Figure~\ref{fig:anglespoterror}.
We exclude all alignment spots from the calibration procedure because they do not fit the uniform grid of spots anyway, and as such this did not disturb the process.\par

Furthermore, the normal map calibration procedure produces inconsistent results at the edge of the MCU.
Unfortunately, the MCU is physically limited in size, such that the spine accessible area is going beyond the edge of the MCU at the corners of the science hexagon.
As a result, poor centroiding performance can be expected at the edge of the field, see the right-hand side of Figure~\ref{fig:anglespoterror}.\par

In total, we achieved a centroiding performance of better than 5.0 $\mu m$ in almost all cases, with an average of about 4.5 $\mu m$.
Also, the majority of the systematic errors come from the residual on L1 that was not possible to fully characterize.

\subsection{Fibre Focus Calibration} \label{sec:calibration:focus}

\begin{figure}[h!]
\begin{center}
\includegraphics[height=7cm]{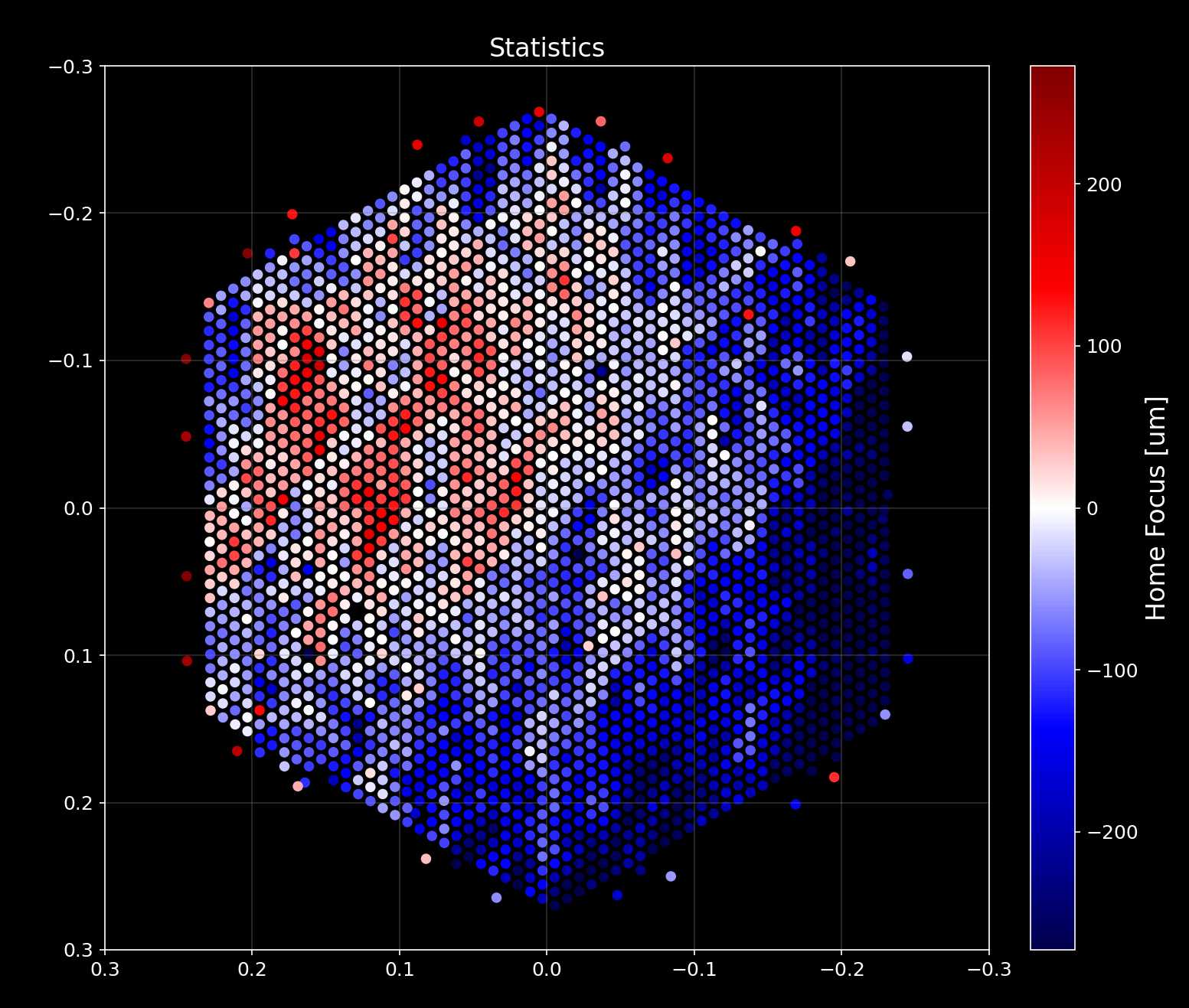}
\end{center}
\caption[Fibre Focus]{
    Focus measurement of the science fibres relative to fiducials, expressed relative to the focal surface of the telescope. There might be a global shift and tilt of the entire field.
    \label{fig:fibfocus}
}
\end{figure}

We discussed in Section~\ref{sec:parallax}, AESOP fibres can rotate and move in- and out of focus.
However, that is only half the story, as fibres also have a manufacturing induced focus position that does not change.
With 4 cameras and careful measurements, we can use the parallax effect with the 4 cameras to measure the fibre focus relative to the fiducial fibres.
That focus measurement relative to fiducial fibres is shown in Figure~\ref{fig:fibfocus}

Currently, we do not precisely know the focus position of AESOP as a whole (i.e. including fiducials), which remains an area of concern as all attempts to measure the focus relative to VISTAs optical system have failed.
One would measure the focus position of AESOP by executing focus runs with VISTA, however this was also not successful.
Our best explanation is, that the failure to measure the focus is due to the optical properties of VISTA.
The surface irregularities we observe (see Figure~\ref{fig:heightmap}) likely extend over the entire mirror surface.
This in turn broadens the parallel bundle of rays at the focus point, i.e. blurs the image.
Defocussing the telescope does not degrade the image quality strong enough compared to the surface irregularities to be observed with fibres.

\subsection{Spine Movement Calibration} \label{sec:calibration:spinecal}

AESOP routes fibres \citenum{brzeski2022aesop} from their metrology reported position to their target position, such that they are not colliding with each other.
This strategy depends on how precisely AESOP is able to predict spine movements.
There are 2 different fibre step sizes, COARSE and FINE.
COARSE is used for large movements, while FINE only in the last $100 \mu m$.\par

To improve the spine movement predictability, spines movements are calibrated.
A spine calibration consists of a sequence of moves in the spines 4 cardinal directions a fix number of steps.
The calibration is done by minimizing the difference between the predicted and observed movement.
So spine calibration does not effect the spine moves, rather the predictability of spine moves.\par

\begin{figure}[h!]
\begin{center}
\includegraphics[height=7cm]{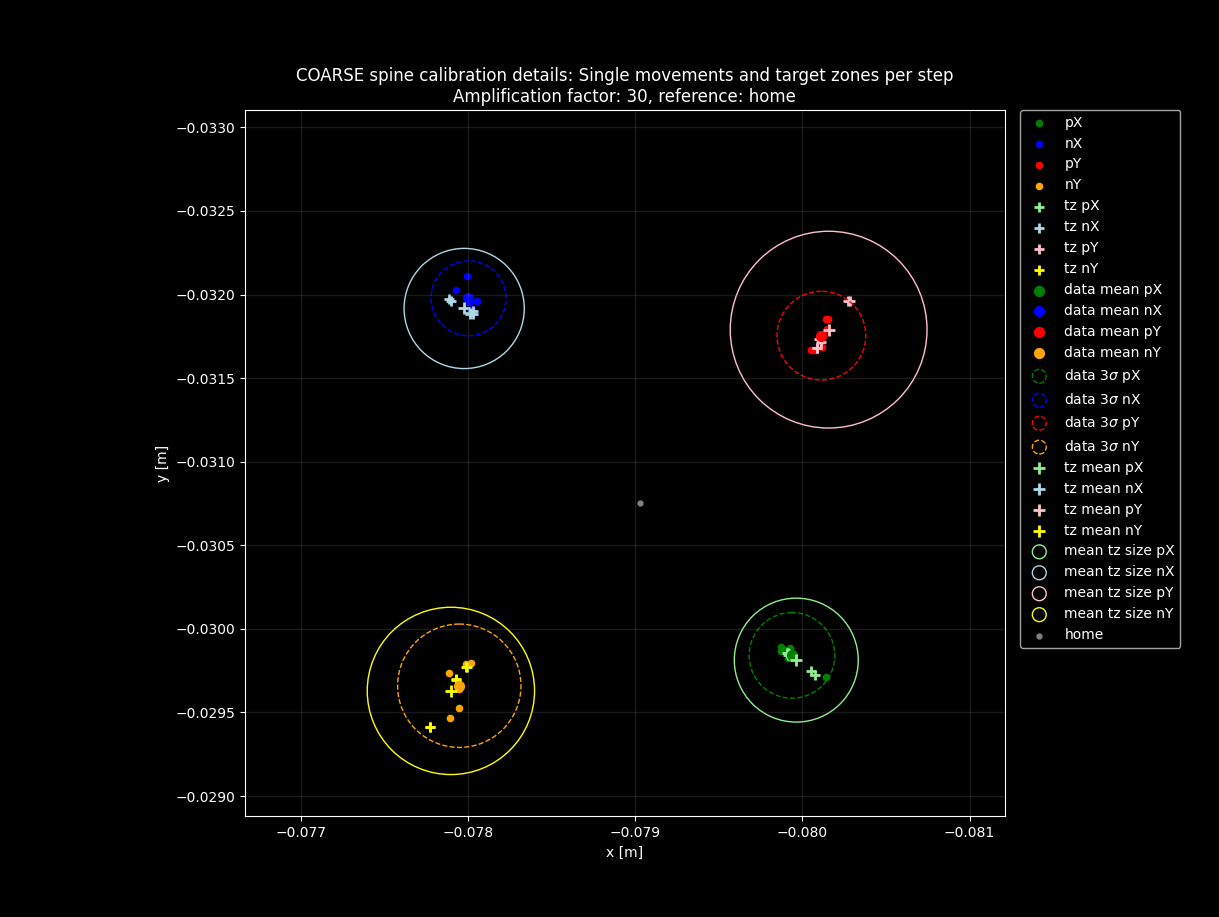} 
\includegraphics[height=7cm]{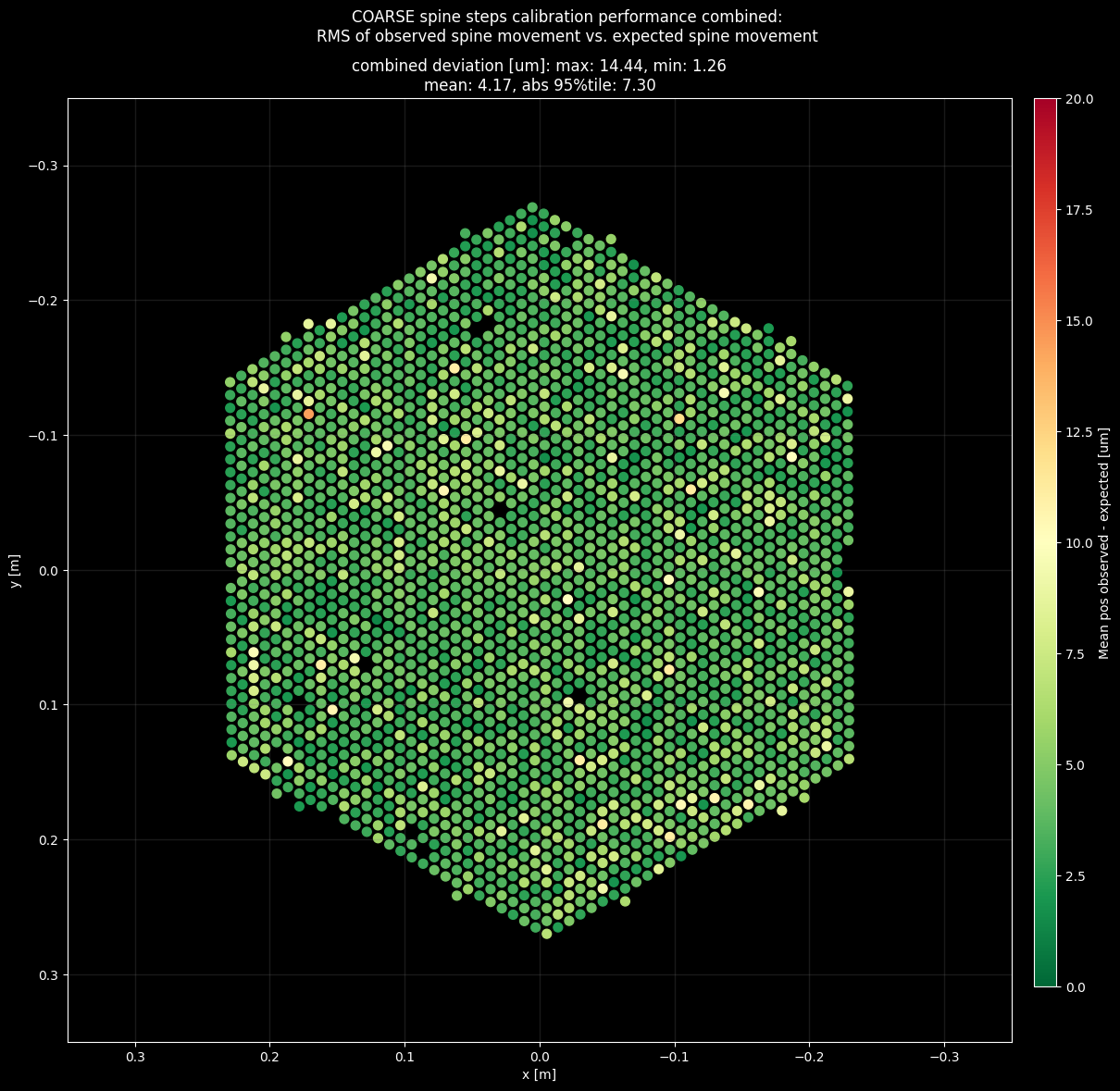}
\end{center}
\caption[Spine Calibration]{
    COARSE spine calibration performance. Left: single spine with predictions in pastel colors and observations in saturated colors. Right: performance of entire AESOP.
    \label{fig:spinecal}
}
\end{figure}

A spine moves in a COARSE step approximately $30 \mu m$ to $70 \mu m$, where the value is fix for each spine.
The left side of Figure~\ref{fig:spinecal} shows the prediction vs. observation of movements of one spine.
The right side of that figure shows the performance of all spines, where the observed deviation from the predicted movement is approx. $4.2 \mu m$ on average.
As the movement error scales linearly with the moved distance, that means that spine movement prediction has an inaccuracy of approx. $5$ to $10 \%$ during COARSE moves.\par

FINE moves are much smaller, only $5$ to $20 \mu m$.
In FINE steps, the mean position prediction differs from the observed step by approx. $50 \%$ of the movement distance.
While this number is much higher than for COARSE steps, it is also much harder to predict small steps since the movement is typically only 1 or 2 steps per spine and the predictions are heavily impacted by measurement limitations during calibration and operation.

\section{SKY REFERENCE WITH RASTER SCANS} \label{sec:rasterscans}

We said initially that the instrument has no access to an absolute reference except the sky.
This is still true and needs to be addressed.
Though to explain the calibration of fibres on sky, we need to assume for a moment that fibres are positioned close to their intended position without going into details.
A proper explanation is given in Section~\ref{sec:fta}, together with final performance numbers. \par

Fiducial fibres are used as references for measuring spine-moved fibres, though their location is only known from lab measurements which are very difficult to conduct and are performed in a completely different environment.
Using the sky as reference is our only option to maximize flux coupling into the fibres, and refining the fiducial positions that match observations.
This is achieved by using raster scans, explained in Section~\ref{sec:rasterscans:process}.\par

In order to access the sky, fibres have to be positioned fairly close to their targets.
Initially, the system was approximately $0.5 mm$, or $9$ arc sec, out of alignment, which we fixed using Secondary Guiding spines, see Section~\ref{sec:rasterscans:sg}.
After that, science fibres were close enough, with an error of approx. 2 arc seconds, to use science fibre based raster scans, see Section~\ref{sec:rasterscans:science}.

\subsection{The Raster Scan Process} \label{sec:rasterscans:process}

First of all, the term 'raster scan' refers to a process to measure the position of fibres relative to the sky.
Usually, an astronomer would use a camera to measure in the focal plane, take an image and then find the location of the stars in this image.
The 4MOST instrument does not have a camera, it has optical fibres.
One can think of each optical fibre as one pixel.
By moving the telescope a tiny bit, one can scan the sky with this fibre/pixel.
For each pointing, the flux is recorded and at the end an image is reconstructed from pointing offsets and flux measurements.\par

\begin{figure}[h]
\begin{center}
\includegraphics[height=6cm]{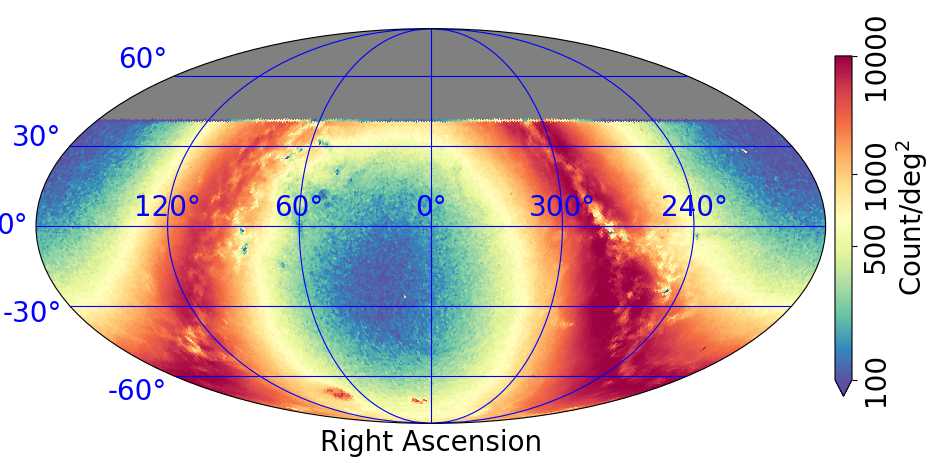}
\end{center}
\caption[Stellar Density of isolated stars]{
    Visualization of the isolated stars density. With 4MOST covering about 4 $deg^2$, there are always at least 400 targets in each pointing.
    \label{fig:isolated}
}
\end{figure}

For this to work, targets that are used for raster scans must be isolated on the sky, i.e. cant have any neighbors.
They also need to be relatively bright, so exposure times are low.
We constructed a catalogue of these specific stars, with brightness more than 18th magnitude in the optical and without any other targets closer than 5 arc seconds in the GAIA catalogue.
There are many stars like that, even near the galactic poles and within the dense regions of the milky way, see Figure~\ref{fig:isolated}.\par

Our reference is the location of stars, provided by the GAIA satellite, which measured the position on sky better than 1 milli arc second for all targets we use for raster scans.
The 4MOST plate scale is approximately 60 $\mu m$ per arc second, i.e. the errors of star locations are negligible for our purpose.\par

There are two kinds of raster scans.
The first was done using secondary guiding fibres, the other with the science fibres.

\subsection{Initial AESOP Alignment with Secondary Guiding fibres} \label{sec:rasterscans:sg}

We have so far not said much about secondary guiding fibres.
These are part of the instrument but used to refine the telescope pointing on sky.
There are 12 of them at the edges of AESOP and are used with raster scans to initialize the AESOP position within the telescope, see \ref{sec:hardware}.
The important part is, flux from the SG system is easier and much faster accessible than flux from spectrographs. \par

After replacing the MCU with AESOP, the fiducial fibres are now the reference.
Metrology cameras are using the fiducials as reference, but due to mounting inaccuracies, we dont know their exact location relative to the telescope optics.
It turned out that we had an error of approx. $0.5 mm$, which is enough to be completely blindsided with the fibre system.\par

\begin{figure}[h]
\begin{center}
\includegraphics[height=6cm]{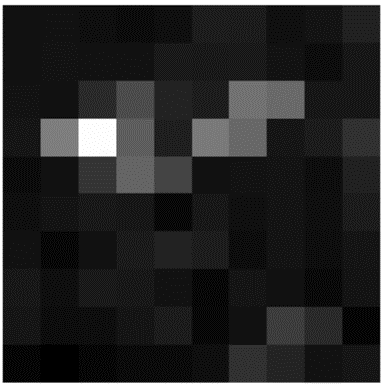}
\end{center}
\caption[SG gridscan]{
    Flux measurements of a single SG bundle with a grid scan, pitch 2 arc seconds. Each pixel represents the accumulated flux of all 7 fibres in one SG probe.
    \label{fig:sggrid}
}
\end{figure}

In a first calibration step, we used the SG fibre bundles as tracers for grid scans.
We configured the field in a way where fibres should see light (but didnt) and then started to scan the sky in a grid pattern.
For each pointing of the telescope, the flux of all 7 fibres in the SG bundles were recorded, sort of like a pixel in an image.
The pitch between pointings were 2 arc seconds.
At the end, we had 12 coarse pixelated images, which were then used in conjunction to find a correction of AESOP location in x, y and rotation in the xy-plane.\par

Due to flux inaccuracies, a gaussian fit to a centroid was not productive and we used the brightest pixel as tracers, see the left side of Figure~\ref{fig:sggrid}.
Each SG probe defines a 2d vector, which in conglomerate are used to compute the xyr offset between AESOP and FTA model.
As a result, the optical model became within 2-3 arc seconds, which is enough to start coarse raster scans with science fibres.
There was one more issue to resolve, the SG bundle rotation state.

\subsection{Secondary Guiding Fibres Rotation Calibration} \label{sec:fta:sgcotcal}

At the beginning of 4MOST instalation, we were faced with a chicke and egg situation.
Due to the spine movement, AESOP spines can (slowly) rotatate inside their mechanical holding.
This is usually inconsequencial, but for the SG probes, the rotations state must be known.\par

In order to measure the rotation state, the bundles need to be centered on a star.
The telescope is moved in a pattern (i.e. a raster scan, see below), the observed pattern on the guide probes is then used to extract the rotation angle of each individual probe.
However, without performing raster scans, the errors for the initial positioning are still large and the stars cannot be centered on the bundles.
At the same time, raster scans cannot be executed without refining the telescope pointing first.\par

\begin{figure}[h]
\begin{center}
\includegraphics[height=5.9cm]{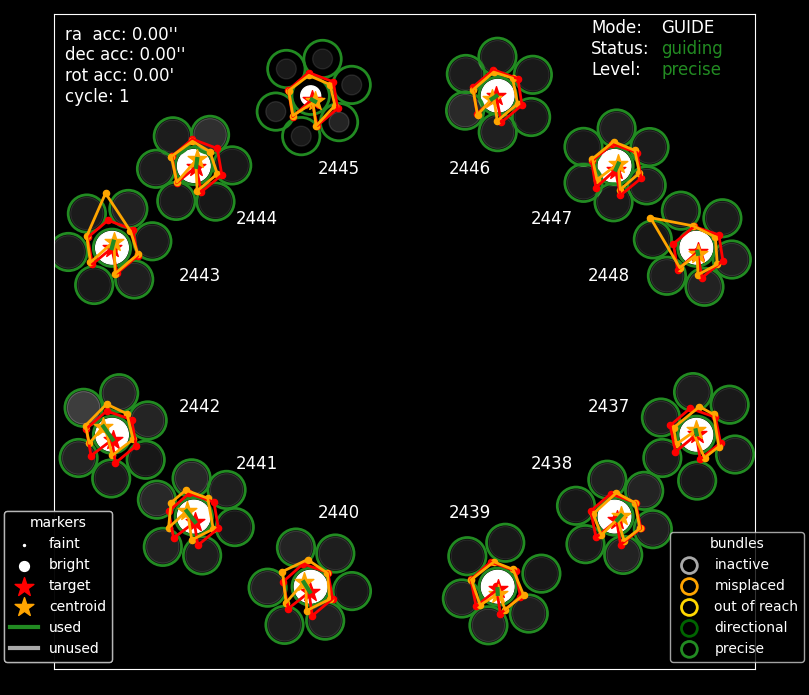}
\includegraphics[height=5.9cm]{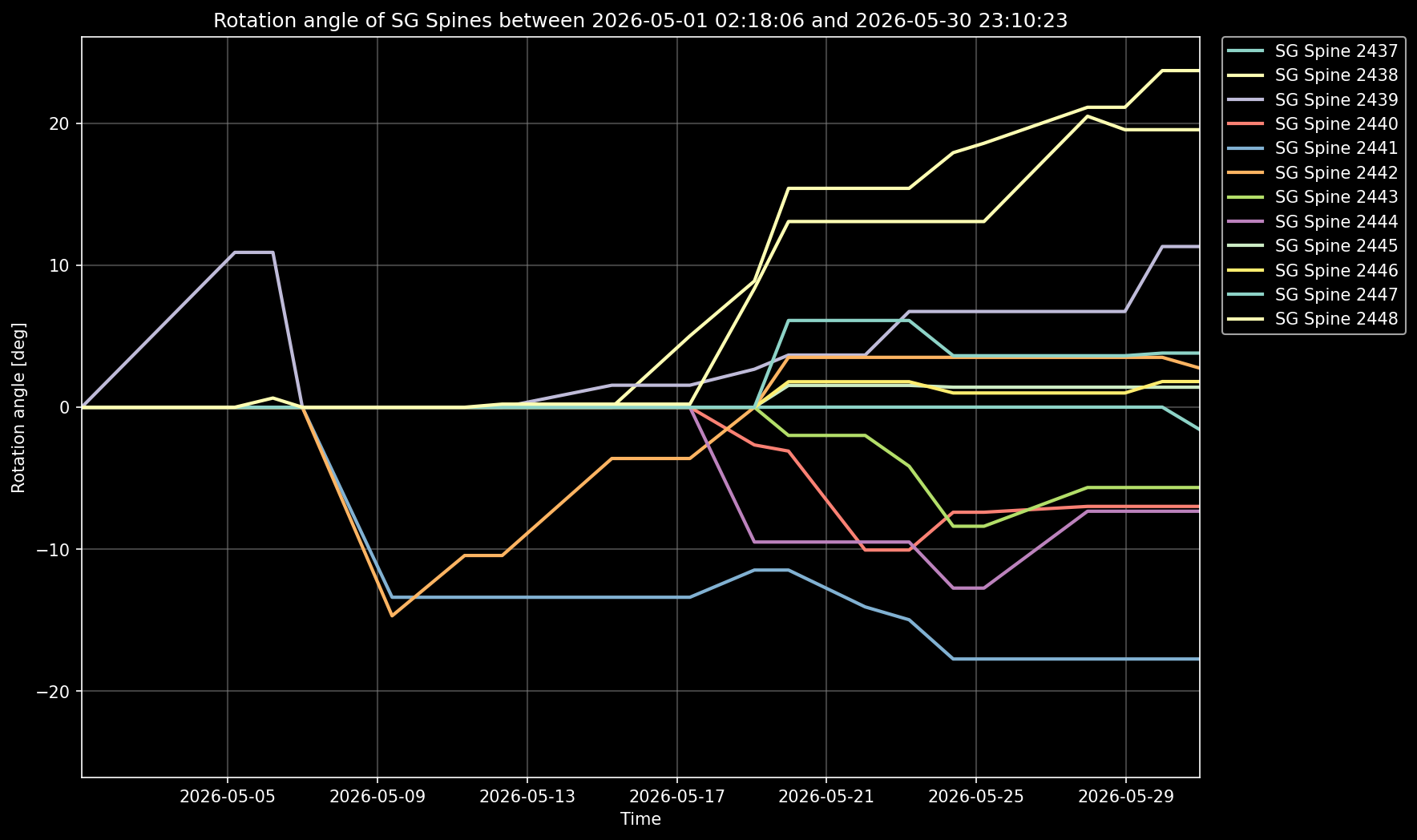}
\end{center}
\caption[SG raster]{
    Left: raster scan pattern on SG probes, pitch: 1 arc seconds. Yellow: centroids, red: expected telescope movement.
    Right: change of SG spine rotation angle over the month of May
    \label{fig:sgraster}
}
\end{figure}

We ended up manually manipulating the position of SG probes after the fibres were configurated to center the stars on the bundles.
Then we were able to perform a raster scan (see Section~\ref{sec:rasterscans:science}) and measure the motion pattern on SG probes, and find the initial rotation.
The left side of Figure~\ref{fig:sgraster} shows the pattern we generated, the telescope pitched 1 arc second between each pointing in a hexagonal shape.\par

We repeat a raster scan each night to track 4MOST performance and measure SG fibre rotation state.
Though to save time, we only execute a 4-pointing raster scan (see Figrue \ref{fig:pattern}), which is still sufficient to measure and track the SG probe rotation state.
The right side of Figure~\ref{fig:sgraster} shows the change in rotation angle of all 12 probes over the month of May.

\subsection{Science Fibre Raster Scans} \label{sec:rasterscans:science}

\begin{figure}[h]
\begin{center}
\includegraphics[width=\textwidth]{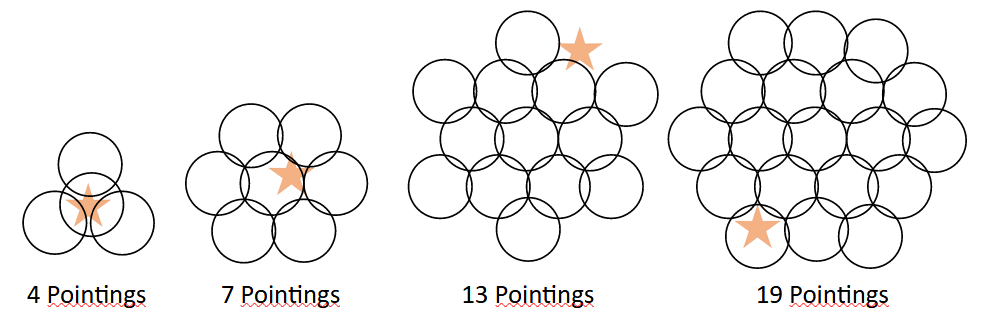}
\end{center}
\caption[Raster Scan Patterns]{
    When doing raster scans, these patterns are available for scanning the sky. Circles represent fibre coverage including dither offsets of the telescope.
    \label{fig:pattern}
}
\end{figure}

Once the instrument model is roughly aligned and SG rotation calibration is done, we used raster scans with the science fibres to calibrate the fiducial position.
While the fiducial fibres provide a reference for the metrology cameras, the fibres cannot collect flux from the sky.
Instead, the science fibres are used and errors are attributed to fiducials by indirectly.
The telescope is dithered by small increments and the flux in each dither point is extracted from the spectrograph measurements.
Figure~\ref{fig:pattern} shows a range of scanning pattern that are available for raster scans.
All but the 4-pointing pattern repeat the first pointing (i.e. the center) at the end.
The center is observed first and forms the reference point for measurement offsets.\par

\begin{figure}[h]
\begin{center}
\includegraphics[height=7cm]{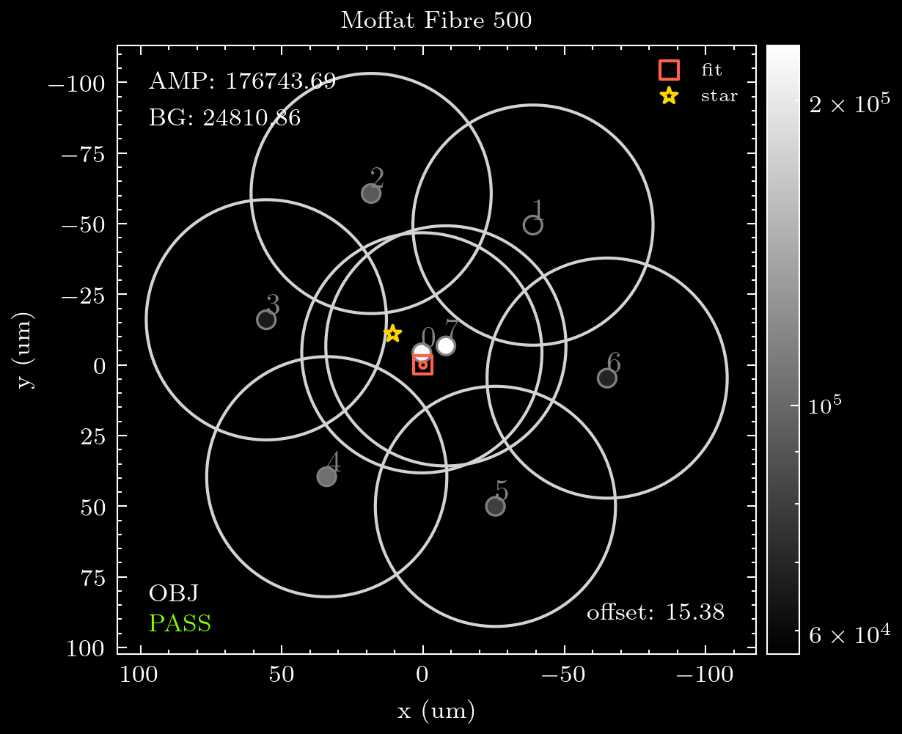}
\includegraphics[height=7cm]{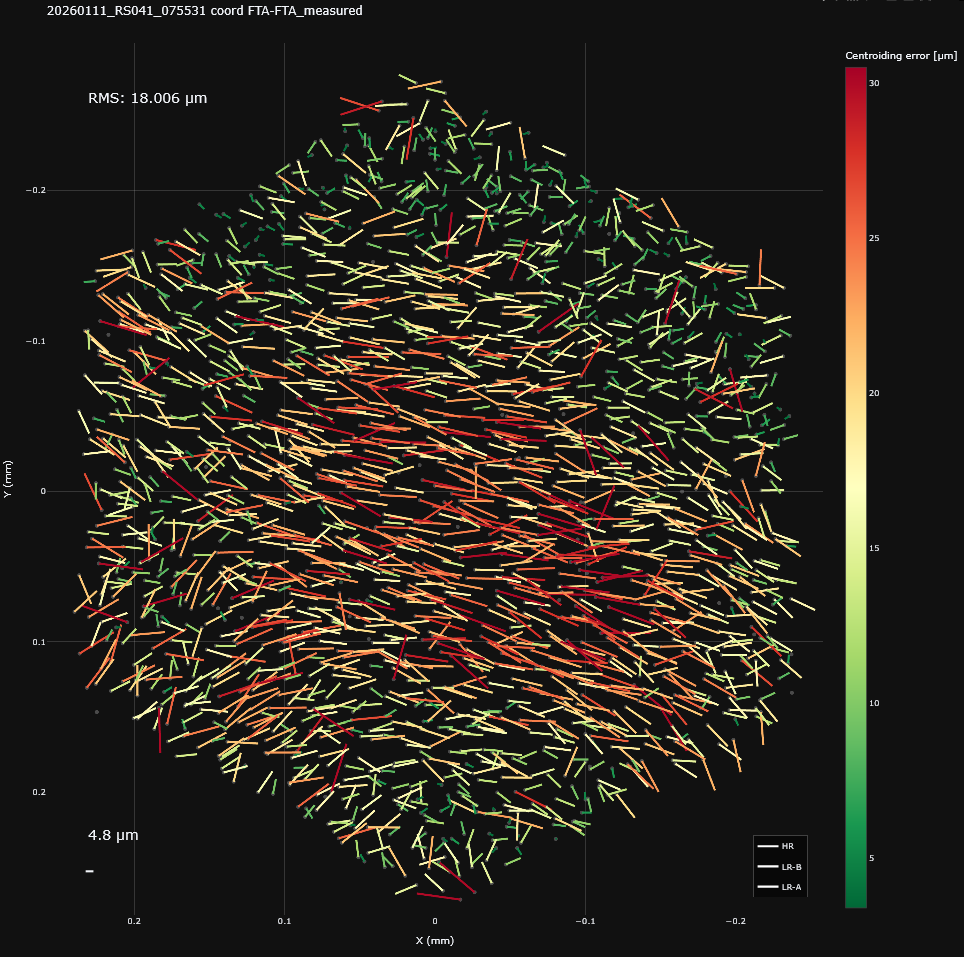}
\end{center}
\caption[Raster Scan Data]{
    Left: the data representation of a single fibre in a raster scan.
    Right: visualization of the raster scan vector field of offsets (representative example).
    \label{fig:rs}
}
\end{figure}

Once a raster scan is complete, the data from the science CCDs is collected and processed offline.
We use the L1 pipeline to extract the fibre flux, ignoring sky background calibration and precise wavelength calibration as only broad wavelength bands are of interest.
Once the flux is extracted, a centroiding is done by fitting a Moffat profile to the flux values and their location.
The left panel in Figure~\ref{fig:rs} shows the flux and offset of a single fibre.
The circles represent the flux collecting area on sky of that fibre, as computed by our software, the centroid of the flux measured in each position shows the offset from our calculations to the data collected.\par

When taking all fibres into account, we receive a 2-dimensional vector field that shows the difference between expected- and measured sky position.
Many such vector fields are accumulated and processed such that the location of the fiducials can be adapted to reduce the measured error.
In the first iteration, the static fiducial fibre positions were adapted to minimize the observed errors.
For this step, we used the 19-pointing raster scan as errors were still very large.\par

\begin{figure}[h]
\begin{center}
\includegraphics[width=\textwidth]{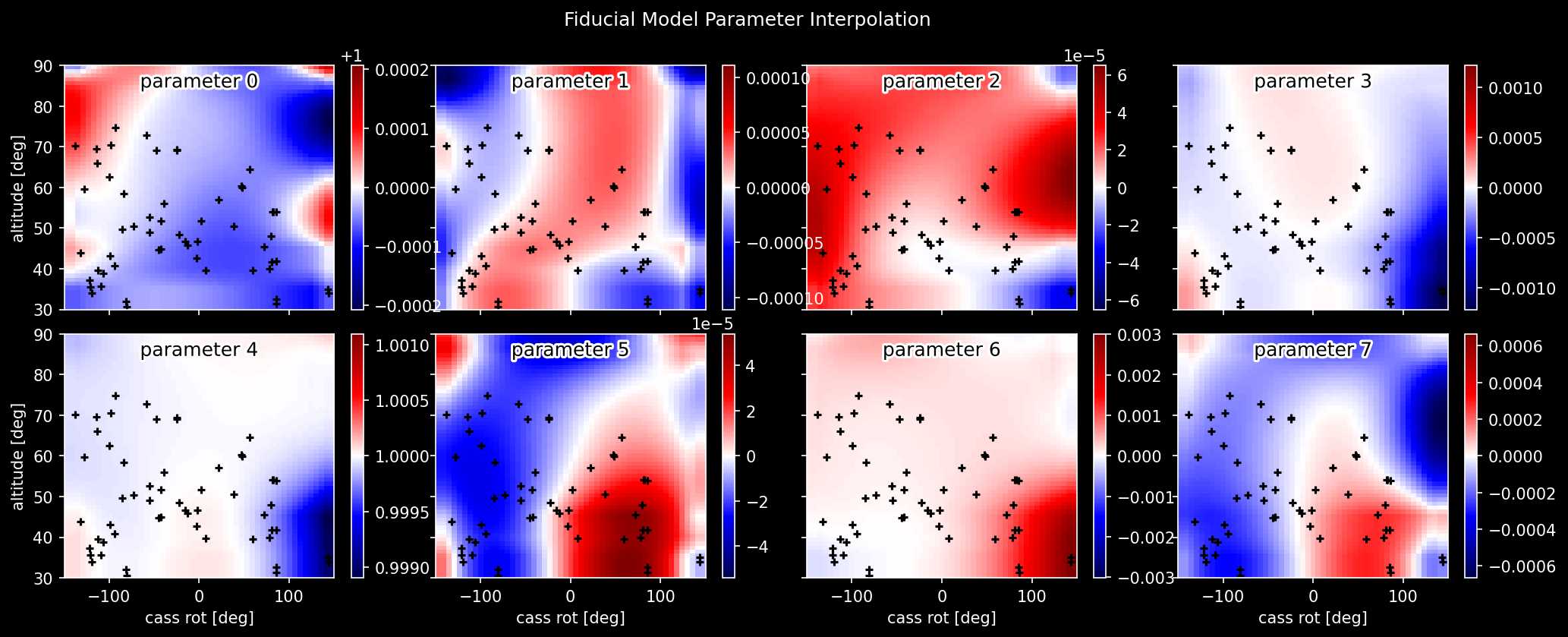}
\end{center}
\caption[Raster Scan parameter]{
    Visualization of the parameter adjustment for the dynamic fiducial model. Black crosses represent support vectors, i.e. telescope state of each measurement. The color represents the direction and strength of the trapezoid function parameter relative to its neutral state (either 0 or 1).
    \label{fig:parameter}
}
\end{figure}

Subsequently, the static fiducial positions are fixed, however we build a dynamic model that alters the fiducial positions depending on the altitude and rotation angle of the telescope.
The model is already discussed in Section~\ref{sec:dynfiducial}, we used 7-pointing raster scans for this part.\par

Performing a single, 7-pointing raster scan takes about 15 minutes.
In order to sufficiently cover the parameter space of telescope altitude, and rotation, we spend approx. 2 nights to build a data set large enough for our dynamic correction model.\par

\begin{figure}[h]
\begin{center}
\includegraphics[width=\textwidth]{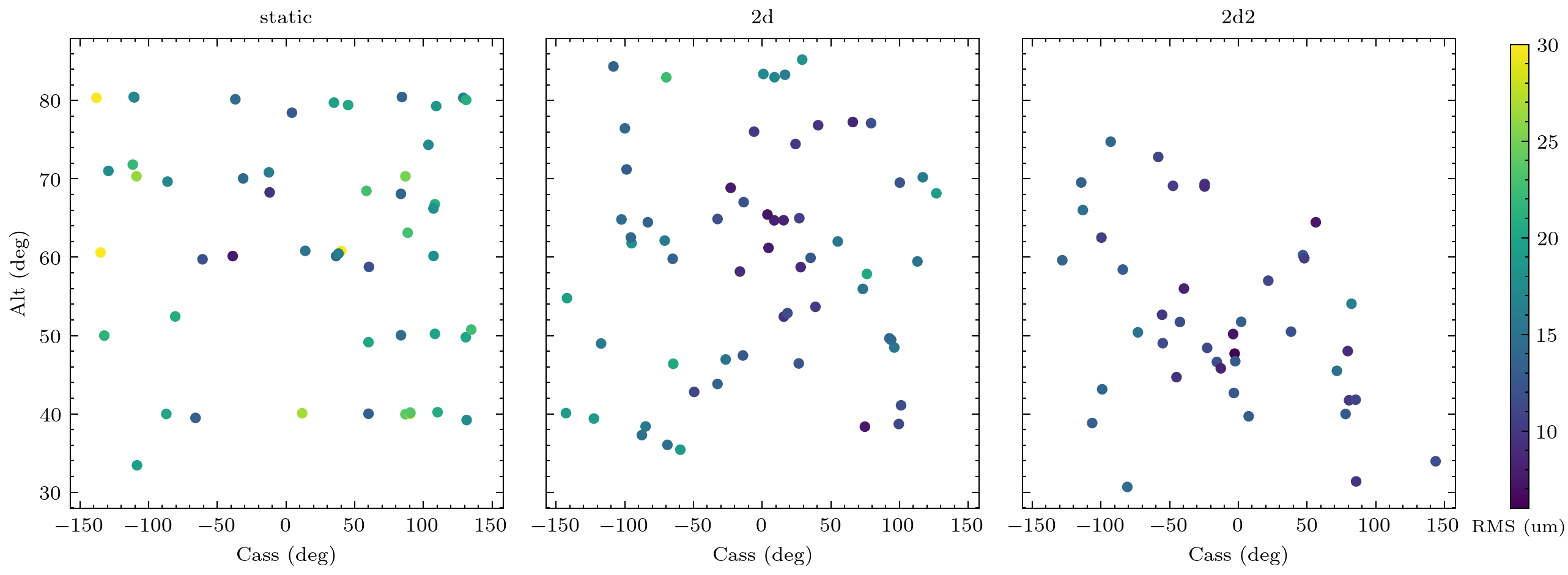}
\end{center}
\caption[Raster Scan progression]{
    Progression of raster scan model performance. Left: performance after static calibration of fiducials. Center: performance after implementation of the 2D fiducial model. Right: Performance after refining the 2D fiducial model.
    \label{fig:rsprogression}
}
\end{figure}

It is hard to get data for the extremes of the parameter space, i.e. very high- or low altitude or large Cassegrain angles.
The telescope tracking is not reliable close to zenith without secondary guiding, which is not available due to the raster scan offset pattern.
Measurements at the extreme ends of the Cassegrain rotation angle are difficult because the field rotation tracks the rotation of the sky and measurements have to be timed very accurately, otherwise the rotator would run into its limit.\par

Nevertheless, the data model extracted from approx. 50 raster scans is shown in Figure~\ref{fig:parameter}.
The data on the right-hand side of Figure~\ref{fig:rs} shows the performance of the system after all calibration steps are done.
We selected here an example that represents the performance at the time of measurement, which is approx. $19 \mu m$ ($0.32 ''$) RMS on sky.
Though, the performance can be as good as $12 \mu m$ ($0.2 ''$) RMS on Sky if the FTA seeing conditions are good (see Section~\ref{sec:fta:seeing}).
A progression of model performance is shown in Figure~\ref{fig:rsprogression}, where the last update of the model was in January 2026.

\section{FIBRE TO TARGET ALIGNMENT} \label{sec:fta}

Above, we showed how to measure the position of fibres and how to calibrate the measurements.
When describing Raster Scans, we implied that fibres would automagically appear near their targets without describing the process in detail.
In this section, we show how this process is done, i.e. how the fibre position are computed and explain the coordinate transformation from sky to focal surface.
We also show how to use AESOP in conjunction with Telescope acquisition in order to minimize overheads and optimize for performance and the use of the secondary guiding system.

\subsection{Sky to Focal Surface} \label{sec:fta:tarfoc}

To project target positions to the focal surface, multiple steps are necessary.
First, the fantastic astropy library is used to handle all astrometric related processes \citenum{astropy2022v5}.
The targets to be observed are selcted by the 4MOST planning SW \citenum{Tempel2020assign}.
They are provided in ICRS coordinate frame with attached parallax, proper motion and epoch.
The astropy library handles all transformation tasks and transformed to altitude and azimuth with VISTA Telescope as reference position and the observation time.
Usually, astropy automatically downloads the latest earth orbital parameters to accomodate earth precession and leap seconds.
But that is not possible on Paranal, since the SW has no access to the internet.
Instead, the latest files are loaded in a separate process and injected on the IWS during startup each night.\par

When transforming to alt/az coordinate frame, astropy also takes care of the wavelength dependent atmospheric refraction.
This information is transformed to Cartesian coordinates on a sphere of 100m radius with the vertex of VISTAs M1 as center of the sphere.
From here, the raytracer mentioned in Section~\ref{sec:raytracing} to transfer to focal surface coordinates.
The optical model is setup according to the telescope parameters, including field rotation and ADC position, extracted at runtime.\par

When the telescope state is loaded into the model, it is important which primary guide camera is used.
Due to the interaction of the telescope pointing model and the FTA pointing model, this is needed to reduce the initial offsets for secondary guiding (see Section~\ref{sec:fta:sg}).
The primary guiding camera and location of the guide star on that camera are used to align the optical model with the sky.
The guiding information for optical model alignment impacts Fibre target positions by approximately $100 \mu m$ to $150 \mu m$.\par

For telescope altitudes above 35 degrees, the wavelength of the simulated rays has no big impact due to the use of 4MOSTs ADC.
However, the optical design of the ADC limits its use to 35 degrees altitude, and target light is allowed to be dispersed for lower altitudes.
Due to the explicit modelling of the atmosphere and optical system of the telescope, the induced dispersion effects are automatically included.
In low altitudes, target definition needs to include a preferred wavelength such that FTA targets the position of that wavelength in the dispersed image.\par

Once the raytracing is done, the coordinates are expressed in focal surface coordinate frames, at 0 focus.
Individual fibre focus deviations is handled on the metrology projection SW, described in Section~\ref{sec:data}

\subsection{COARSE Spines Moves and Telescope Acquisition} \label{sec:fta:coarse}

Once target positions for fibres are known, AESOPs spines need to move the fibres where they are supposed to go.
This is done iteratively in approximately 7 iterations, interlaced with metrology frames as feedback.
Once a move is done, AESOP will tell metrology where to expect fibres for identification, see Section~\ref{sec:aesopspotidentification}.
The prediction of fibre positions needs to be calibrated and is described in Section~\ref{sec:calibration:spinecal}.\par

\begin{figure}[h!]
\begin{center}
\includegraphics[width=\textwidth]{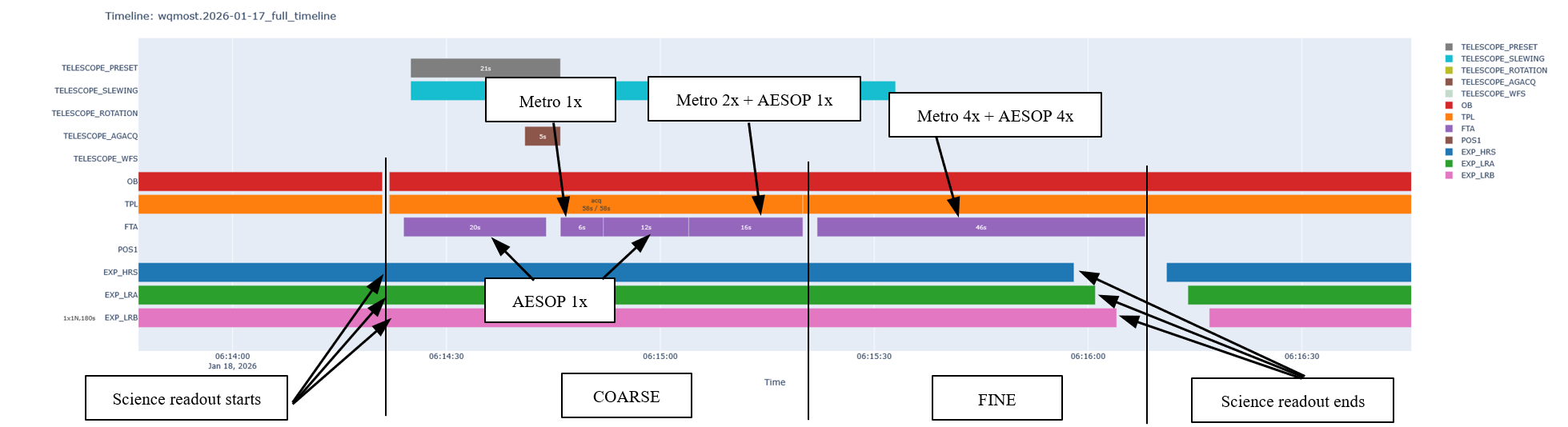}
\end{center}
\caption[Ideal ACQ]{
    Timeline of an ideal acquisition sequence with only small Telescope movement and 7 iterations FTA. Only approx. 10 seconds overhead for FTA.
    \label{fig:idealacq}
}
\end{figure}

In order to minimize overheads, the coarse part of fibre positioning is done in parallel to telescope acquisition.
However, the acquisition sequence is disruptive to FTA and vice versa where parallelized actions are not possible.
The exact process is done as follows:
\begin{enumerate}
    \item The initial fibre position is known from the last measurement of the last position sequence
    \item Spines are moved blind for one iteration in parallel of telescope axis movement.
          Metrology cannot be used while the Cassegrain rotator is moving, as it blurs the images of fibres.
          Altitude and azimuth movements can be done in parallel, but the interaction of the telescope and instrument is not fine grained enough to allow for this level of condition.
          This move is done with pre-computed fibre positions as primary guiding is not yet active.
    \item After Telescope acquisition, one frame of metrology measurement is done.
    \item While the telescope performs a WFS frame (which takes approx. 20 seconds), AESOP performs another blind positioning.
          It would be possible to run multiple loops of AESOP at this stage, but the back-illumination light disturbs the WFS exposures as the light raises the background level above tolerable levels.
          Hence FTA waits for the telescope to finish the WFS frame.
    \item Once WFS is done, FTA performs a final coarse fibre position, which is frames by a metrology measurement before and after the move.
\end{enumerate}

At the end of the process, the telescope is guiding on its primary guide probe and mirrors are in the correct shape.
In total, 3 coarse positioning iterations are performed.\par

\subsection{FINE Spine Moves} \label{sec:fta:fine}

During precise movements, AESOP decides on a spine-per-spine basis whether to move it COARSE or in FINE steps.
In this phase, metrology needs to be very precise, hence this can only be done after the telescope is finished with its acquisition sequence.
Typically, COARSE steps are used automatically for spines further than 100um from their target, while FINE steps are used below.
We consider between 7 and 9 total iterations and are currently experimenting with trading spine iterations (which take time) vs. positioning performance to find the sweet spot.\par

\begin{figure}[h!]
\begin{center}
\includegraphics[width=\textwidth]{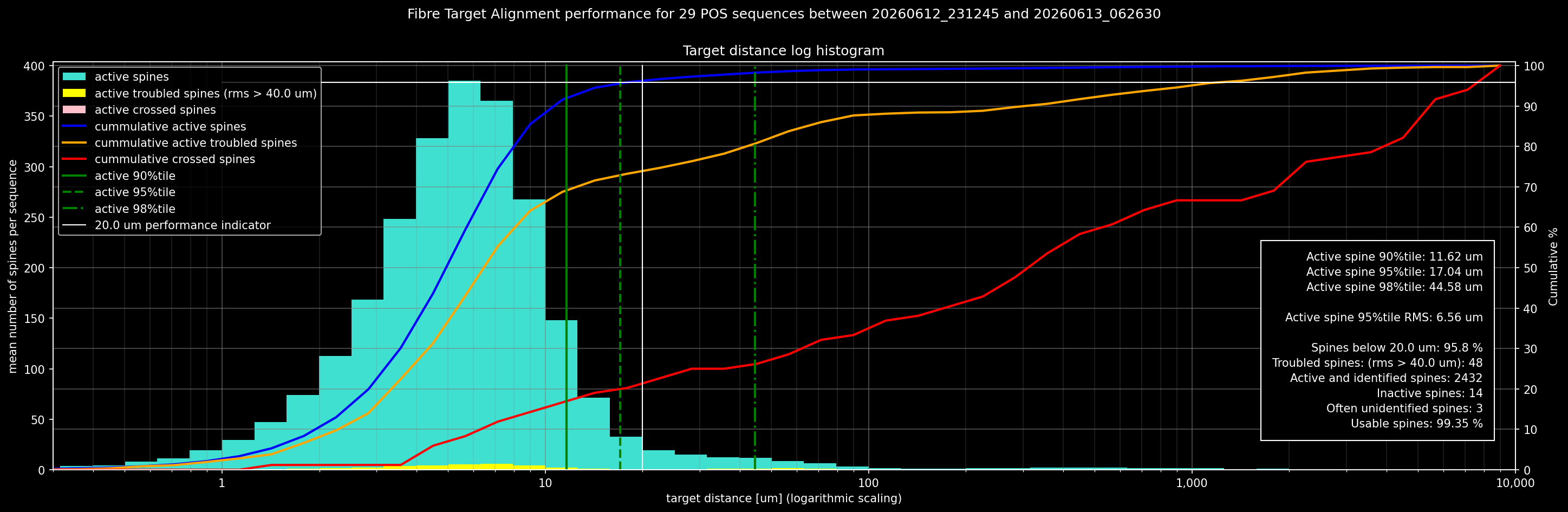}
\end{center}
\caption[FTA Performance]{
    Statistics of FTA performance reached after 9 iterations, accumulated over 29 positioning sequences early June 2026.
    \label{fig:ftaperformance}
}
\end{figure}

The statistics shown in Figure~\ref{fig:ftaperformance} show some of the best results we achieved so far at low seeing conditions (see Section~\ref{sec:fta:seeing} for more information on seeing conditions).
Notable, we consider the RMS performance on the best $95\%$ of the spines as a key performance metric, which achieves approx. $6.5 \mu m$.
Another metric is the fraction of fibres below $20 \mu m$, which in these 2 days of data were at approx. $96 \%$.
There are 14 inactive spines, of which 3 are often not identified.\par

Ideally, the FINE spine positioning finishes somewhat before the readout of the detectors finish.
However, this is only possible if FTA is using 7 iterations and the telescope acquisition is done in such a way, that it does not interfer with FTA.
For 7 iterations, FTA requires approx. $105 s$ from any random configuration to any other random configuration of spines, safely below the requirement of $120 s$.
The performance in 7 iterations is reaching approx. $7.5 \mu m$ to $8 \mu m$ RMS on the $95 \%$ile, depending on seeing conditions.
Which is also safely within the margin of $10 \mu m$ RMS on the $95 \%$ile.\par

In bad seeing conditions, FTA might not quite reach $10 \mu m$ RMS on the $95 \%$ile.
But also here, we are currently implementing improvements, by taking more metrology frames, which reduces the impract of seeing at the cost of approx. $5 s$ per additional metcam frame per camera.\par

Due to the tilting spine technology, used by AESOP, 4MOST is able to place up to 7 spines into a very tight space of only $20$ arc sec radius without loss of performance.

\subsection{Secondary Guiding} \label{sec:fta:sg}

During science observations, 4MOST utilizes its 12 secondary guiding probes to refine the telescope pointing (see Section~\ref{sec:hardware}).
The telescope control system and the 4MOST FTA system use different pointing models.
There is a small residaul of typically $1$ arc secend between the two systems, which are eliminated using the secondary guiding probes.
Since the SG fibre bundles are positioned in the same way as science fibres but also are able to see the sky, they are used for final pointing refinement.
In addition, the SG system also provides rotation guiding in addition to alt/az, which is important when the telescope passes close to zenith.\par

There are 12 guide probes, which are used to measure star positions relative to expected star positions.
It takes approx. $8$ SG iterations of $10$ seconds each to refine the pointing enough such that no further improvement can be made.\par

\begin{figure}[h!]
\begin{center}
\includegraphics[height=7cm]{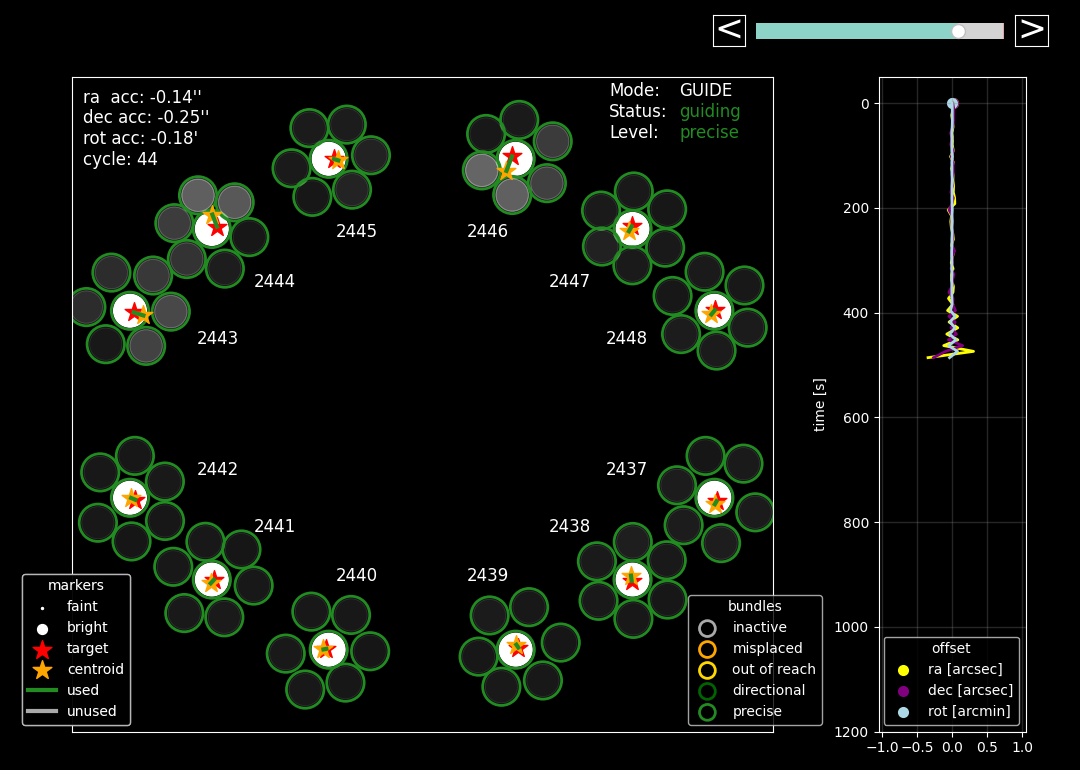} 
\includegraphics[height=7cm]{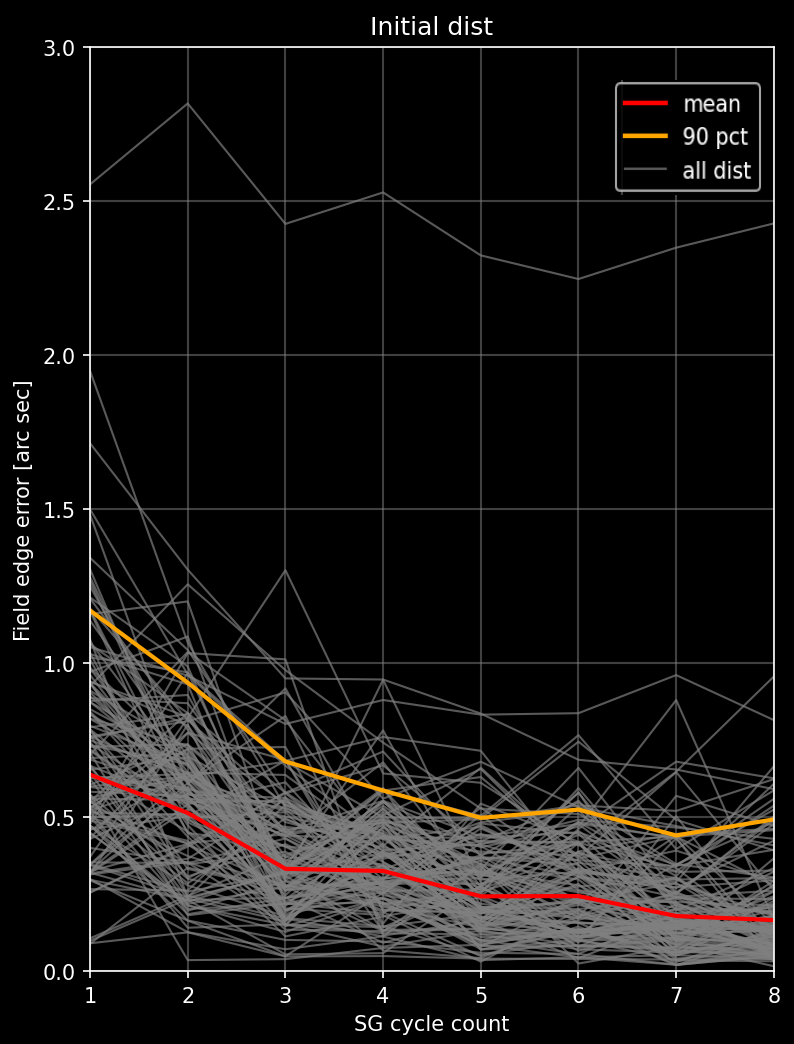}
\end{center}
\caption[SG Panel]{
    4MOST Secondary Guiding Panel during operation.
    \label{fig:sgpanel}
}
\end{figure}

The left image in Figure~\ref{fig:sgpanel} shows the panel during telescope operation.
The panel shows the history of corrections on its right-hand side, indicating whether the system is settling to a refined position.
The left 2/3 of the panel is a visualization of the reconstructed fluxes in the fibres, the centroiding of targets and corrections that are made by the system.
Yellow stars represent centroids, red stars represent expected centroid positions, errors show the intended correction.
Each probe is approx. 3 arc seconds in diameter and has a range of star detection of approx. 4 arc seconds diameter.\par

The right panel in Figure~\ref{fig:sgpanel} shows historic data of offsets of the first 8 iterations over multiple SG sequences.
Since the correction is applied in 3 axis (alt, az and rot), the error vector length provided in this image is the mean length of the correction vector at the edge of the field, combining linear offsets with rotation in a meaningful way.
The data presented in this panel is taken after the latest refinement discussed in the next subsection, middle of June 2026.

\subsection{Coordinate Refinement} \label{sec:fta:refinement}

During operation of the 4MOST, we track many parameters associated with offsets that the secondary guiding system imposes on the telescope in the first 8 to 10 SG iterations.
Based on the observed telescope state parameters and tracked offsets, we identify systematic effects that can be exploited to refine the initial fibre positioning and minimize the initial offset.
The route cause for the observed offsets is that the Telescope uses a difference pointing model than FTA.
We pursue a data driven optimization here, because modeling the differences between the 2 systems, including telescope axis behavior and flexure is too cumbersome and error prone.\par

\begin{figure}[h!]
\begin{center}
\includegraphics[width=0.49\textwidth]{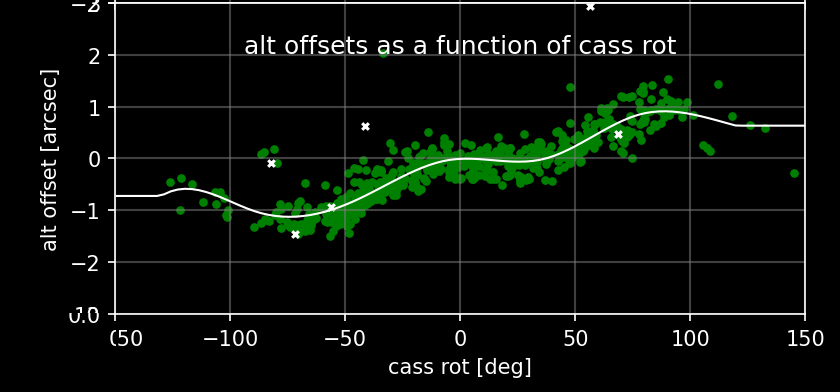} 
\includegraphics[width=0.49\textwidth]{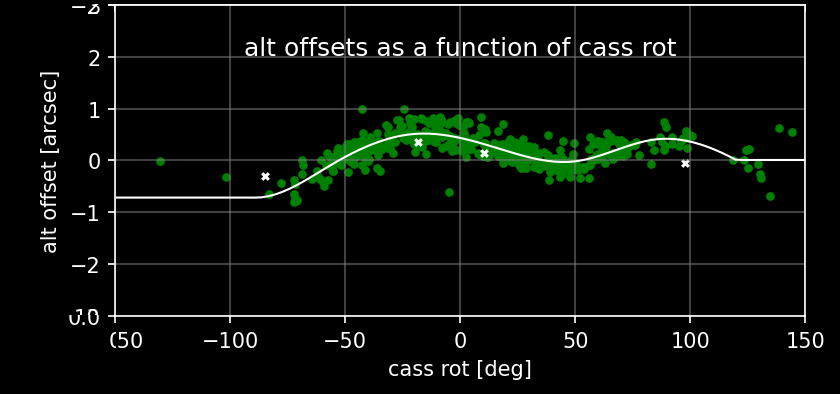}
\end{center}
\caption[Refinement]{
    Refinement of initial fibre pointing in direction of altitude as a function of Cassegrain rotation angle, left when guiding on AG1, right when guiding on AG2.
    \label{fig:refinement}
}
\end{figure}

By using data that is collected during normal operation, we do not interrupt the system and do not spend extra observation time for calibration.
Figure~\ref{fig:refinement} shows the refinement model we used to reduce the initial offset in one occasion.
Statistically, the refinement improved the pointing by approx. 0.2 arc seconds.
In the presented case, we use a spline interpolation to model the altitude offset of the system as a function of Cassegrain rotation angle and which primary guiding camera was used.

\subsection{FTA Seeing} \label{sec:fta:seeing}

Air turbulence can impact the centroid measurements by the metrology system, since the optical path towards the cameras is fairly long.
The most impactful section on centroid measurements is the airspace between M1 and M2.
We call this seeing 'FTA seeing' and was a concern from the beginning of the project.\par

Concerns about FTA seeing is one of the reasons we build 4 metrology cameras instead of just 1, and it turns out that was the right call.
We already presented the shape and nature of turbulence inside the telescope optical path between cameras and focal surface in Figure~\ref{fig:metcam_sequence}.
However, this uses the MCU to identify the seeing pattern, which is not available during normal operation.\par

Spines are a bad tracer for FTA seeing, since they move in between metrology frames.
We also do not want to spend extra observing time with the metrology system in order to measure the seeing using science fibres explicitly.
Instead, we trace seeing only based on the measurements of fiducial fibres.
This gives us 24 tracers at the edge of the field in each metrology frame.\par

\begin{figure}[h!]
\begin{center}
\includegraphics[height=7cm]{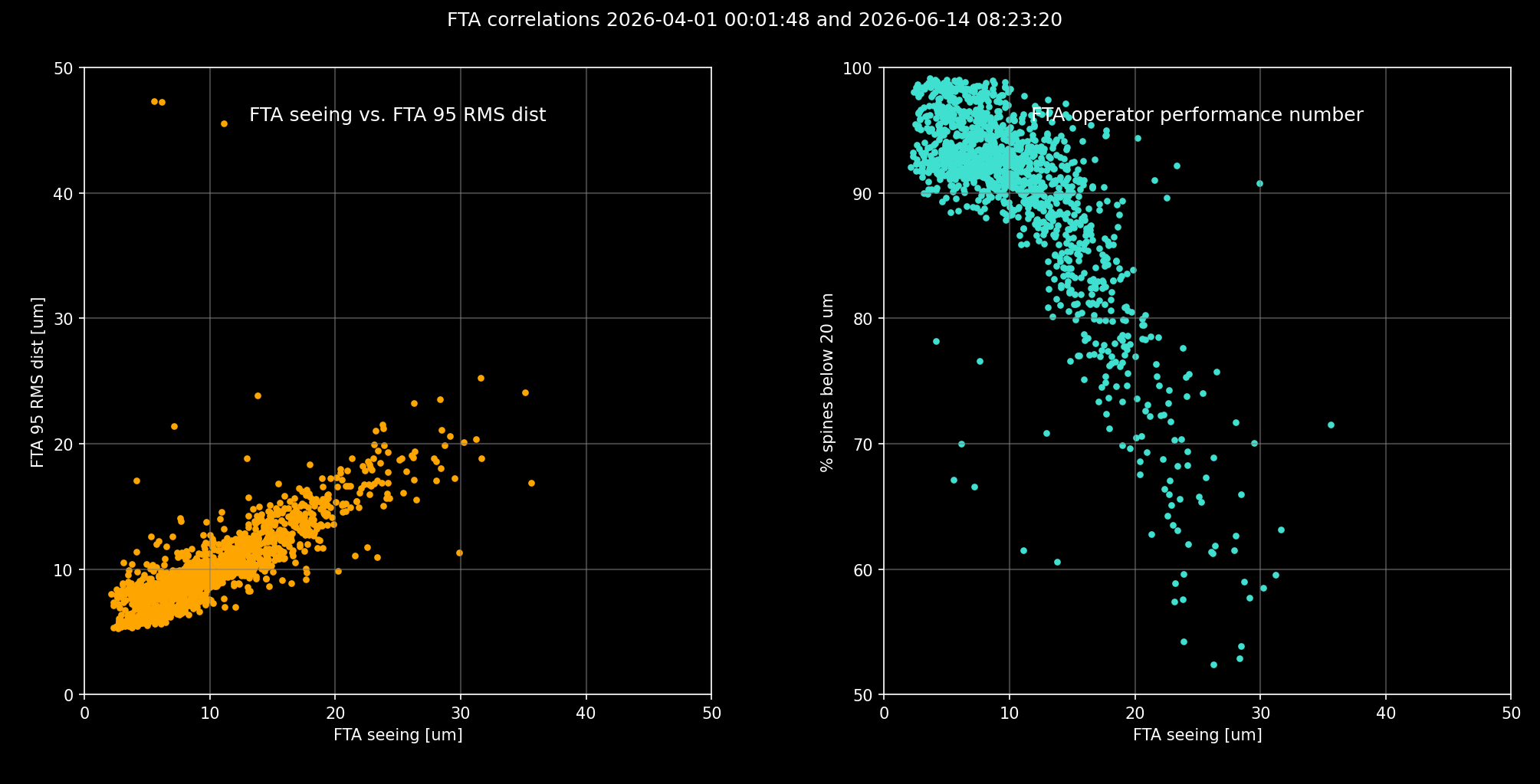}
\end{center}
\caption[FTA seeing]{
    Historic data of FTA seeing vs. RMS95 (left) and fibre positioning success (right) from April to now. The data is a bit skewed towards worse performance since the system experienced multiple improvements in FTA performance in that time frame. The correlation is still very obvious.
    \label{fig:ftaseeing}
}
\end{figure}

FTA seeing is measured by each MetCam individually.
The value is computed by measuring the difference of fiducial fibre  measurements between the current and previous observation.
We use the fully corrected result here, eliminating any linear component, meaning FTA seeing only measures the uncorrected non-liner component at the location of the fiducials.
The global FTA seeing is computed as the mean of individual FTA seeing values of camera, for each individual frame.
For instance, when we take 2 measurements with each MetCam, we have 8 individual FTA seeing measurements, which are then averaged.\par

FTA seeing is tracked during normal operation of the instrument, and there is a strong correlation between FTA seeing and fibre position performance, see Figure~\ref{fig:ftaseeing}.
Because of that correlation, we do a dynamic decision to take 2 MetCam frames instead of 1 if the fibres are closer than $30 \mu m$ RMS in the $95 \%$ile and the FTA seeing measurement is above $20 \mu m$.
We might tweak these parameters over time in order to maximize performance of 4MOST.
Statistically, assuming 20 minutes of exposure time, each $1 \%$ improvement of spine positions is worth approx. 15 seconds of invested overhead.
The dynamic decision to take more Metrology frames costs approx. 5 seconds per frame, but may improve positioning by a noticeable fraction.\par

Currently, the system is often limited by the seeing conditions, the team in Potsdam and Paranal are working on temperature control inside the dome during the day, which will improve performance during the night.

\subsection{No-Go-Zones} \label{sec:fta:nogo}

Approximately 20 spines of AESOP (i.e., approx. $1 \%$) show some unusual behavior, in that they fail to access certain areas of their patrol area.
These regions are always towards the edge of the individual spine patrol range.
Typically, each spine has a patrol radius of $11.5 mm$, with a physically reachable radius of approx. $14 mm$.
For comparison, the pitch between spines is $9.5 mm$.\par

\begin{figure}[h!]
\begin{center}
\includegraphics[width=0.3\textwidth]{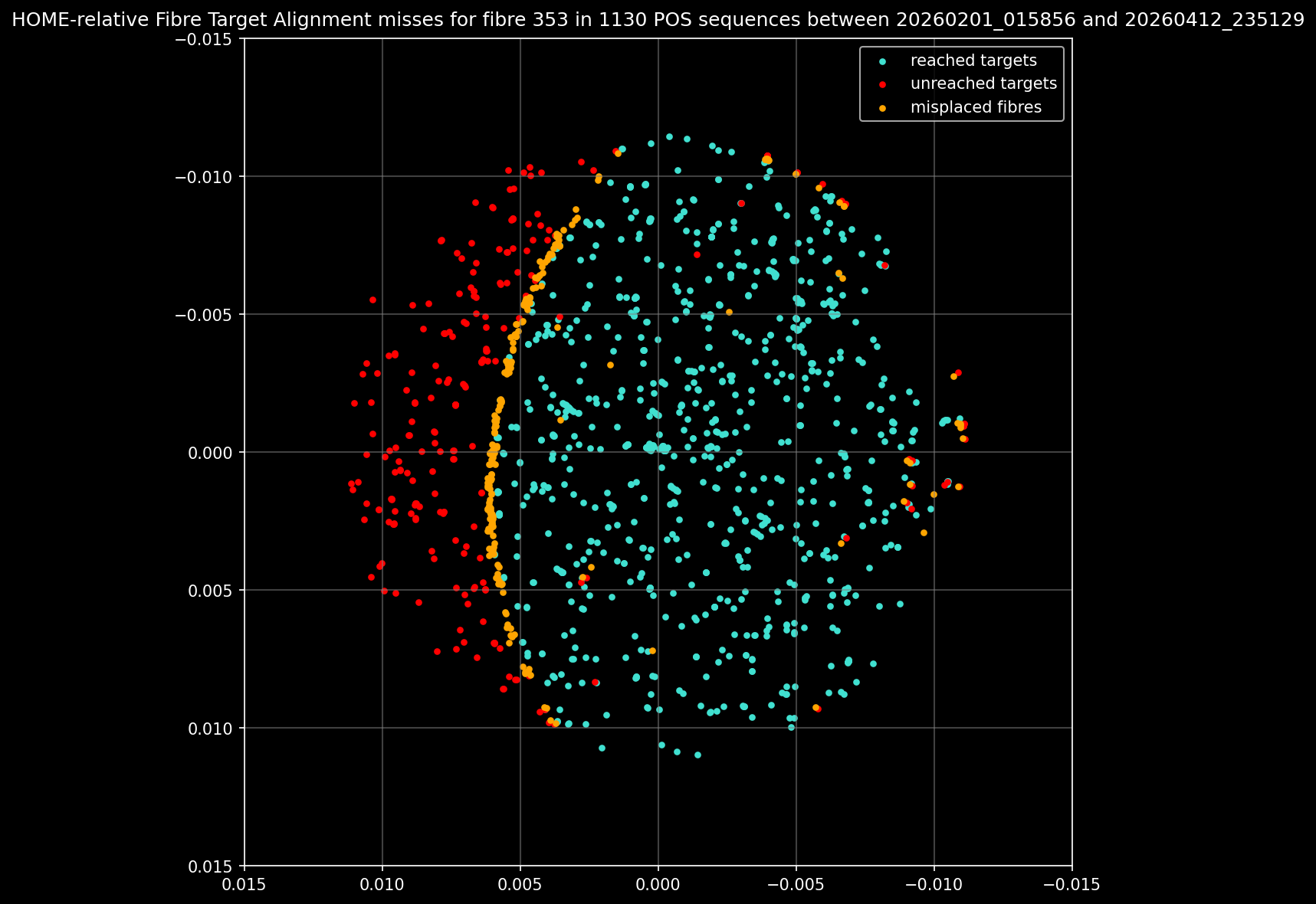} 
\includegraphics[width=0.3\textwidth]{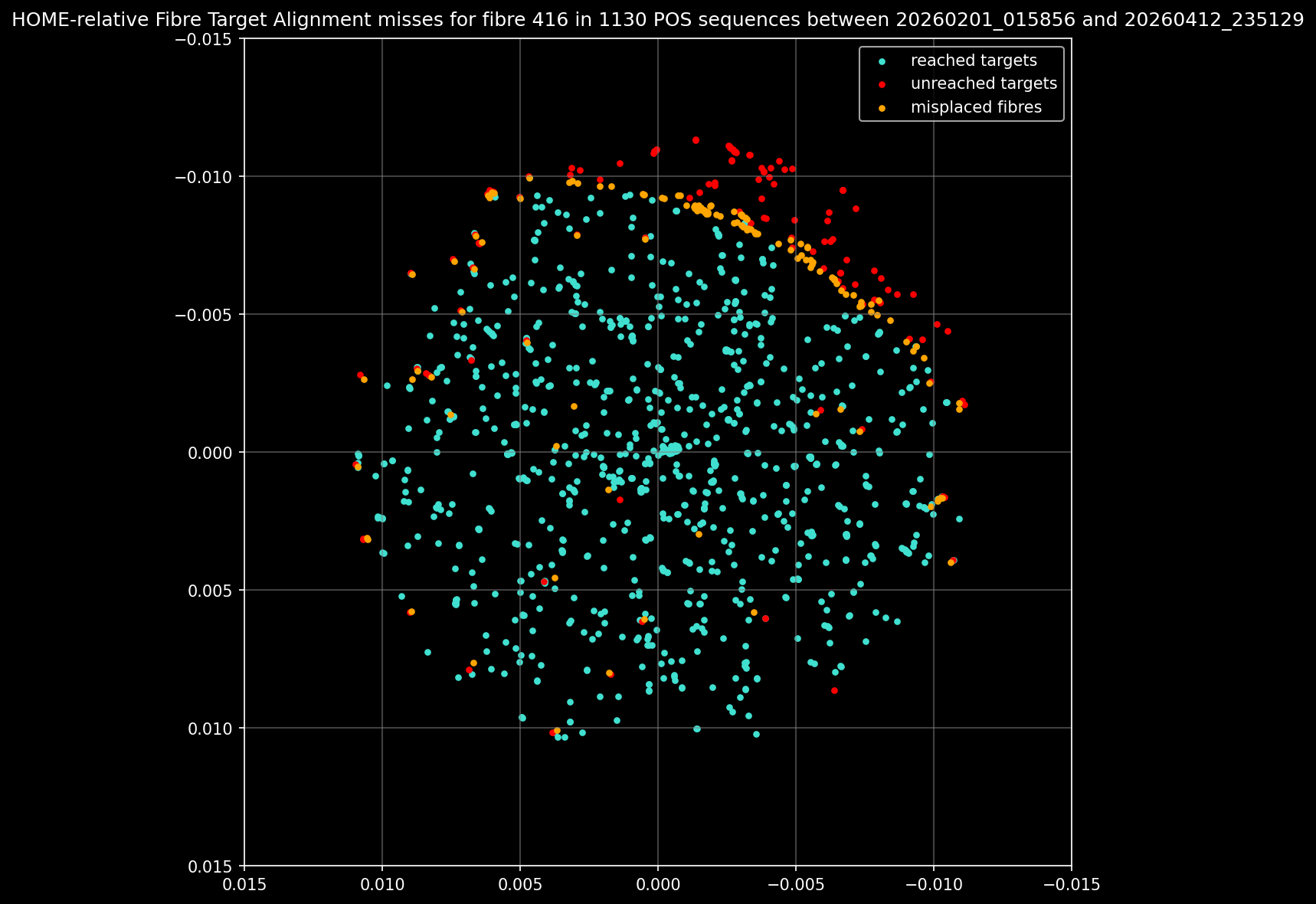}
\includegraphics[width=0.3\textwidth]{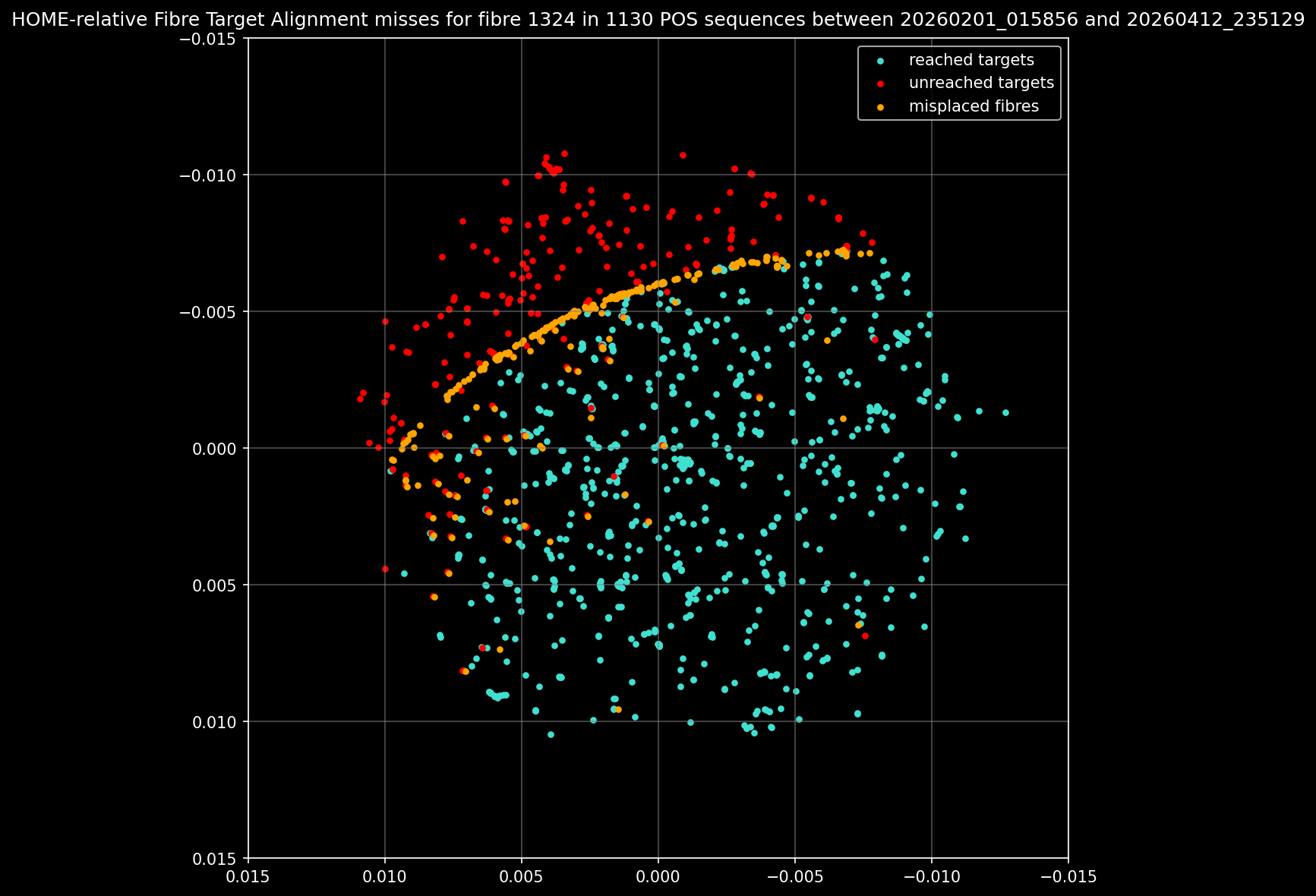}
\end{center}
\caption[no go zones]{
    No-go zones of 3 spines.
    \label{fig:nogo}
}
\end{figure}

However, some spines cannot enter small regions of their patrol area, which we named no-go-zones.
We can only speculate as to why these regions are not accessible.
There might be dust grains in the system, or mechanical defects on the spine components.
Whatever the case may be, during normal operations, we measured the no-go zones by observing spines systematically not reaching their targets.
Figure~\ref{fig:nogo} shows the region of no-go zones for $3$ of the $20$ spines we identified.
In this figure, green spots indicate target locations, that a spine successfully reached, red spots indicate target positions that a spine failed to reach.
Orange represents locations a spine ended up, when it fails to reach a target position.\par

There are many reasons, why a spine fails to reach a target.
It could have been collided with another spine, ran out of time, routing around other spines in its vicinity or it might have been misidentified by the metrology system.
However, it typically cannot show the systematic behavior that is presented here.
This issue is solved during the planning stage, by not moving spines into the no-go-zones.

\section{SUMMARY} \label{sec:summary}

In this paper, we have shown how to measure fibres or calibration spots, identify them and project their measured location from camera coordinates to physical space using a custom ray-tracer.
We showed how to calibrate the optical system of VISTA with 4MOST, and how to account for all relevant effects that impact measurement performance.
We achieved a centroiding performance of better than $5 \mu m$ RMS relative to the local reference frame of the fiducial fibres.\par

While we reached our requirements for fibre target alignment just 3 weeks after installation of the system $24 \mu m$ ($0.4 ''$) RMS vs. the sky), we continued and improved the performance over time.
In January, we installed the 2D-fiducial model, which improved position performance vs. sky to approx. $19 \mu m$ ($0.32 ''$) RMS and with improvements to fibre positioning and sequencing, we now reach approx. $16 \mu m$ ($0.27 ''$) RMS vs. the sky.
Currently, the system performance is limited by air turbulence inside the dome.
On good condition days, we reach a on-sky performance of below $12 \mu m$ ($0.2 ''$), with only a small number of straggler spines that form a long tail of misplaced spines.

\acknowledgments

The 4MOST project received funding from the BMFTR under grants 05A20BA1 and 05A23BA1.

\bibliography{spie_fta_bib} 

\begin{thebibliography}{10}

\bibitem{deJong2024}
de~Jong, R.~S., Bellido-Tirado, O., Brynnel, J.~G., Amini, A.~E., Frey, S.,
  F{\"u}{\ss}lein, C., Giannone, D., Johl, D., Kuba, S., Lemke, U., Micheva,
  G., Saviauk, A., Steinmetz, M., Walcher, J.~C., Winkler, R., Lind, K.,
  Loveday, J., Mainieri, V., Pirard, J.-F., Gaessler, W., Laurent, F., Merloni,
  A., Navarro, R., Remillieux, A., Rothmaier, F., Smedley, S., and Walton, N.,
  ``{4MOST: the 4-metre Multi-Object Spectroscopic Telescope Project at the
  start of commissioning},'' in [{\em Ground-based and Airborne Instrumentation
  for Astronomy X}{\nolinebreak\hspace{0.1em}]},  Bryant, J.~J., Motohara, K.,
  and Vernet, J. R.~D., eds.,  {\bf 13096},  130960J, International Society for
  Optics and Photonics, SPIE (2024).

\bibitem{brzeski2022aesop}
Brzeski, J., Adams, D., Baker, G., Baker, S., Brown, R., Case, S., Chin, T.,
  Coyne, J., Farrell, T., Gillingham, P., Houston, E., Klauser, U., Kripak, Y.,
  Kunwar, N., Lawrence, J., Mali, S., Maslak, W., McGregor, H., Muller, R.,
  Nichani, V., Pai, N., O'brien, E., Saunders, W., Smedley, S., Venkatesan, S.,
  Waller, L., Zahoor, J., and Zheng, J., ``{Overall performance of AESOP: the
  4MOST fibre positioner},'' in [{\em Ground-based and Airborne Instrumentation
  for Astronomy IX}{\nolinebreak\hspace{0.1em}]},  Evans, C.~J., Bryant, J.~J.,
  and Motohara, K., eds.,  {\bf 12184},  121846M, International Society for
  Optics and Photonics, SPIE (2022).

\bibitem{cirasuolo2020crescent}
Cirasuolo, M., Gonzalez, O., Rees, P., Bryson, I., Fairley, A., Taylor, W.,
  Afonso, J., Lilly, S., Evans, C., Flores, H., et~al., ``Crescent moons: an
  update on the ongoing construction of the new vlt's multi-object
  spectrograph,'' in [{\em Ground-based and Airborne Instrumentation for
  Astronomy VIII}{\nolinebreak\hspace{0.1em}]},   {\bf 11447},  222--232, SPIE
  (2020).

\bibitem{cunningham2022wfc}
Cunningham, M.~H., Doel, P., Brooks, D., Brynnel, J., Frey, S., deJong, R.,
  G\"abler, M., Schr\"ock, M., Lehmitz, M., Sablowski, D., and Barden, S.,
  ``The assembly and alignment of the 4most wide field corrector,'' in [{\em
  Proc. SPIE}{\nolinebreak\hspace{0.1em}]},   {\bf 12184},  Paper 257 (2022).

\bibitem{BardenSPIE2016optomechanic}
Barden, S.~C., Saviauk, A., and Winkler, R., ``{4MOST Metrology System Optical
  and Mechanical Design},'' {\em {Proc. SPIE}} {\bf 9908},  331 (2016).

\bibitem{HaynesSPIE2018fibrefeed}
Haynes, D.~M., Saviauk, A., Kelz, A., Jahn, T., Haynes, R., Plüschke, D.,
  Zscheyge, F., Brown, R., Piotrowski, J., Winkler, R., and Barden, S., ``4most
  fibre feed: performance and final design,'' {\em Proc. SPIE} {\bf 10702}
  (2018).

\bibitem{Winkler2024FTA}
Winkler, R., Stilz, I., Pramskiy, A., Sun, W., Roje, P., Liebner, T.,
  Sablowski, D., Frey, S., Zins, G., and Smedley, S., ``{The fibre target
  alignment process of the 4MOST instrument},'' in [{\em Software and
  Cyberinfrastructure for Astronomy VIII}{\nolinebreak\hspace{0.1em}]},  Ibsen,
  J. and Chiozzi, G., eds.,  {\bf 13101},  131010L, International Society for
  Optics and Photonics, SPIE (2024).

\bibitem{MandelSPIE2016control}
Mandel, H.~G., Pramskiy, A., Rothmaier, F.~M., Stilz, I., and Winkler, R.,
  ``{The 4MOST facility control software},'' in [{\em Software and
  Cyberinfrastructure for Astronomy IV}{\nolinebreak\hspace{0.1em}]},   {\bf
  9913},  991336, International Society for Optics and Photonics (2016).

\bibitem{CrouseDavid2016}
Crouse, D.~F., ``On implementing 2d rectangular assignment algorithms,'' {\em
  IEEE Transactions on Aerospace and Electronic Systems}~{\bf 52}(4),
  1679--1696 (2016).

\bibitem{astropy2022v5}
{Astropy Collaboration} and {Astropy Project Contributors}, ``The astropy
  project: Sustaining and growing a community-oriented open-source project and
  the latest major release (v5.0) of the core package,'' {\em apj}~{\bf 935},
  167 (Aug. 2022).

\bibitem{Tempel2020assign}
{Tempel, E.}, {Norberg, P.}, {Tuvikene, T.}, {Bensby, T.}, {Chiappini, C.},
  {Christlieb, N.}, {Cioni, M.-R. L.}, {Comparat, J.}, {Davies, L. J. M.},
  {Guiglion, G.}, {Koch, A.}, {Kordopatis, G.}, {Krumpe, M.}, {Loveday, J.},
  {Merloni, A.}, {Micheva, G.}, {Minchev, I.}, {Roukema, B. F.}, {Sorce, J.
  G.}, {Starkenburg, E.}, {Storm, J.}, {Swann, E.}, {Thi, W. F.}, {Traven, G.},
  and {de Jong, R. S.}, ``Probabilistic fibre-to-target assignment algorithm
  for multi-object spectroscopic surveys,'' {\em A\&A}~{\bf 635},  A101 (2020).

\end{thebibliography}
\bibliographystyle{spiebib} 

\end{document}